\documentclass[%
   12pt,
   twoside,
   tipotesi=scudo,
   numerazioneromana,
   ]{toptesi}

\ifPDFTeX
    \usepackage[utf8]{inputenc}%
    \usepackage[T1]{fontenc}
\fi

\DeclareMathOperator{\erfc}{erfc}
\DeclareMathOperator{\rect}{rect}
\DeclareMathOperator{\sinc}{sinc}
\DeclareMathOperator{\var}{var}
\DeclareMathOperator{\sign}{sign}
\DeclareMathOperator{\expt}{\mathbb{E}}
\DeclareMathOperator{\hilb}{\mathcal{H}}
\DeclareMathOperator{\kurt}{kurt}
\DeclareMathOperator{\clip}{\mathcal{Q}}

\preto\fullcite{\AtNextCite{\defcounter{maxnames}{99}}}

\ifPDFTeX %
    \usepackage{lmodern} %

\else %
    \usepackage{fontspec}
    \defaultfontfeatures{Ligatures=TeX}
    \usepackage{unicode-math}%
\fi

\makeindex[intoc]%

\usepackage{kantlipsum} %
\usepackage[%
    pdfpagemode={UseOutlines},
    bookmarksopen,
    pdfstartview={FitH},
    colorlinks,
    linkcolor={blue},
    citecolor={blue},
    urlcolor={blue}
  ]{hyperref}

\ifPDFTeX \usepackage{indentfirst}\fi
\begin{document}

\begin{ThesisTitlePage}
\PhDschoolLogo{Logo-ScuDo-blu} %
\ProgramName{Electrical, Electronics and Communications Engineering\\}
\CycleNumber{31\textsuperscript{th}}
\author{Dario Pilori}
\title{Advanced Digital Signal Processing Techniques for High-Speed Optical Communications Links}
\SupervisorNumber{2}
\SupervisorList{%
    Prof.~Gabriella Bosco, Supervisor\\
    Prof.~Roberto Gaudino, Co-supervisor}
\ExaminerList{%
Prof.~Magnus Karlsson, Referee, Chalmers Tekniska H\"{o}gskola AB\\
Dr.~Robert Killey, Referee, University College of London\\
Prof.~Maurizio Magarini, Politecnico di Milano\\
Dr.~Marco Secondini, Scuola Universitaria Superiore Sant'Anna Pisa\\
Prof.~Renato Orta, Politecnico di Torino}
\ExaminationDate{2019}

\Disclaimer{%
\noindent This dissertation has been submitted in partial fulfillment of the requirements for the degree of ``Dottore di Ricerca'' (Ph.D.)	at the Politecnico di Torino.
\\
\\
\noindent I hereby declare that, the contents and organization of this dissertation constitute my own original work and does not compromise in any way the rights of third parties, including those relating to the security of personal data.	
}

\Signature{%
\begin{flushright}
\parbox{0.5\textwidth}{\centering
Dario Pilori\\
Torino, 22 March 2019
}
\end{flushright}}

\CClicence{%
	This thesis is licensed under a Creative Commons Attribution 4.0 International License; see \url{https://creativecommons.org/licenses/by/4.0/}.
	The text may be reproduced, provided that credit is given to the original author.
}

\end{ThesisTitlePage}

\nomenclature[A]{$B_0$}{Reference optical bandwidth for OSNR}
\nomenclature[A]{$\mathrm{I}$}{In-Phase}
\nomenclature[A]{$f$}{Frequency}
\nomenclature[A]{$\mathrm{Q}$}{Quadrature}
\nomenclature[A]{$R_\textup{s}$}{Symbol rate (Baud)}
\nomenclature[A]{$T$}{Symbol duration}

\nomenclature[G]{$\alpha$}{Fiber attenuation (dB/km)}
\nomenclature[G]{$\beta_2$}{Fiber chromatic dispersion (ps/THz$\cdot$km)}
\nomenclature[G]{$\gamma$}{Fiber non-linear coefficient (1/W$\cdot$km)}
\nomenclature[G]{$\lambda$}{Wavelength}
\nomenclature[G]{$\xi$}{Carrier-to-Signal Power Ratio}

\nomenclature[Z]{ADC}{Analog-to-Digital Converter}
\nomenclature[Z]{AIR}{Achievable Information Rate}
\nomenclature[Z]{APSK}{Amplitude Phase Shift Keying}
\nomenclature[Z]{ASE}{Amplified Spontaneous Emission}
\nomenclature[Z]{AWGN}{Additive White Gaussian Noise}
\nomenclature[Z]{BER}{Bit Error Ratio}
\nomenclature[Z]{BICM}{Bit-Interleaved Coded Modulation}
\nomenclature[Z]{BPD}{Balanced Photodiode}
\nomenclature[Z]{BPS}{Blind Phase Search}
\nomenclature[Z]{CCDM}{Constant-Composition Distribution Matcher}
\nomenclature[Z]{CD}{Chromatic Dispersion}
\nomenclature[Z]{CMA}{Constant Modulus Algorithm}
\nomenclature[Z]{CMOS}{Complementary Metal-Oxide Semiconductor}
\nomenclature[Z]{CPE}{Carrier Phase Estimator}
\nomenclature[Z]{CUT}{Channel Under Test}
\nomenclature[Z]{CSPR}{Carrier-to-Signal Power Ratio}
\nomenclature[Z]{DAC}{Digital-to-Analog Converter}
\nomenclature[Z]{DC}{Data-Center}
\nomenclature[Z]{DCF}{Dispersion-Compensating Fiber}
\nomenclature[Z]{DCI}{Data-Center Interconnects}
\nomenclature[Z]{DD}{Direct Detection}
\nomenclature[Z]{DFB}{Distributed-FeedBack laser}
\nomenclature[Z]{DFE}{Decision Feedback Equalization}
\nomenclature[Z]{DM}{Distribution Matcher}
\nomenclature[Z]{DML}{Directly Modulated Laser}
\nomenclature[Z]{DMT}{Discrete Multitone}
\nomenclature[Z]{DP-NLSE}{Dual-Polarization Non-Linear Schr\"{o}dinger Equation}
\nomenclature[Z]{DS}{Dispersion Shifted fiber}
\nomenclature[Z]{DSB}{Dual Side-Band}
\nomenclature[Z]{DSL}{Digital Subscriber Line}
\nomenclature[Z]{DSP}{Digital Signal Processing}
\nomenclature[Z]{EAM}{Electro-Absorption Modulator}
\nomenclature[Z]{ECL}{External Cavity Laser}
\nomenclature[Z]{EDFA}{Erbium Doped Fiber Amplifier}
\nomenclature[Z]{EGN}{Enhanced Gaussian-Noise model}
\nomenclature[Z]{ER}{Extinction Ratio}
\nomenclature[Z]{FBG}{Fiber Bragg Grating}
\nomenclature[Z]{FDM}{Frequency Division Multiplexing}
\nomenclature[Z]{FEC}{Forward Error Correction}
\nomenclature[Z]{FFE}{Feed-Forward Equalizer}
\nomenclature[Z]{FIR}{Finite Impulse Response}
\nomenclature[Z]{GEQ}{Gain Equalizer filter}
\nomenclature[Z]{GMI}{Generalized Mutual Information}
\nomenclature[Z]{GN}{Gaussian-Noise model}
\nomenclature[Z]{GPGPU}{General-Purpose GPU computing}
\nomenclature[Z]{GS}{Geometrical Shaping}
\nomenclature[Z]{IM}{Intensity Modulation}
\nomenclature[Z]{LAN}{Local Area Network}
\nomenclature[Z]{LLR}{Log-Likelihood Ratio}
\nomenclature[Z]{LMS}{Least Mean Squares}
\nomenclature[Z]{LO}{Local Oscillator}
\nomenclature[Z]{MI}{Mutual Information}
\nomenclature[Z]{MIMO}{Multiple Input-Multiple Output}
\nomenclature[Z]{ML}{Maximum Likelihood}
\nomenclature[Z]{MLSE}{Maximum Likelihood Sequence Estimator}
\nomenclature[Z]{MPO}{Multi-fiber Push On}
\nomenclature[Z]{MZM}{Mach-Zehnder Modulator}
\nomenclature[Z]{NGMI}{Normalized Generalized Mutual Information}
\nomenclature[Z]{NLI}{Non-Linear Interference}
\nomenclature[Z]{NLPN}{Non-Linear Phase Noise}
\nomenclature[Z]{NRZ}{Non Return to Zero}
\nomenclature[Z]{NZDSF}{Non-Zero Dispersion-Shifted Fiber}
\nomenclature[Z]{OBPF}{Optical Band-Pass Filter}
\nomenclature[Z]{OFDM}{Orthogonal Frequency Division Multiplexing}
\nomenclature[Z]{OH}{Overhead}
\nomenclature[Z]{OMA}{Optical Modulation Amplitude}
\nomenclature[Z]{OOK}{On-Off Keying}
\nomenclature[Z]{OSNR}{Optical Signal-to-Noise Ratio}
\nomenclature[Z]{PAM}{Pulse Amplitude Modulation}
\nomenclature[Z]{PAPR}{Peak-to-Average Power Ratio}
\nomenclature[Z]{PAS}{Probabilistic Amplitude Shaping}
\nomenclature[Z]{PBS}{Polarization Beam Splitter}
\nomenclature[Z]{PC}{Polarization Controller}
\nomenclature[Z]{PD}{Photodiode}
\nomenclature[Z]{PDF}{Probability Density Function}
\nomenclature[Z]{PLL}{Phase-Locked Loop}
\nomenclature[Z]{PM}{Polarization Multiplexed}
\nomenclature[Z]{PMD}{Polarization Mode Dispersion}
\nomenclature[Z]{PON}{Passive Optical Network}
\nomenclature[Z]{PRBS}{Pseudo Random Bit Sequence}
\nomenclature[Z]{PS}{Probabilistic Shaping}
\nomenclature[Z]{PSCF}{Pure-Silica Core Fiber}
\nomenclature[Z]{QAM}{Quadrature Amplitude Modulation}
\nomenclature[Z]{ROP}{Received Optical Power}
\nomenclature[Z]{RIN}{Relative Intensity Noise}
\nomenclature[Z]{RRC}{Root Raised Cosine}
\nomenclature[Z]{RS}{Reed-Solomon code}
\nomenclature[Z]{SBS}{Stimulated Brillouin Scattering}
\nomenclature[Z]{SCM}{Subcarrier Multiplexing}
\nomenclature[Z]{SDM}{Spatial Division Multiplexing}
\nomenclature[Z]{SFP}{Small Form-factor Pluggable transceiver}
\nomenclature[Z]{SMF}{Single-Mode Fiber}
\nomenclature[Z]{SNR}{Signal-to-Noise Ratio}
\nomenclature[Z]{SOI}{Silicon On Insulator}
\nomenclature[Z]{SRO}{Symbol Rate Optimization}
\nomenclature[Z]{SRS}{Stimulated Raman Scattering}
\nomenclature[Z]{SSB}{Single Side-Band}
\nomenclature[Z]{SSBI}{Signal-Signal Beating Interference}
\nomenclature[Z]{SSFM}{Split-Step Fourier Method}
\nomenclature[Z]{TDECQ}{Transmitter and Dispersion Eye Closure Quaternary}
\nomenclature[Z]{VOA}{Variable Optical Attenuator}
\nomenclature[Z]{VSB}{Vestigial Side-Band}
\nomenclature[Z]{WDM}{Wavelength Division Multiplexing}
\nomenclature[Z]{XPM}{Cross Phase Modulation}

\subject{Advanced Digital Signal Processing Techniques for High-Speed Optical Communications Links.}

\summary%

The main topic of this work is the application of advanced Digital Signal Processing (DSP) techniques to high data-rate optical links.

The first part of the thesis is devoted to direct-detection systems, with a specific focus on data-center (DC) communications. These links, until recently, used
very simple On-Off Keying modulation. However, the strong push towards higher data-rates introduced the use of some DSP algorithms in the DC. 
First, this thesis introduces a novel architecture for 400-Gbit/s Intra-DC (shorter than 2 km) applications. This architecture uses each fiber pair in both directions, doubling
the per-laser and per-fiber capacity. Crosstalk is reduced by using a short feed-forward adaptive equalizer and a small frequency shift between lasers transmitting
over the same fiber. This system, after an initial theoretical evaluation, was successfully tested with an experiment. Then,
Intra-DC links were analyzed, i.e. links interconnecting different DCs in the same region. The target distance was 80 km. For this purpose, Single Side-Band (SSB) modulation
was compared with Intensity Modulation/Direct Detection schemes over dispersion-uncompensated links with single-photodiode receivers. Using, as an application example,
the DMT modulation format, the comparison was carried over different span lengths. It was found that, for distances longer than 10 km, SSB has a significant advantage over IM/DD systems.

The second part of the thesis focuses on coherent systems. In particular, it focused on constellation shaping techniques, which have been recently
proposed for optical communications. To perform a fair analysis, shaped constellations were compared, at the same net data rates, with standard lower-cardinality
QAM constellations. It was found that, over a linear channel, the gain of constellation shaping is approximately 25\%. Afterwards, this thesis discusses the
non-linear properties of the optical channel, which are known to be enhanced by constellation shaping. In particular, it focuses on the generation of
Non-Linear Phase Noise (NLPN) at different system conditions, using both standard QAM and shaped constellations. It was found that, in some scenarios, such as low symbol rates
or low dispersion fibers, NLPN significantly degrades the performance of shaped constellations. Then, this thesis proposes some compensation techniques, which
are able to partially mitigate the effect of NLPN in those conditions.

\emptypage %
\acknowledgements%

The development of any thesis, especially a PhD thesis, cannot be done without the support of many people, in any stage of my research career.
First of all, I would not be even close to getting a PhD without the constant support of my family.

For my PhD path, I would first like to thank my advisors Gabriella Bosco and Roberto Gaudino, for their invaluable teachings and support throughout the entire path. Then, I would
like to thank the entire OptCom research group, for the numerous discussions that allowed me learning many things. In particular, I would like to thank
Andrea Carena, Vittorio Curri and Pierluigi Poggiolini. Then, a big thank to all the PhD students and post-docs that were (and are) part of the group. From every discussion
with them I learned many small (and big) new ideas. In particular, I thank Mattia Cantono, Alessio Ferrari, Fernando P. Guiomar, Marianna 
Hovsepyan and Emanuele Virgillito.
Outside the OptCom group, I would like to thank the technical staff of the Politecnico. By collaborating with them, I got to learn a different (but very useful) aspect
of the work inside a University. In particular, I thank Marco Bertino and Paolo Squillari.

During my PhD, I had precious collaborations with the industry and the academia. 
In particular, I would like to thank Fabrizio Forghieri from Cisco Photonics for the discussions during the weekly conference calls. I would also want to thank the 
IPQ institute at the Karlsruhe Institute of Technology for hosting me during my three-month abroad research period. In particular, I thank 
M.M.H. Adib, Christoph F\"{u}llner and Sebastian Randel.

Going backwards, I would not have started a PhD without the precious teachings of the Nokia Bell Labs staff in Crawford Hill, NJ, during my two internships. In particular, I would
like to thank my advisor, Sebastian Randel, and Damiano Badini, Sergi Caelles, Junho Cho, S. Chandrasekhar, Ren\'{e} Essiambre, Bob Jopson, Greg Raybon, Peter Winzer.
Then, last but not the least, Maurizio Magarini from Politecnico di Milano for giving me the opportunity to do my Master thesis in that amazing place.

And, of course, the biggest thank of all goes to Serena.

\tablespagetrue\figurespagetrue %

\allcontents

\chapter*{List of publications}
During the Ph.D. program, the following scientific contributions were published on journals and conferences:
\section*{International peer-reviewed journals}
\begin{enumerate}
	\item \fullcite{Pilori:2018}
	\item \fullcite{Pilori:JLT2018} [invited from top-scoring OFC 2017 contribution]
	\item \fullcite{Pilori_JLT:16}	
	\item \fullcite{Cantono:JOCN:2019} [invited from top-scoring OFC 2018 contribution]
	\item \fullcite{Cantono:2018}
	\item \fullcite{Kashef:2018}
\end{enumerate}

\section*{International conferences}
\begin{enumerate}
	\item \fullcite{Pilori:OFC2019}
	\item \fullcite{Ferrari:OFC2019}
	\item \fullcite{Adib:2018}
	\item \fullcite{Poggiolini:ECOC2018}
	\item \fullcite{Pilori:ICTON2018} [invited]
	\item \fullcite{Pilori:OFC2018}
	\item \fullcite{Cantono:OFC2018}
	\item \fullcite{Pilori:IPC2017}
	\item \fullcite{Guiomar:2017}
	\item \fullcite{Pilori:FFSS2017}
	\item \fullcite{Pilori:2017}
	\item \fullcite{Bertignono:2017} [top-scoring paper]	
	\item \fullcite{Bosco:2017} [invited]
\end{enumerate}

\mainmatter %

\chapter{Introduction}

    \graphicspath{{Chapter1/}}
In the past 20 years, the total world Internet traffic has grown by an enormous rate ($\sim1\,000\times$) \cite{Winzer:2018}. 
To sustain this growth, all components of the Internet infrastructure have steadily evolved. Focusing on the raw speed of point-to-point optical links, it is interesting
to see the compound annual growth rate \cite[Fig. 4]{Winzer:2017} of the per-$\lambda$ interface rate and its relative symbol rate. While the interface rate
has experienced a $\sim20\%$ annual growth rate, the symbol rates have grown by just $\sim10\%$. This means that, to sustain this growth, also the spectral efficiency
had to steadily evolve.

Among the different technological improvements, the introduction of Digital Signal Processing (DSP) techniques, initially taken from the wireless communications community,
has given a tremendous boost to spectral efficiency. This has been possible thanks to the improvements in the speed of CMOS circuits \cite[Sec. 2.3]{Winzer:2018}. The first
DSP application in optical communications was on high-speed long-haul links. In this scenario, DSP enabled coherent detection and polarization multiplexing, allowing a dramatic
increase in spectral efficiency. More recently, DSP has been also applied to shorter links, such as inter data-center communications, or even intra data-center links \cite{Zhong:2018}.
Therefore, introduction of DSP in optical communications have been one of the key technological improvements that allowed such speed increase.

Consequently, the purpose of this thesis is to investigate the application of DSP to different optical communications applications, from short-reach ($<2$ km) data-center 
interconnects, to long-haul links. These links can be divided into two large categories, based on the receiver structure: direct-detection and coherent
detection. Indeed, the structure, the requirements, and (most importantly) the DSP algorithms radically differ in these categories. 

For this reason, this thesis is divided into two Parts. 
Part \ref{part:dd} is devoted to direct-detection systems. After an introduction in Chapter \ref{ch:dcarchitect}, Chapter \ref{ch:bidir} will present a novel DSP-enabled architecture
for $<2$ km links, while Chapter \ref{ch:ssb} will focus on $\sim80$ km inter data-center links. Part \ref{part:coh} will be instead devoted to coherent systems. In particular, it will
be focused on the application of the constellation shaping DSP technique to long-haul links. After an introduction in Chapter \ref{ch:coherent}, Chapter \ref{ch:shaping} will be focused
on constellation shaping on an Additive White Gaussian Noise (AWGN) channel, while Chapter \ref{ch:phnoise} will instead focus on the generation (and compensation) of non-linear phase noise by
fiber non-linear effects.

\part{Direct-Detection Systems}\label{part:dd}

\chapter{Introduction to Direct-Detection Systems}\label{ch:dcarchitect}

\graphicspath{{Chapter2/}}

This Chapter introduces direct-detection systems, which are widely-adopted in short reach ($<120$ km) links, 
typically present in data-centers (DC). After a general introduction to DC networking, presenting the most common standards, 
attention will be focused on two different scenarios: connections within a DC (Intra-DC) and between different DC (Inter-DC).

Afterwards, two Chapters (\ref{ch:bidir} and \ref{ch:ssb}) will present the contribution of this thesis to these topics, presenting a novel spatial-multiplexing architecture for Intra-DC (Chapter \ref{ch:bidir}) and a comparison between intensity modulation and single-sideband transmission for Inter-DC (Chapter \ref{ch:ssb}).

\section{Introduction}\label{sec:2:intro}
During past decade, research in high-speed optical communications was focused on long-distance links, 
which used to be the network segment where traffic was steadily growing \cite{Desurvire:2006}. This lead to the development of coherent transmission and detection schemes, which were able to dramatically increase network capacity.
At that time, almost all short distance links employed simple NRZ intensity modulation / direct-detection
at relatively low data-rates \cite{std:ethernet2002}, with very simple (if any) digital signal processing involved.

Nowadays, new applications such as cloud computing, video-on-demand, virtual and augmented reality demand a large capacity increase on
links in the DC \cite{Cheng:2018}. According to \cite[Appendix B]{CiscoGCI:2018}, in 2021 93.9\% of the total IP traffic on the Internet will be between end users and DC. Moreover, this traffic accounts only for 14.9\% of the total DC-related traffic, while the rest will be between DC (13.6\%) and inside each DC (71.5\%). Therefore, more sophisticated architectures, modulation formats and signal processing are required to cope with this traffic increase.

In this particular scenario, there are challenges that are quite unique in the field of optical communications \cite{Zhong:2018}. 
As opposed to long-distance links, single interface data-rate is only one of the factors that needs to be improved. 
Other elements, such as cost, form factor, power consumption and latency are
extremely important for this scenario, and they need to be carefully taken into account when developing new algorithms and architectures for short-reach applications.

In this Chapter, it will be given an overview of current standard and interfaces for short-reach direct-detections applications, focusing
on the DC environment, which can be divided into two main groups:
\begin{itemize}
\item Intra-DC: connections inside a data-center, usually shorter than $2$ km.
\item Inter-DC: connections between different data-centers in a region, in the range of $2-120$ km. This application can also be called Data-Center Interconnect (DCI).
\end{itemize}

Solutions adopted for these two groups are quite different.
Intra-DC links, given the limited distance, do not need optical amplification,
therefore they can transmit in the O-band and avoid chromatic dispersion. 
Inter-DC links, instead, require optical amplifications, forcing the use of C-band and thus facing the problem of chromatic dispersion. 
Because of this, current (and future) solutions for these applications will radically differ. Consequently, 
this Chapter will be divided into two Sections: \ref{sec:2:intra}, and \ref{sec:2:inter}, dealing (respectively) with Intra-DC and Inter-DC links.

\section{Intra-DC links}\label{sec:2:intra}

\subsection{Ethernet standards and form factors}\label{sec:2:ethernetstd}
\begin{figure}
	\begin{subfigure}[b]{0.48\textwidth}
		\centering \includegraphics[width=\textwidth]{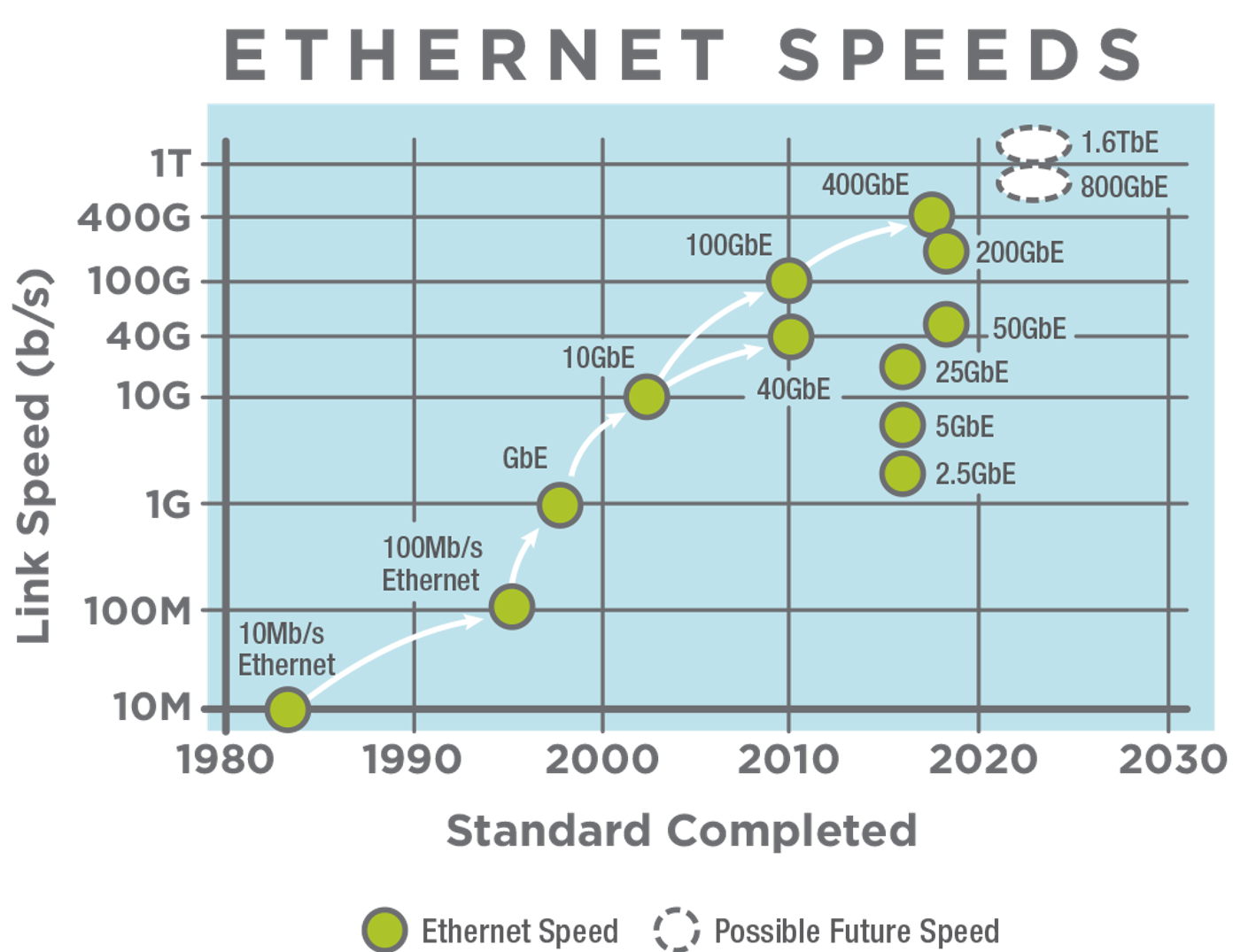}
		\caption{}\label{subfig:ethspeedstd}
	\end{subfigure}
	\begin{subfigure}[t]{0.48\textwidth}
		\centering \includegraphics[width=\textwidth]{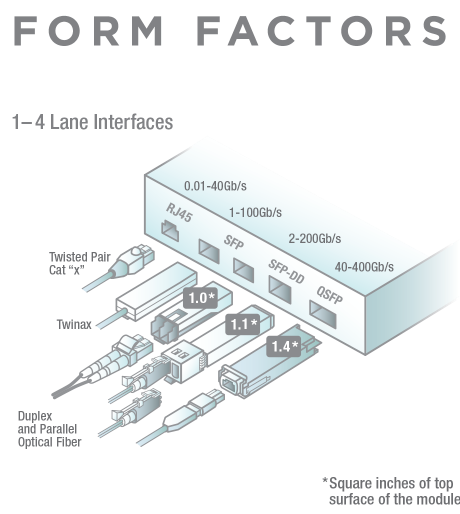}
		\caption{}\label{subfig:ethformfactor}
	\end{subfigure}
	\caption{Evolution of Ethernet interfaces data-rate (a) and current form factors (b). Power consumption of a form factor is approximately proportional
		to its top surface. Source: Ethernet Alliance.}\label{fig:ethspeedstd}
\end{figure}
Inside the DC, most links are based on well-defined standards. The most common 
standards are Ethernet\footnote{\url{https://ethernetalliance.org/}}, Fibre Channel\footnote{\url{https://fibrechannel.org/}} and
InfiniBand\footnote{\url{https://www.infinibandta.org/}}. In general, the main difference between these standards is the application:
InfiniBand is mainly adopted for high-performance computing, Fibre Channel for storage and Ethernet for IP
traffic. However, the differences primarily lay at higher networking layers, while in the physical layer the standards are similar.

Recently, adoption of Ethernet has been steadily growing, even outside its standard use cases.
Since a study of the networking aspects of these standards is out of the scope of this thesis, we will choose
Ethernet as a reference standard, keeping in mind that the physical layer differences with the other standards
are small. 

Fig. \ref{subfig:ethspeedstd} shows the evolution, over the years, of the data-rates of the Ethernet standards. 
The oldest standard, released in 1983, used a single copper coaxial cable at a data-rate of 10 Mbit/s. 
Since then, line-rate increased exponentially, up to the most recent standard of 400 Gbit/s, released at the end of 2017. 
At the same time, technology improvements were able to reduce the form factor (hence, power consumption) of the transceiver. 
For instance, the first 100GBASE-LR4 interfaces, which achieve a data-rate of 100 Gbit/s over 10 km of Single-Mode Fiber (SMF), used the CFP form factor, whose size is $82\times13.6\times144.8$ mm and maximum power consumption $20$ W\cite{Cisco:100GBASECFP}.
Modern 100GBASE-LR4 transceivers use the QSFP28 form factor, whose size is  $71.78\times 8.50 \times 18.35$ mm, with a power consumption of $4$ W\cite{Cisco:100GBASEQSFP}.

Transceivers are usually pluggable with standardized form factors. 
Current form factors of modern Ethernet standards are shown in Fig. \ref{subfig:ethformfactor}. At speed lower (or equal) than 1 Gbit/s, interfaces mainly use copper twisted-pair cables with the RJ-45 corrector. 
Above 1 Gbit/s, and up to 100 Gbit/s, different variants of the SFP form factor are used. Then, for higher-speed interfaces, the 
SFP-DD\footnote{\url{http://sfp-dd.com/}} and QSFP-DD\footnote{\url{http://www.qsfp-dd.com/}} form factors are used, which are slightly larger (and consume more power) than SFP.

\subsubsection{400-Gbit/s Ethernet}
At the time of writing this thesis, the highest-speed published Ethernet standard is IEEE 802.3bs-2017 \cite{std:ethernet2017}, released
in 2017, which defines several standards at 200 Gbit/s (200GBASE) and 400 Gbit/s (400GBASE), which will be analyzed in the following Section.

\begin{table}
\centering
\begin{tabular}{c c c c c c}
\toprule
Name & Medium & WDM  & Symbol rate & Modulation & Reach \\
     &		  & channels & (GBaud) & format & (km) \\
\midrule
400GBASE-SR16 & $16\times$ MMF & 1 & 26.5625 & NRZ & 0.1  \\
400GBASE-DR4 & $4\times$ SMF & 1 & 53.125 & PAM-4 & 0.5 \\
400GBASE-FR8 & $1\times$ SMF & 8 & 26.5625 & PAM-4 & 2 \\
400GBASE-LR8 & $1\times$ SMF & 8 & 26.5625 & PAM-4 & 10 \\
\bottomrule
\end{tabular}
\caption{Physical Layer specifications for 400 Gbit/s Ethernet \cite{std:ethernet2017}.}\label{tab:400g_eth}
\end{table}
Focusing on the 400 Gbit/s standards, 802.3bs defines different physical layer specifications over optical fiber, depending on the distance,
which are summarized in Table \ref{tab:400g_eth}. Moreover, the list of 400GBASE standards is quickly evolving, and new standards are
expected to be released in the future.
All the standards use parallel channels to obtain 400 Gbit/s, either using multiple media (SR16 and DR4) or WDM (FR8 and LR8). 
Except for SR16, the PAM-4 modulation format is used to increase spectral efficiency and reduce the number of parallel lines. Still, longer-distance standards require 8 parallel paths, which represents a significant source of complexity and cost.

From a signal-processing perspective, the most challenging standards are longer-reach ones (FR8 and LR8), since they have to deal with
tighter power budgets. From an architectural point of view, the two standards are identical, except from the power budget (for details,
we refer the reader to \cite{std:ethernet2017}).

\begin{figure}
\centering \includegraphics[width=0.7\textwidth]{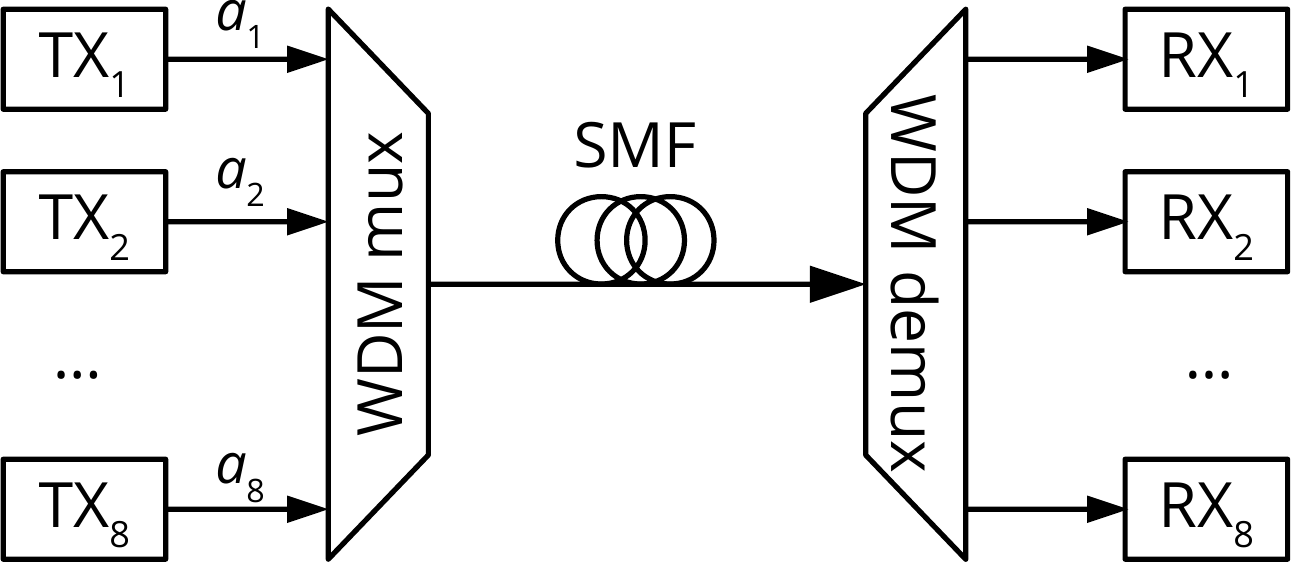}
\caption{Basic block scheme of the 400GBASE-FR8 standard, based on 8 WDM channels, each modulated with 50 Gbit/s PAM-4. For simplicity, the diagram shows transmission on a single direction.}
\label{fig:400gbase_block}
\end{figure}

A basic block scheme of FR8/LR8 standards is shown in Fig. \ref{fig:400gbase_block}. The line speed of 400 Gbit/s is achieved using
8 parallel WDM channels, each using the PAM-4 modulation format at 50 Gbit/s. Channels are spaced by 800 GHz in the O-band,
and each channel has a bandwidth of approximately 367.7 GHz to allow some tolerance in the laser wavelength. Channel central frequencies are taken from the 100-GHz DWDM grid \cite{std:dwdm}. The central freqiencies (and wavelengths) of this grid are shown in Table \ref{tab:lanwdmgrid}; this grid is also called LAN-WDM (or LWDM) grid.
\begin{table}
	\centering
	\begin{tabular}{c c c}
		\toprule
		Channel & Frequency & Wavelength \\
		 & (THz) & (nm) \\
		\midrule
		1 & 235.4 & 1273.54 \\
		2 & 234.6 & 1277.89 \\
		3 & 233.8 & 1282.26 \\
		4 & 233.0 & 1286.66 \\
		5 & 231.4 & 1295.56 \\
		6 & 230.6 & 1300.05 \\
		7 & 229.8 & 1304.58 \\
		8 & 229.0 & 1309.14 \\
		\bottomrule
	\end{tabular}
	\caption{WDM grid for the 400GBASE-FR8 standard (LAN-WDM).}\label{tab:lanwdmgrid}
\end{table}

\subsubsection{Forward Error Correction}
The choice of a proper FEC algorithm for this application is not trivial, since it must combine high performance with low latency, power consumption and silicon footprint. For these reasons, soft-decision FEC, widely deployed in long-haul links, is not adopted, relying
instead on simpler hard-decision FEC.

For 100 Gbit/s Ethernet, the IEEE has standardized two different FEC codes, KR4 and KP4. 
KR4 is a Reed-Solomon RS(528,514) code, specifically targeted for NRZ modulation, while KP4 is a more powerful RS(544,514),
and it is targeted for PAM-4 modulation. For 400 Gbit/s, the IEEE adopted the KP4 code.

This code achieves a post-FEC BER of $10^{-15}$ with a pre-FEC BER of approximately $3\times10^{-4}$, with an electrical
 coding gain of $6.64$ dB \cite{techrep:400gfec}. More details on FEC for short-reach applications can be found in \cite[Chapter 6]{phd:shoaib}.

\subsection{Channel model}\label{sec:2:chmodel}
For these applications, fiber channel model is quite unique, given the use of intensity modulation/direct detection and the very short propagation distances involved.
The most important difference with respect to longer-distance communications is the use of O-band ($\sim1310$ nm) instead of the popular C-band ($\sim1550$ nm), which means
that the attenuation and chromatic dispersion profiles will be different.

In this Section, the main characteristics of the channels are summarized, focusing on the differences with respect to C-band.

\subsubsection{Chromatic dispersion}
The main reason for the use of O-band is the low value of chromatic dispersion, which allows avoiding the need of dispersion compensation, at least
at the symbol rates of interest.

\begin{figure}
\centering
\includegraphics[width=0.6\textwidth]{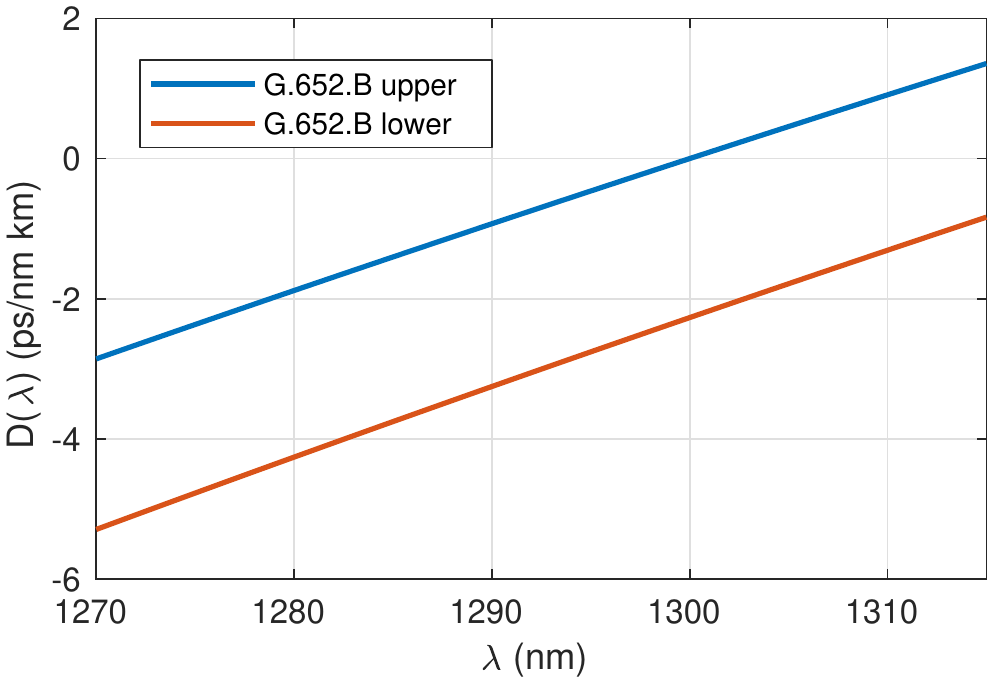}
	\caption{Maximum chromatic dispersion of G.652.B optical fiber (SMF) in the O-band.}\label{fig:smfdispersion}
\end{figure}

However, even if the value of chromatic dispersion is low, it is not zero, and it may affect transmission at high symbol rates. 
For single-mode fiber, the range of chromatic dispersion is defined by ITU-T recommendation G.652 \cite{std:smf}. 
As an example, in Fig. \ref{fig:smfdispersion} it is shown the dispersion range of a G.652.B SMF fiber in the O-band, obtained using \cite[eq. (6-1)]{std:smf}.

For the 400GBASE-FR8 standard (2 km SMF), the first WDM channel (with central wavelength $1373.54$ nm) experiences the strongest chromatic dispersion, which in the worst case
can be $-9.85$ ps/nm. In \cite{Eiselt:2016}, the authors measured the CD tolerance of a $26$-GBaud PAM-4 signal, and found $1$ dB penalty with $180$ ps/nm of dispersion.
Therefore, we can conclude in this scenario dispersion can be completely neglected.

\subsubsection{Attenuation}
In the O-band, fiber attenuation is slightly higher than C-band. According to G.652 recommendation, attenuation in the O-band must be lower than $0.4$ dB/km, while in the C-band
must be lower than $0.35$ dB/km. Modern SMF fibers have smaller values. For instance, measurements of a modern (G.652.D) SMF fiber performed in our lab found an attenuation of
$0.18$ dB/km at $1550$ nm and $0.33$ dB/km at $1310$ nm.

Nevertheless, given the limited propagation distance ($<10$ km for all 400 Gbit/s standards), propagation loss represents a small contribution to the overall link
budget. For instance, the total link power budget for 400GBASE-FR8 ($2$ km SMF) is $7.4$ dB, and fiber propagation accounts only for $1$ dB (which assumes $0.5$ dB/km attenuation).

\subsubsection{Polarization Mode Dispersion}
According to G.625 recommendation, the maximum value of PMD is $0.20~\mathrm{ps}/\sqrt{\mathrm{km}}$, even if modern SMF fibers have smaller values \cite{Breuer:2003}.

Nevertheless, in \cite{Eiselt:2016} the authors measured a $1$-dB penalty in $26$-GBaud PAM-4
for DGD values greater than $15$ ps, which is not realistic for the distances of interest.
Therefore, we conclude that also PMD can be neglected in this scenario. 

\subsubsection{Noise}
Since optical amplification is not employed, the main source of noise is receiver thermal noise.
For this thesis, laser relative intensity noise (RIN), which can be a significant noise source, especially for Directly-Modulated Lasers (DMLs) is neglected.
Regarding thermal noise, the performance metric is the Received Optical Power (ROP).
The variance (in current) of noise can be evaluated from the equivalent-input noise power spectral density $i^2_\textup{n}$, reported on receivers datasheets, with
\begin{equation}\label{eq:pintiasnr}
\sigma^2_n = i_\textup{n}^2 B \qquad [\mathrm{A}^2]
\end{equation}
where $B$ the electrical bandwidth of the receiver.

\subsection{Modulation formats and signal processing}
Intra-DC systems use simple intensity modulation, which, in the optics domain, basically means using Pulse Amplitude Modulation (PAM).
Since legacy NRZ-OOK is equivalent to PAM-2, this Section will only deal with PAM-$M$ modulation without any loss of generality.

\subsubsection{Pulse Amplitude Modulation}
\begin{figure}
	\centering
	\includegraphics[width=0.4\textwidth]{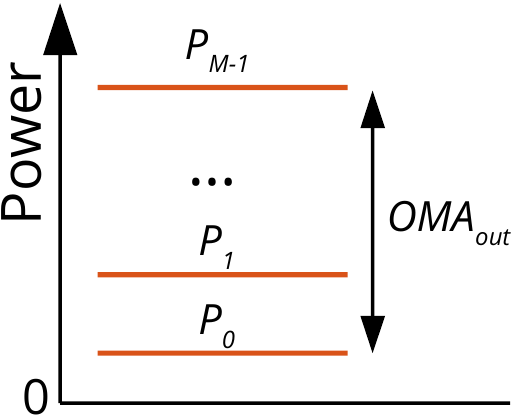}
	\caption{Power levels of a PAM-$M$ transmission with the definition of outer OMA.}\label{fig:pamoma}
\end{figure}

With PAM-$M$, the transmitter encodes bits in $M$ different amplitude levels, where $M=2^b$ is a power of two, and $b$ the number of transmitted bits per symbol.
Fig. \ref{fig:pamoma} shows an example of PAM-$M$ levels, ranging from the first level $P_0$ to the last $P_{M-1}$. From this definition, the average transmit
optical power $P$ is trivial to evaluate
\begin{equation}
P = \frac{1}{M}\sum_{i=0}^{M-1} P_i
\end{equation}
However, optical power does not fully characterize the system, because the Bit Error Ratio (BER) depends on the relative spacing between levels. Therefore, the
Ethernet standards use another parameter, called outer Optical Modulation Amplitude (OMA), defined as the difference (in linear scale) between the lowermost
and uppermost levels:
\begin{equation}
\textup{OMA}_\textup{out} = P_{M-1}-P_0
\end{equation}
This parameter is linked to the average optical power by means of another parameter, called Extinction Ratio (ER), defined as the \emph{ratio} between the lowermost and 
uppermost levels $\textup{ER} = P_{M-1}/P_0$. Assuming equispaced levels, the relation is
\begin{equation}
P = \textup{OMA}_\textup{out} \frac{\textup{ER} +1 }{2(\textup{ER} -1 )}
\end{equation}

\subsubsection{Bit Error Ratio of PAM}
Using the definition of outer OMA and PAM detection theory \cite[Ch. 4]{proakis2007digital}, the Symbol Error Ratio (SER) of an IM/DD PAM system
limited by electrical receiver noise can be evaluated from the distance between the equispaced levels:
\begin{equation}
d = \frac{\textup{OMA}_\textup{out}}{M-1}
\end{equation}
By substituting this definition into the expression of the Symbol Error Ratio of PAM, the SER can be calculated as:
\begin{equation}\label{eq:pamseroma}
P_s = \frac{M-1}{M} \erfc\left[ \frac{  R_d \cdot \textup{OMA}_\textup{out} }{ \sqrt{8 B} (M-1) i_\textup{n}  } \right]
\end{equation}
where $R_d$ is the photodiode responsitivity (expressed in A/W). 
Note that, in this expression, the argument of the \texttt{erfc} function is directly proportional to the received \emph{optical} power.

\begin{figure}
	\centering
	\includegraphics[width=0.6\textwidth]{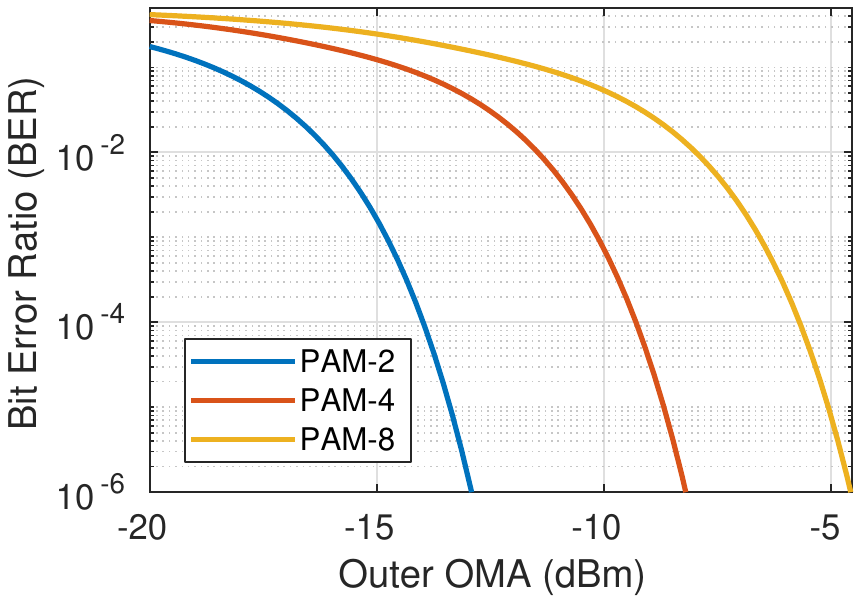}
	\caption{Bit Error Ratio of different PAM modulation formats as a function of the outer OMA.}\label{fig:pamber}
\end{figure}
Fig. \ref{fig:pamber} shows an example of BER (assuming Gray mapping) as a function of the outer OMA for $26.5625$-GBaud PAM transmission. 
Photodiode responsivity was set to $0.5$ A/W and noise current to $16.5~\mathrm{pA}/\sqrt{\mathrm{Hz}}$. Asymptotically, a one-bit
increase in the modulation format requires a $\sim3$ dB increase of the OMA.

\subsubsection{Digital Signal Processing}
IM/DD systems are inherently very simple, and they are used to operate without any DSP algorithm. 
However, with the increase of symbol rate and multilevel modulation, limited bandwidth of components starts to become an issue, requiring
some sort of channel equalization at the receiver.

In literature, there have been presented several equalization techniques \cite{Zhong:2018}. Some of them, such as Maximum Likelihood
Sequence Estimator (MLSE) are able to dramatically improve transmission in presence of strong bandwidth limitations, but its complexity
may be overkill for Intra-DC applications.

To have an overview of the current commercial state-of-the-art, 
the Ethernet standard uses a $5$-tap T-spaced feed-forward equalizer (FFE) for testing the Transmitter and Dispersion Eye Closure Quaternary (TDECQ)
of 400GBASE-FR8 and LR8 standards. For the implementation of the actual receiver, the IEEE does not give any recommendation. 
In presence of strong bandwidth limitations, non-linear Decision Feedback Equalizers (DFE) are commonly employed \cite{Zhong:2018}.

\section{Inter-DC connections}\label{sec:2:inter}
Modern data-centers are based on distributed architectures, where there are several DC inside the same region to provide scalability and redundancy \cite{Nagarajan:2018}.
This architecture represents a radical change with respect to the traditional ``mega data-center'' approach, and it has driven market demand to high-speed
connectivity between those DC. Since this Inter-DC scenario is more similar to the long-haul scenario, standard coherent line-cards may be adopted to deliver the required capacity. However,
Inter-DC connections have stricter requirements in term of latency, cost and power consumption which makes coherent long-haul systems unfeasible. 

In this scenario, as opposed to Intra-DC, there are not well-defined standards. This, on one hand, gives more freedom to the system designer. On the other hand,
it makes it difficult to retrieve the technical specifications of current Inter-DC systems. 
Therefore, this Section will be devoted to the channel model and a brief comparison of the
proposed modulation techniques.

\subsection{Channel model}
As a generally accepted definition, Inter-DC encompasses links between $10$ and $120$ km. However, links that are shorter than $\sim40$ km can be treated as ``long'' Intra-DC links, since they
can employ the same modulation formats and architectures as Intra-DC. For longer distances, link budget become too tight, requiring optical amplification and moving the transmission
wavelengths to the C-band. This means that some forms of dispersion compensation are necessary.

The use of optical amplification changes the type of noise that is impairing an Inter-DC link. While Intra-DC links are mainly limited by receiver thermal noise,
optically-amplified Inter-DC links are limited by ASE noise inserted by the amplifiers. Therefore, the main performance metric becomes the OSNR, defined as power
of the signal (in two polarizations) divided by the power of ASE. Consequently, this scenario becomes similar to a long-haul transmission system.
The main difference is PMD, which, as for the same reasons explained in Sec. \ref{sec:2:chmodel}, can be neglected.

\subsection{Coherent vs. intensity modulation}\label{sec:2:cohint}
Given the channel model, for Inter-DC both coherent and IM/DD systems can be adopted. The goal of this Section is to perform a comparison on the key differences
between the two schemes, focusing on Inter-DC applications. Part of this Section has been presented at ICTON 2018 conference as an invited presentation \cite{Pilori:ICTON2018}.

\begin{figure}
	\centering
	\begin{subfigure}[b]{.45\linewidth}
		\centering\includegraphics[width=0.9\textwidth]{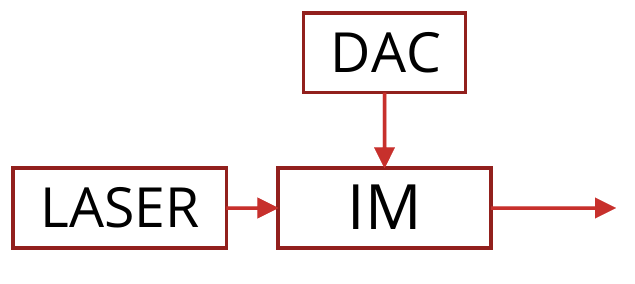}
		\caption{IM/DD transmitter}\label{fig:imddcoh:imddtx}
	\end{subfigure}%
	\begin{subfigure}[b]{.45\linewidth}
		\centering\includegraphics[width=0.9\textwidth]{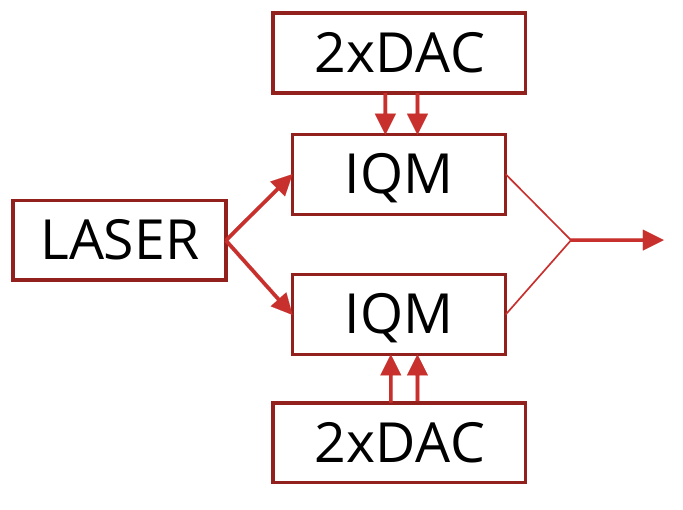}
		\caption{Coherent transmitter}\label{fig:imddcoh:cohtx}
	\end{subfigure}
	\hfill
	\begin{subfigure}[b]{.45\linewidth}
		\centering\includegraphics[width=0.9\textwidth]{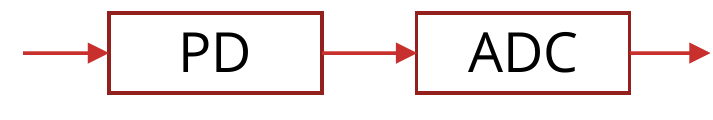}
		\caption{IM/DD receiver}\label{fig:imddcoh:imddrx}
	\end{subfigure}%
	\begin{subfigure}[b]{.45\linewidth}
		\centering\includegraphics[width=0.9\textwidth]{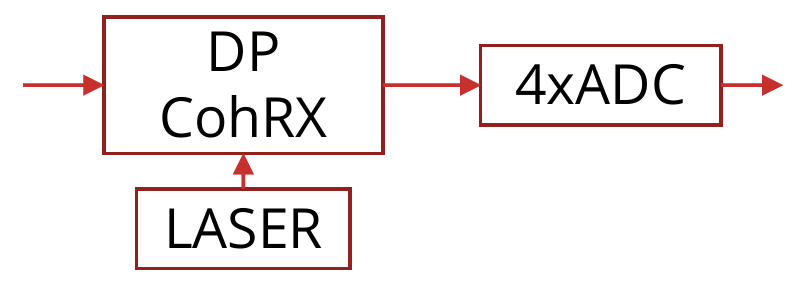}
		\caption{Coherent receiver}\label{fig:imddcoh:cohrx}
	\end{subfigure}%
	\caption{High-level architecture of a typical IM/DD and coherent transmitter and receiver. IM: Intensity Modulator, PD: Photodiode, IQM: I/Q Modulator, DP CohRX: Dual-Polarization Coherent Receiver.}\label{fig:imddcoh}
\end{figure}
A high-level comparison between IM/DD systems and coherent architectures is shown on Fig. \ref{fig:imddcoh}. Starting with the transmitter, an IM/DD transmitter
(Fig. \ref{fig:imddcoh:imddtx}), other than a laser, only needs an intensity modulator, which can be either a Mach-Zehnder Modulator (MZM) or an Electro-Absorption
Modulator (EAM). In some cases, the laser is even directly modulated. The electrical signal is generated by a DAC with a small bit resolution, since pulse-shaping
is generally not applied. A coherent system, instead, (Fig. \ref{fig:imddcoh:cohtx}), needs a low-linewidth laser, a dual-polarization I/Q MZM, four DACs with high-enough
resolution to apply pulse shaping. At the receiver, while an IM/DD receiver (Fig. \ref{fig:imddcoh:imddrx}) just needs a photodiode followed by an ADC, a coherent
receiver (Fig. \ref{fig:imddcoh:cohrx}) needs a laser, a dual-polarization coherent receiver, which includes two $90^\circ$ hybrids and $4$ balanced photodetectors, and $4$ ADCs.

\begin{table}
\centering
\begin{tabular}{c c c}
\toprule
& PAM-$M$ (IM/DD) & PM-$M^2$-QAM (coherent) \\
\midrule
Dispersion compensation & Optical & Electrical \\	
Spectral efficiency & $\sim 0.5\log_2 M$ (bit/s/Hz) & $\sim 2\log_2 M^2$ (bit/s/Hz) \\
\bottomrule
\end{tabular}
\caption{Drawbacks of IM/DD compared tt coherent with multilevel modulation with $M$ real-valued levels and Nyquist pulse shaping.}\label{tab:imdddrawbacks}
\end{table}
By looking at the architecture, complexity and thus cost of a coherent receiver is enormously higher than IM/DD systems. While technology advancements in integrated photonics can
reduce the cost difference, IM/DD systems are inherently much simpler, which makes them the preferred solution for short distance links.

However, IM/DD systems have also some serious drawbacks. In the context of Inter-DC connections, the main drawbacks are two, and are summarized in table \ref{tab:imdddrawbacks},
where IM/DD PAM-$M$ is compared with coherent polarization-multiplexed (PM) $M^2$-QAM. Other drawbacks (OSNR sensitivity, PMD, \dots) are not relevant in this scenario. 
The biggest issue is chromatic dispersion. Since dispersion is an all-pass filter on the electric field, it can be fully compensated with (almost)
no penalty by applying an inverse filter on the electric field. While on coherent systems it can be done electrically, on IM/DD systems it can only be performed optically.
On an IM/DD system, uncompensated dispersion has an electrical transfer function that is not all-pass \cite{Wang:1992}, and it can be only partially compensated with DSP
algorithms \cite{Agazzi:2005}. Optical dispersion compensation, either done using Dispersion-Compensating Fiber (DCF) or Fiber Bragg Gratings (FBG), other than introducing additional latency and insertion loss, reduces the flexibility of the system. 

Another issue is spectral efficiency. IM/DD PAM-$M$ has a \texttt{sinc}$^2$-shaped power spectral density. Considering only the main lobe, the spectral efficiency is
$0.5\log_2 M$ bit/symb/Hz. On the other hand, PM-$M^2$-QAM is usually shaped with a Root Raised Cosine (RRC) filter, which reduce the spectral occupancy of the signal, depending
on the roll-off factor. Therefore, assuming Nyquist pulse shaping (e.g. $0\%$ roll-off), the spectral efficiency is $2\log_2 M^2$ bit/symb/Hz, where the $2$ factor takes into account
the two polarizations. Consequently, the spectral efficiency of coherent PM-$M^2$-QAM compared to IM/DD PAM-$M$ is between $4$ and $8$ times larger, depending on the roll-off factor.

\subsubsection{Outlook}
Looking at this comparison, there is not any clear winner for this Section. At the time of writing this thesis, IM/DD is still widely deployed for this application. 
For instance, Inphi has recently presented a PAM-4 silicon-photonics transceiver able to achieve $100$-Gbit/s up to $120$ km \cite{Nagarajan:2018}. 
Nevertheless, for the future, it is inevitable a switch to coherent transmission. Coherent systems, as opposed to IM/DD systems, do not need optical
dispersion compensation. Therefore, there is a strong interest towards the development of ``hybrid'' solutions, which try to combine the gains of coherent with the simplicity of IM/DD for Inter-DC links, without needing optical dispersion compensation. The use of these systems will allow installation of dispersion-uncompensated links, and this will
ease the future transition to coherent. Chapter \ref{ch:ssb} will be devoted to an example of those systems, based on Single-Sideband modulation.

\subsection{Digital signal processing}
The selection of DSP algorithms for Inter-DC applications strictly depends on the choice between coherent and IM/DD. With coherent transmission, the DSP chain is approximately
the same as any standard long-haul coherent transceiver. Usually, it is even simpler, since PMD is negligible, and the amount of chromatic dispersion is small.
Therefore, for details on DSP for coherent systems, we refer the reader to Part \ref{part:coh} of this thesis.

On the other end, for IM/DD systems, the DSP chain can be very complex, since it has to deal with strong impairments, such as chromatic dispersion and bandwidth
limitations. Since the actual choice of the algorithms depend on the architecture, in Chapter \ref{ch:ssb} the DSP chain of a SSB system will be detailed.
For a more general overview of the possible algorithms for this application, the reader is referred to \cite{Zhong:2018}.

\chapter{Spatial Multiplexing for Intra-DC}\label{ch:bidir}

\graphicspath{{Chapter3/}}
 
In Sec. \ref{sec:2:intra}, intra data-center links were introduced, with
the main channel characteristics and the current state-of-the-art techniques. 
Focusing on these applications, this chapter presents a novel bi-directional architecture, coupled with receiver adaptive equalization,
which is able to double the per-laser (and per-fiber) data-rate for $<2$ km Intra-DC links. 
This architecture has been first presented at ECOC 2017 conference \cite{Nespola:2017}, extended in the IEEE Photonics Journal \cite{Pilori:2018} and then
presented at IPC 2018 conference \cite{Pilori:IPC2018}.

\section{Introduction}

\subsection{Options to increase capacity}\label{sec:2:capacity}
The newest Ethernet standard, presented in Sec. \ref{sec:2:ethernetstd}, is able to achieve 400 Gbit/s
over $2$ km using $8$ WDM channels, each transmitting at 50 Gbit/s (400GBASE-FR8).
Thus, a 400GBASE-FR8 transceiver needs 8 lasers, each tuned at a different wavelength in the LWDM grid, as shown in Table \ref{tab:lanwdmgrid}. 
Lasers represent one of the largest sources of power consumption, and, on Silicon-on-Insulator (SOI) platforms, require
expensive heterogeneous integration techniques.
Therefore, the performance metric for this link is the per-laser capacity, i.e. the total data-rate divided by the number of lasers \emph{in a transceiver}. 

In general, to increase capacity of a generic communication systems, 
one needs to act on one (or more than one) physical dimensions \cite{Winzer:2014}:
\begin{itemize}
	\item Frequency
	\item Polarization
	\item Quadrature
	\item Space
	\item Time
\end{itemize}
For Intra-DC links, some dimensions can be exploited more easily than others. In details:
\paragraph{Quadrature}
Using intensity modulation/direct detection schemes, quadrature cannot be easily exploited to increase capacity. 
Chapter \ref{ch:ssb} will show possible techniques to achieve this result, but the additional complexity makes these solutions feasible only for longer links.

\paragraph{Frequency}
The use of frequency spatial dimension means the adoption of WDM, which is already used in this scenario. However, this does not increase the per-laser data-rate, 
unless special hardware, such as frequency combs \cite{Marin-Palomo:2017,Hu:2018}, is adopted. At the time of writing this thesis, frequency combs are at the early stage
of research, and their possible commercial deployment is still far in time.

\paragraph{Time}
Exploiting time means, substantially, increasing either the symbol rate or the number of levels.
With respect to the symbol rate, there have been reported OOK symbol rates up to 204 GBaud \cite{Mardoyan:2018}, which represents a $10\times$ improvement compared to current standard ($\sim25$ GBaud). However, increasing the symbol rate requires a huge effort in terms of hardware
development. In fact, according to \cite[Fig. 4]{Winzer:2017} symbol rates of commercial technologies have been increasing only at a $\sim10\%$ compound annual growth rate.
Moreover, chromatic dispersion may start to become an issue with very large symbol rates, even in O-band.
Increasing the number of levels, instead, requires substantial improvements in the transceiver. 
For instance, switching from PAM-4 to PAM-8 bears a theoretical power penalty of $3.7$ dB (see Fig. \ref{fig:pamber}). 
In addition to that, PAM-8 may be more sensitive to several impairments (multipath interference, quantization, filtering, \dots) with respect to PAM-4, which further increase the required sensitivity.

In conclusion, with current technologies, quadrature, time and frequency cannot be easily exploited to provide substantial improvements in terms of per-laser data-rate. 
The remaining dimensions, space and polarization are the most promising candidates to provide this improvement, which will be detailed in this thesis.

\subsection{Spatial multiplexing}
Exploiting spatial dimensions means finding $N$ parallel channels, separated in space, which give an overall more efficient solution than $N$ independent transceivers.
Comparison with parallel independent transceivers is very important, since fiber, inside the DC, is not a scarce resource.
Efficiency is usually obtained by array integration at various levels (optical, electrical, or both). These techniques take the generic name of SDM - Spatial Division Mutiplexing, which have already been adopted in other optical communications scenarios \cite{Winzer:OFC2018,Dar:2018}.

SDM can be achieved with many methods, depending on the requirements in terms of reach, data-rate, cost and power consumption. 
Inside the data-center, this is usually achieved by using parallel fibers connected to the same transceiver using an MPO (Multi-fiber Push On) connector. Cost and power reduction
is achieved by integration and laser sharing inside the transceiver. An example of this application is the 400GBASE-DR4 standard (see Tab. \ref{tab:400g_eth}), which uses
four fiber pairs to achieve 400 Gbit/s over 500 meters. The most serious limitation of MPO, other than the additional loss of the connector, is the need of installation
of fiber ribbons, i.e. cables with multiple fiber pairs that are terminated in an MPO connector.
In the DC, especially in longer links (connecting different racks or sections), there are deployed standard duplex (one for each direction) SMF cables.
While shorter links (connecting servers to rack switches) can be easy replaced, changing longer links can be an expensive procedure. 

Since most of deployed cables are duplex, a possible solution would be using each cable in two directions. This option, which is the main contribution of this Chapter,
will be discussed from Sec. \ref{sec:3:bidirarch} onwards.

\subsection{Polarization multiplexing}
The last dimension that can be exploited is polarization, which is potentially able to double spectral efficiency. 
In fact, coherent systems already use polarization multiplexing. However, due to random fiber birefringence and PMD, polarization multiplexing requires
a receiver that is able to detect the full electric field (amplitude and phase) and polarization in order to recover the transmitted data.

In the literature, there have been presented several methods to allow polarization multiplexing with IM/DD. The most important are:
\begin{itemize}
\item Stokes-vector receivers, which can detect a polarization-multiplexed intensity-modulated transmission with lower complexity than a coherent receiver \cite[Sec. V]{Zhong:2018}.
\item Kramers-Kronig receivers, which can detect the full electric field only with two direct-detection receivers \cite{Antonelli:2017}, at the expense
of a larger receiver analog bandwidth.
\item External polarization controllers, that align polarization of the incoming signal \cite{Nespola:2018} to two IM/DD receivers.
\end{itemize}
Each of these solutions has its advantages and disadvantages, and the ``best'' solution will strongly depend on the cost of the optical devices involved. 
It is also noteworthy that none of these options have, to the best of my knowledge, been implemented in a commercial transceiver.

In \cite{Nespola:2018}, we used a silicon-photonics circuit to perform polarization de-multiplexing at the receiver. Polarization was controlled using a micro-controller connected
to $5$ heater pads that apply a phase shift in the device. Using this device, we were able to perform real-time reception of a polarization-multiplexed PAM-2 (NRZ) signal. 
While this result is quite promising, there are many issues that needs to be solved, mainly related to the device itself and the polarization control algorithm, which will
be dealt in future research.

\section{Bi-directional architecture}\label{sec:3:bidirarch}
\subsection{Introduction}
As discussed in Sec. \ref{sec:2:capacity}, spatial multiplexing is one of the most viable options to increase per-laser capacity in an Intra-DC system, since it allows laser sharing between parallel paths.
Since most of the fibers deployed in DC are duplex (pairs of SMF, one for each direction), a possible solution is to use each fiber of the pair in both directions. 
In fact, from a physical point of view, each fiber can be used bidirectionally.

Nevertheless, almost every optical communication system use fiber only in one direction. 
The only exceptions are access networks based on Passive Optical Network (PON) standards. 
Strictly speaking, since fiber diameter is very small, the choice of using mono-directional propagation is mainly done for convenience. 
For Intra-DC applications, without optical amplification, the biggest issue is the separation at the transceivers of data in the two directions.
The optimal method would be the use of optical circulators, which unfortunately are not easy to realize in integrated optics \cite{Doerr:14,Pintus:13,Mitsuya:13}, which make them unsuitable to the space requirements of Intra-DC transceivers. PONs solve this issue by using large wavelength separation between the two directions.
For instance, the GPON standard \cite{std:gpon} uses $1480-1500$ nm for the downstream direction while it uses $1260-1360$ nm for the upstream
direction. In this case, separation is performed using WDM couplers. However, due to the symbol rates
used in Intra-DC (much higher than PONs), O-band is mandatory to avoid chromatic dispersion. Therefore, this solution is also feasible.

In conclusion, while bi-directional schemes are a promising solution to increase capacity with laser sharing, there is the important issue of separating light in the transceivers. 
Since the most common methods are not applicable for Intra-DC scenarios, a novel method should be developed, which will be discussed in next sections.

\subsection{General schematic}
\begin{figure}
	\centering
	\includegraphics[width=0.95\textwidth]{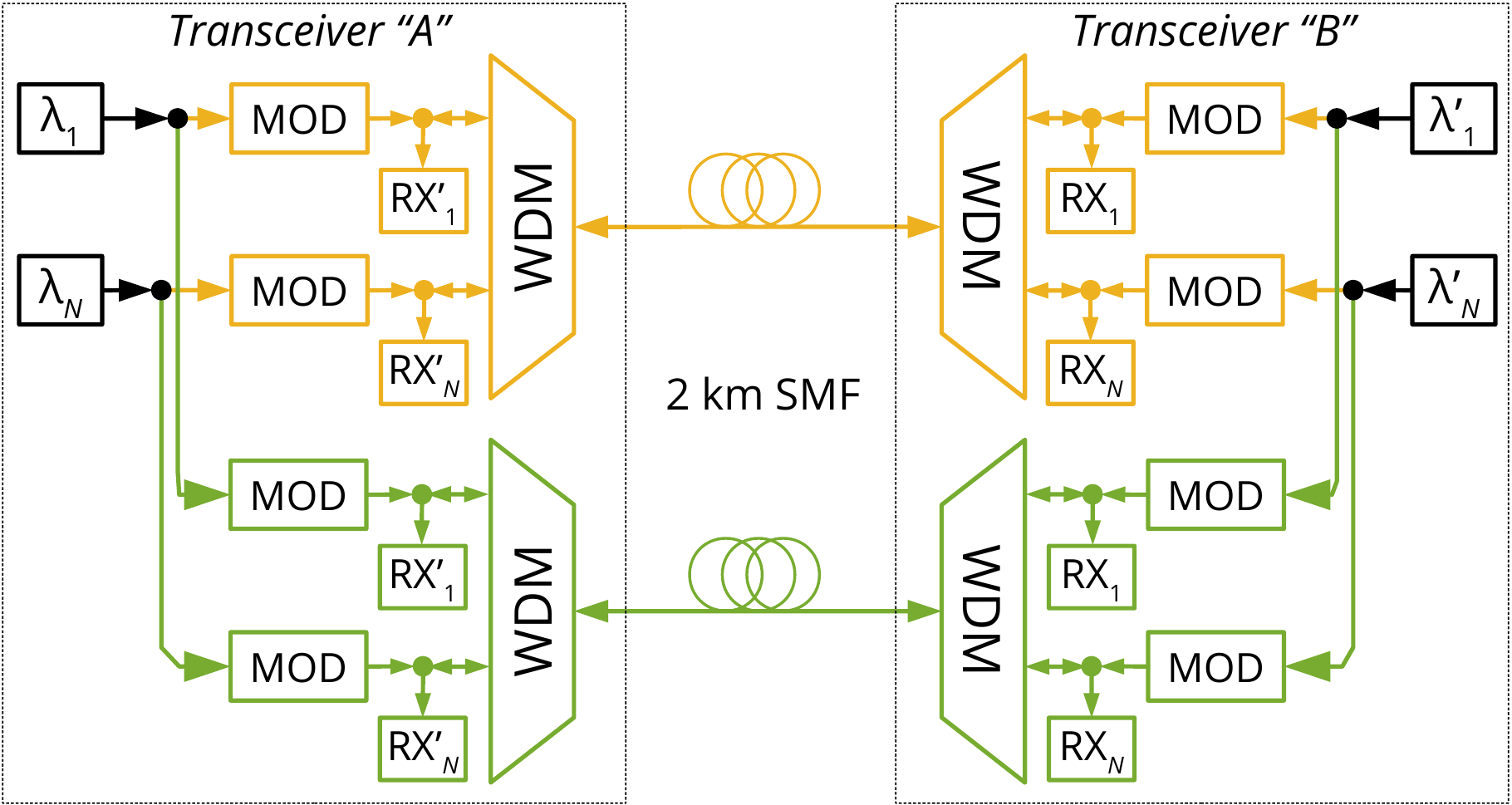}
	\caption{High-level schematic of the proposed bi-directional transmission scheme.}\label{fig:bidirscheme}
\end{figure}
A high-level schematic of the proposed bi-directional architecture is shown in Fig. \ref{fig:bidirscheme}. Inside each transceiver, lasers are shared between the two directions,
realizing a higher level of integration than two independent transceivers, reducing overall cost and power consumption. 
Then, each laser (operating in the O-band) is divided by $3$-dB splitters and sent to two modulators, one for each fiber. 
A WDM coupler is used both to combine data from different channels and to separate the incoming WDM channels. Half of the incoming (using another
3-dB coupler) light is sent back to the modulator, which is suppressed by the isolators at the output of lasers. 

\begin{figure}
	\centering
	\includegraphics[width=0.5\textwidth]{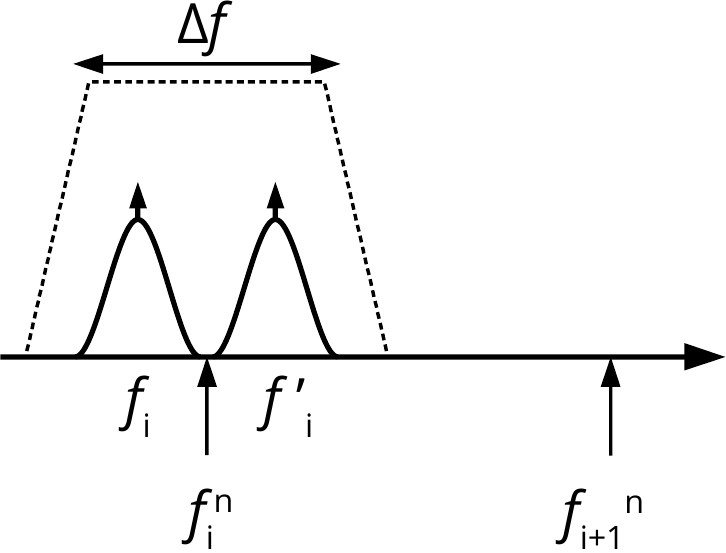}
	\caption{Position of WDM channels of both directions in the WDM grid.}\label{fig:bidirdetuning}
\end{figure}
This architecture achieves separation of data in two directions with $3$-dB splitters, which reduces the power budget (compared to an MPO-based solution) 
by approximately $6$-dB. This solution, which is simpler with respect to solutions previously discussed (circulator
or different wavelengths), suffers from back-reflections. Assuming that a back-reflection occurs, data will interfere with the signal
transmitted in the other direction, which has the same (nominal) frequency, so that it cannot be suppressed by WDM couplers.

To avoid this issue, we propose a \emph{slight} frequency shift of the lasers inside one of the transceivers, called transceiver ``B'' in Fig. \ref{fig:bidirscheme}.
This shift would be small enough to keep lasers at the same nominal wavelengths of the WDM grid. This operation is feasible, since --
for Intra-DC applications -- WDM grids allow wider tolerances than long-haul applications. For instance, the LAN-WDM grid has a frequency spacing
of $800$-GHz, and it is the most dense grid used in Intra-DC standards. The other grid widely used in Intra-DC is the CWDM grid, which has a wavelength
spacing of $20$ nm, corresponding to $\sim3.5$ THz in the O-band. Fig. \ref{fig:bidirdetuning} shows an example of this frequency shift. 
For the $i$-th channel, the standard specifies a nominal central frequency $f_i^\textup{n}$ and
a range $\Delta f$. This range is wider than the spectral occupancy of a channel to allow some tolerances. In
this architecture, this tolerance is exploited by changing the central frequency of the transceivers to $f_i$ and $f_i^\prime$.
Those values are close enough to the nominal frequency such that both transceivers are compliant to the standard (which defines
the selectivity of WDM filters). However, this approach requires a more precise wavelength control with respect to standard Intra-DC applications, since -- as it
will be shown later -- this frequency shift must be greater than a minimum value.

We will then prove that this small frequency shift is sufficient to strongly reduce the penalty of back-reflections. This will be done
first with a theoretical model, explained in Sec. \ref{sec:3:theomodel}. Then, it will be validated with an experiment, which
will be shown in Sec. \ref{sec:3:experiment}.

\section{Impact of reflection crosstalk}\label{sec:3:theomodel}
\begin{figure}
	\centering
	\includegraphics[width=0.5\textwidth]{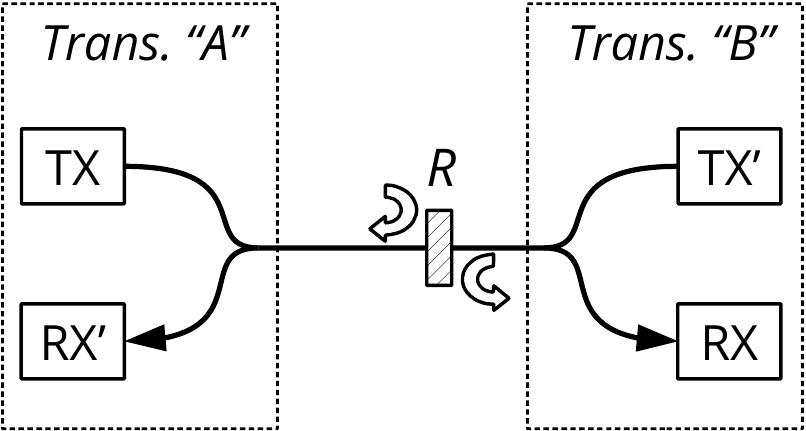}
	\caption{Generic block diagram of crosstalk caused by a spurious back-reflection.}\label{fig:cohinterferencescheme}
\end{figure}
Let us assume the generic bi-directional architecture of Fig. \ref{fig:bidirscheme}, which is depicted (in a simplified form)
in Fig. \ref{fig:cohinterferencescheme}. In case of a back-reflection, a portion $P_\textup{rifl}$ of the transmitted optical power $P$ is reflected back. The ratio between 
the incoming power and the reflected power $R=P_\textup{rifl}/P$ is called reflectivity.
In a bi-directional transmitter scheme, data generated by transceiver ``A'' gets reflected back to its receiver, which is tuned to receive
data from transceiver ``B'' (indicated by $^\prime$ since it has a slightly different frequency). The same may happen
in the other direction.

Usually, other than the reflectivity, another parameter is used, called Signal to Interference power Ratio (SIR),
defined as the ratio between the optical power of the wanted signal and the optical power of the interfering (reflected) signal, measured immediately
before the receiver. In this section, we will assume a simplified scenario where the line is lossless, i.e. fiber and connector losses are zero. In this case,
assuming that $R\ll1$, the SIR is simply the opposite of the reflectivity:
\begin{equation}
\mathrm{SIR} = \frac{1}{R}
\label{eq:sirreflection}
\end{equation}

This issue has already been studied in details for NRZ signals, and takes the general name of \emph{coherent crosstalk}, since the
frequency of the interfering signal is very close to the frequency of the wanted signal. However,
past studies \cite{Gimlett:89} studied this effect in a different scenario, which is multi-path interference. In that context,
multiple reflections ($2,4,\dotsc$) generate an interfering signal which is the sum of delayed copies of the signal itself, which have exactly the same frequency.
Moreover, most of the models assume that the interfering signal is Gaussian-distributed. 
However, in \cite{Attard:05}, it was shown that for a few reflections, the ``Gaussian'' approximation is
overly pessimistic. Since, in the considered architecture, the main source
of interference is a single reflection, the model must not use the Gaussian approximation.

\subsection{Theoretical model}\label{sec:3:theomodelderiv}
To develop an analytical model, some simplifying assumptions have to be made. 
For this model, we adopted the same assumptions as \cite[Sec. IV]{Attard:05}, which are here summarized:
\begin{enumerate}
\item Signal and crosstalk are bit-aligned in time. This is a worst-case assumption, since PAM-$M$ transitions have lower power than the levels.
\item Phase difference, due to laser phase noise, is constant within one symbol.
\item Signaling use purely-rectangular PAM-$M$ pulses, without any bandwidth limitation at the transmitter and receiver.
\item Receiver uses an ideal matched filter, i.e. a rectangular filter, without equalization.
\item Signal and interferer are polarization-aligned. Other than being a worst-case approximation, in \cite{Goldstein:95} it was shown that this situation
happens more frequently than expected.
\end{enumerate}

Using these assumptions, a fully analytical result can be obtained. Assuming that the transmitted PAM-$M$ symbol $a$ is interfering with the PAM-$M$
symbol $b$ transmitted in the opposite direction (and reflected), 
the two lasers are freqency-separated by $\Delta f$ and have a phase difference $\phi$, and symbol duration is $T$, the received signal before decision is:
\begin{equation}
y = a + \frac{b}{\mathrm{SIR}} + 2\sqrt{\frac{ab}{\mathrm{SIR}}}\cos(\phi)\frac{\sin(\pi\Delta f T)}{\pi\Delta f T}
\label{eq:paminterf}
\end{equation}
Note that, in this equation, there is no time index, since this effect is not time-dependent. Also note that $a$ and $b$ refer to the symbols' optical
\emph{power}.
Full derivation of this formula is reported in Appendix \ref{app:paminterf}.

In this equation, interference is generated by two different sources:
\begin{itemize}
\item The first source, $\frac{b}{\mathrm{SIR}}$, depends only on the SIR. This is called \emph{incoherent} crosstalk, since it does not depend
on the frequency separation. Unless $\Delta f$ is larger than the bandwidth of the WDM coupler, this term cannot be suppressed. 
\item The second source depends on $\Delta f$ and $\phi$. This term is (for small $\Delta f$) larger than the first one, which means that crosstalk can be effectively
reduced by acting on $\Delta f$.
\end{itemize}

Equation \eqref{eq:paminterf} can be then used to compute the Bit Error Ratio (BER). Assuming that the received signal is impaired by additive white
 Gaussian noise with standard deviation $\sigma_n$ (see Sec. \ref{sec:2:chmodel} for details), 
 the probability of a symbol error, given the transmitted and interfering symbols $a$ and $b$, is
\begin{equation}
P\left( e|a, b \right) = 
\frac{1}{2} \sum_i \expt \left\{
\erfc\left(
\frac{ \left| a- \left[  
V_{\textup{th},i} + 
\frac{b}{\mathrm{SIR}} +
2\sqrt{\frac{a b}{\mathrm{SIR}}}\frac{\sin(\pi\Delta f T)}{\pi\Delta f T} \beta
\right]  \right| }{\sqrt{2}\sigma_n}
\right)
\right\}
\label{eq:pamser}
\end{equation}
The expectation is performed over $\beta=\cos(\phi)$, which, assuming that $\phi$ is uniformly distributed between $0$ and $2\pi$, it has the following
probability density function \cite{Bertignono:16}
\begin{equation}
p(\beta) = \frac{1}{\pi\sqrt{1-\beta^2}}
\end{equation}
$V_{\textup{th},i}$, instead, is/are the decision threshold/thresholds of symbol $a$, which can be one or two, depending on the position of the symbol.
Then, from \eqref{eq:pamser}, calculation of the BER is straightforward.

\subsection{Simulation tool}
\begin{figure}
	\centering
	\includegraphics[width=0.7\textwidth]{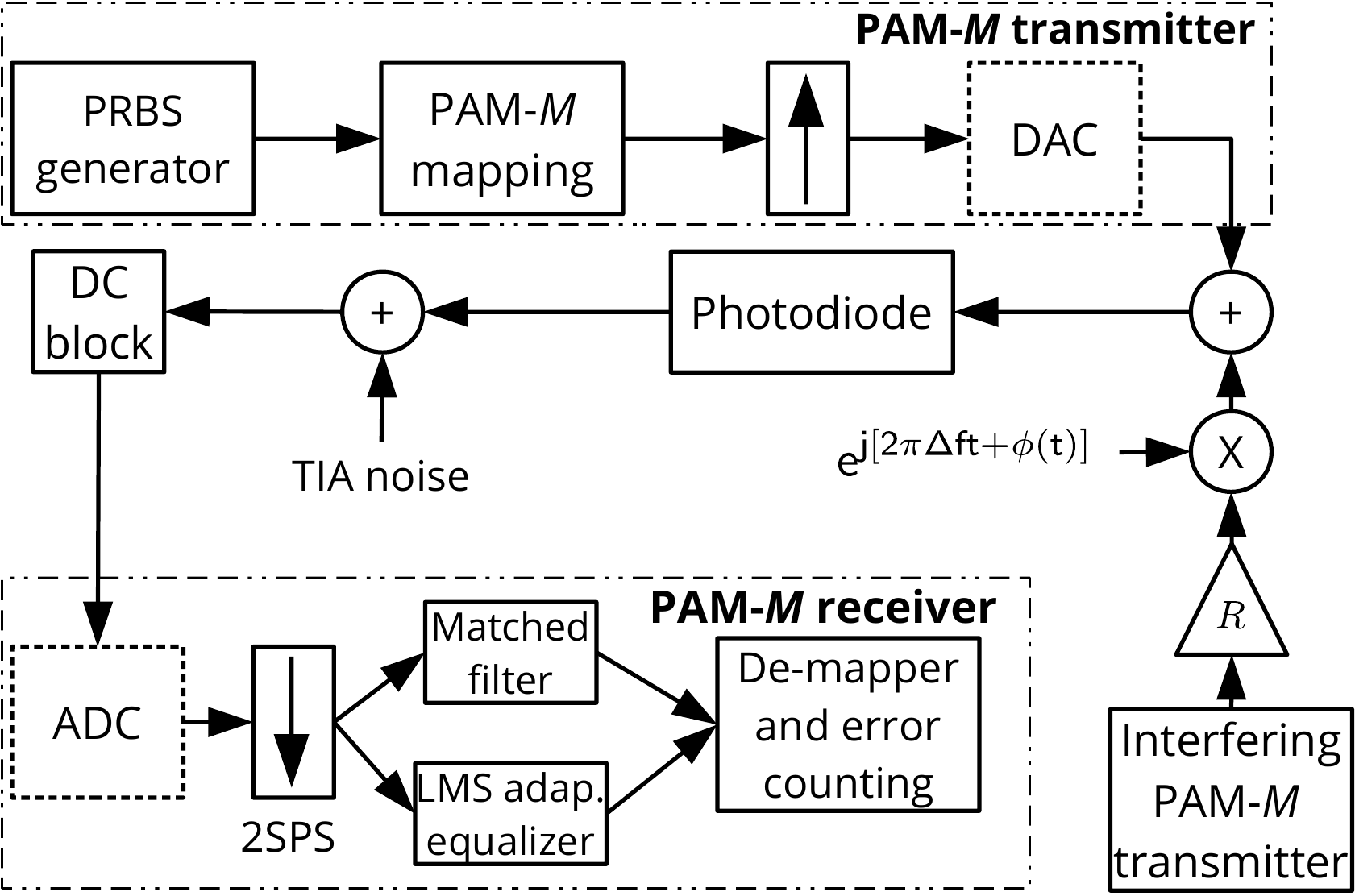}
	\caption{Block scheme of the PAM-$M$ simulator tool.}\label{fig:pamsimsetup}
\end{figure}
The theoretical model presented in Sec. \ref{sec:3:theomodelderiv} needs to be validated. 
To do so, we set up a PAM-$M$ time-domain simulation tool, whose schematic is shown in Fig. \ref{fig:pamsimsetup}. 
A Pseudo Random Bit Sequence generator generates a bit sequence, that is mapped into a PAM-$M$ sequence with a finite extinction ratio. The signal is then upsampled with
ideal rectangular pulses, and then low-pass filtered and quantized by a DAC. Then, another interfering PAM-$M$ signal is generated with random bits, 
multiplied by a complex exponential to apply a frequency and phase deviation, scaled by the reflectivity $R$ and added to the signal. As discussed in Sec. \ref{sec:3:theomodel},
link is assumed ideal, therefore $R$ is the inverse of SIR \eqref{eq:sirreflection}. $\phi(t)$ represents laser phase noise, and it is assumed to have a Lorentzian shape, with
a $2$-MHz linewidth.

At the receiver, the signal is low-pass filtered and quantized by an ADC and sampled at two samples per symbol. Then, it can be either filtered with an ideal matched filter, 
or by a fractionally-spaced Least-Mean Squares (LMS) adaptive equalizer. At the end, a slicer performs hard decision, followed by BER computation. 
Considering the KP4 code (see Sec. \ref{sec:2:ethernetstd}), we choose a conservative BER threshold of $2\times10^{-4}$.

This simulator will be used both to validate the theoretical model and, in Sec. \ref{sec:3:experiment}, to validate the experimental results.

\subsection{Simulation results}\label{sec:3:simres}
\begin{table}
	\centering
	\begin{tabular}{c c}
		\toprule
		Parameter & Value \\
		\midrule
		Symbol rate & $R_\textup{s}=53$ GBaud \\
		Extinction ratio & $\mathrm{ER}=10$ dB \\
		DAC/ADC & Ideal \\
		Receiver & Matched filter \\
		\bottomrule
	\end{tabular}
	\caption{Simulation parameters used to validate the theoretical model.}\label{tab:simvalidparams}
\end{table}
\begin{figure}
	\begin{subfigure}[b]{0.48\textwidth}
		\centering \includegraphics[width=\textwidth]{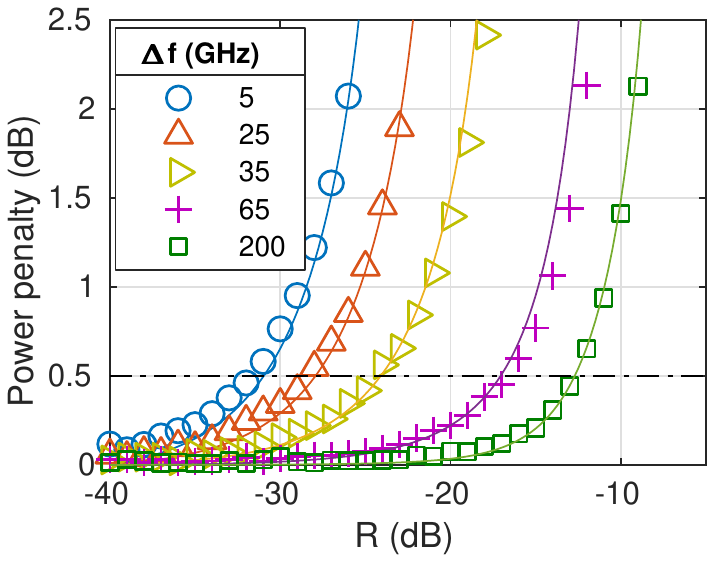}
		\caption{}\label{subfig:simvalidr}
	\end{subfigure}
	\begin{subfigure}[b]{0.48\textwidth}
		\centering \includegraphics[width=\textwidth]{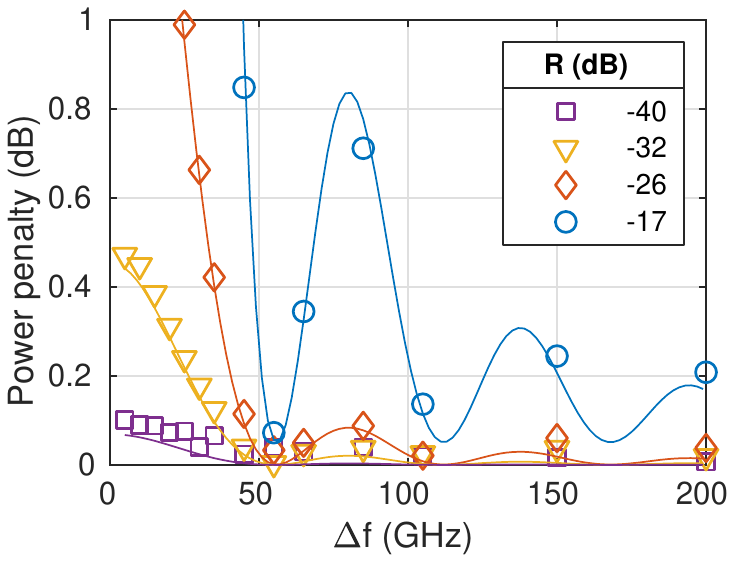}
		\caption{}\label{subfig:simvaliddeltaf}
	\end{subfigure}
	\caption{Optical power penalty as a function of the reflectivity $R$ and laser frequency separation $\Delta f$ for $53$ GBaud PAM-$4$. Markers
	are obtained with simulations, while solid lines are obtained with theoretical formula \eqref{eq:pamser}.}\label{fig:pamsimvalidation}
\end{figure}
The simulator was set up using the parameters summarized in Table \ref{tab:simvalidparams}. 
A $53$ GBaud PAM-$4$ signal is generated with an extinction ratio of $10$ dB and measured with a receiver 
with an equivalent noise current of $17~\mathrm{pA}/\sqrt{\mathrm{Hz}}$ and unit responsivity ($1$ A/W).
Since the model does not take into account bandwidth limitations, ADC and DAC are assumed ideal, and the receiver uses a simple
matched filter.
First, it is calculated the minimum ROP to transmit at the BER threshold without interference. 
Then, this ``baseline'' ROP is subtracted from the ROP of the other results with interference, obtaining the optical power penalty
that is shown in Fig. \ref{fig:pamsimvalidation}. This operation has been adopted for all subsequent results of this chapter.
Note that this result does not depend on the receiver noise current, thanks to the normalization to the no-interference ROP (``baseline'').

Results are shown in Fig. \ref{fig:pamsimvalidation} as a function of the reflectivity $R$ and the laser spacing $\Delta f$, where
markers represents numerical simulations and solid lines were obtained with the analytical formula. 
It can be seen that the theoretical formula can accurately predict performance in a wide range of values of $R$ and $\Delta f$. 

\subsubsection{Discussion}
Fig. \ref{subfig:simvalidr} shows the power penalty as a function of the reflectivity, for different values of $\Delta f$. 
Regardless of the frequency spacing, power penalty keeps low for increasing reflectivity $R$, up to a point where penalty suddenly
increase. An increase of $\Delta f$ moves this ``threshold'' to stronger values of $R$. 
By setting an arbitrary threshold at $0.5$ dB power penalty, some system-level consequences can be derived.

For small values of $\Delta f$ (e.g. $5$ GHz, blue circles in the figure), the threshold is approximately at $R=-31.5$ dB. This value is too demanding; 
in fact, a single TIA-568 
LC connector, widely used in data-centers, has a maximum specified reflectivity of $-26$ dB. 
However, if $\Delta f$ is increased above the symbol rate (e.g. $65$ GHz, purple crosses in the figure),
the threshold increases to $-17$ dB, which is a reasonable value for real-world conditions.
Fig. \ref{subfig:simvaliddeltaf} shows the same results as Fig. \ref{subfig:simvalidr}, but as a function of the spacing. It is interesting to see that, for strong
reflectivities, power penalty fluctuates, which is due to the $\sinc(.)$ function in \eqref{eq:paminterf}. 

To summarize, these results give two very important conclusions. First, the theoretical model is valid over a wide range of parameters. Second, and most importantly, the proposal
of a \emph{slight} change in $\Delta f$ to reduce back-reflection penalties, presented in Sec. \ref{sec:3:bidirarch}, is feasible.
A frequency shift in the order of the symbol rate is well within the ranges of the WDM grids adopted for Intra-DC applications.

Then, the last step to definitely prove the feasibility of the architecture is an experimental validation, which will be performed in next Sec. \ref{sec:3:experiment}.
 
\section{Experimental validation}\label{sec:3:experiment}
\subsection{Experimental setup}
\begin{figure}
	\centering
	\includegraphics[width=0.9\textwidth]{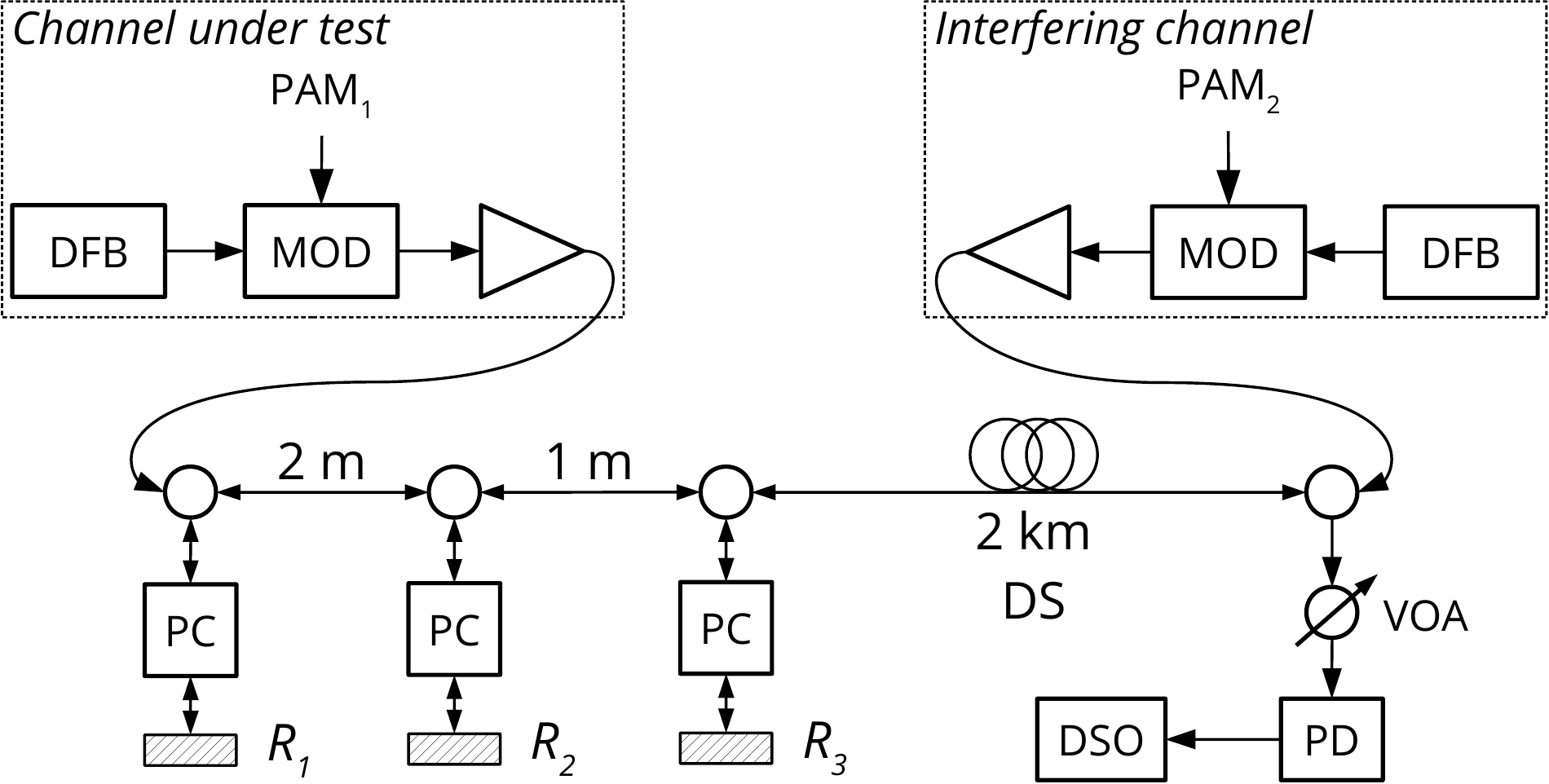}
	\caption{Block scheme of the experimental setup.}\label{fig:pamexpsetup}
\end{figure}
A high-level block scheme of the experiment is shown in Fig. \ref{fig:pamexpsetup}. For simplicity, only one WDM channel was transmitted, 
since the purpose of the experiment is the evaluation of the back-reflection penalty at the same nominal wavelength. 
Moreover, due to the unavailability of O-band components, the experiment was performed in the C-band, which allowed
the use of Erbium Doped Fiber Amplifiers (EDFAs) to recover the extra losses of the lumped components. 
Nevertheless, given the limited propagation distance, ASE noise inserted by the EDFAs was negligible with respect
to receiver thermal noise. For propagation, we used a $2$-km span of Dispersion-Shifted (DS) fiber, which has approximately the same dispersion curve in C-band as SMF in the O-band.

In the setup there are two transmitters. One is generating the channel under test with pattern PAM$_1$ and the other one generates the interfering channel with a different pattern
PAM$_2$. Each of them is made of a Distributed Feedback (DFB) laser ($\sim2$ MHz linewidth), 
connected to a $33$-GHz lithium-niobate optical modulator, which modulates the intensity of light based on a PAM-$4$ signal generated by a pattern generator. 
Before the interfering channel, a $3$-dB coupler sends light from the channel under test (and reflections of the interfering channel) 
to a Variable Optical Attenuator (VOA), used to set the ROP, a photodiode and
a real-time oscilloscope. Receiver Digital Signal Processing, performed offline, consists in a $10$-tap ($T/2$-spaced) adaptive equalizer, followed by hard decision and error counting.

Back-reflections are inserted using three mirrors $R_1$, $R_2$ and $R_3$, connected to the fiber using $3$-dB couplers and polarization controllers (PCs) to align
polarization of the reflected light to the incoming signal. For the experiment, we either used one single reflector ($R_3$) or all of them, to emulate multiple reflections.

\subsection{Results with single reflector}
\begin{figure}
	\begin{subfigure}[b]{0.48\textwidth}
		\centering \includegraphics[width=\textwidth]{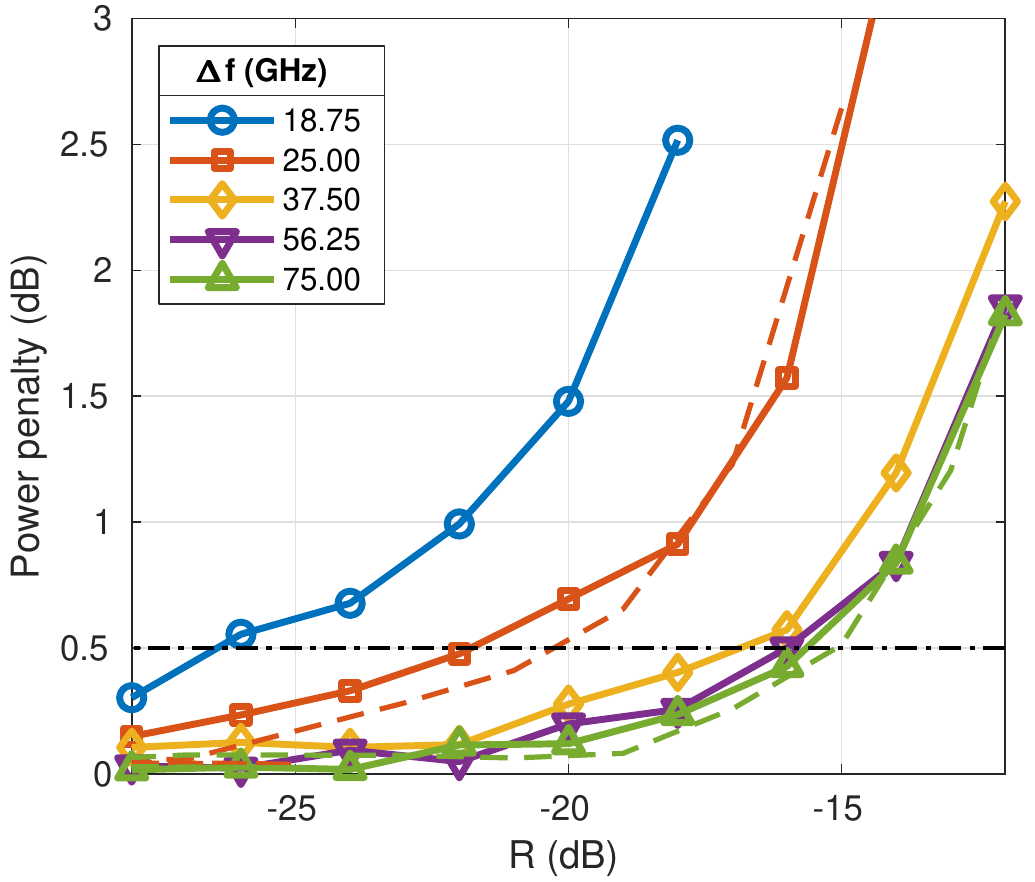}
		\caption{28 GBaud}\label{subfig:singleref28}
	\end{subfigure}
	\begin{subfigure}[b]{0.48\textwidth}
		\centering \includegraphics[width=\textwidth]{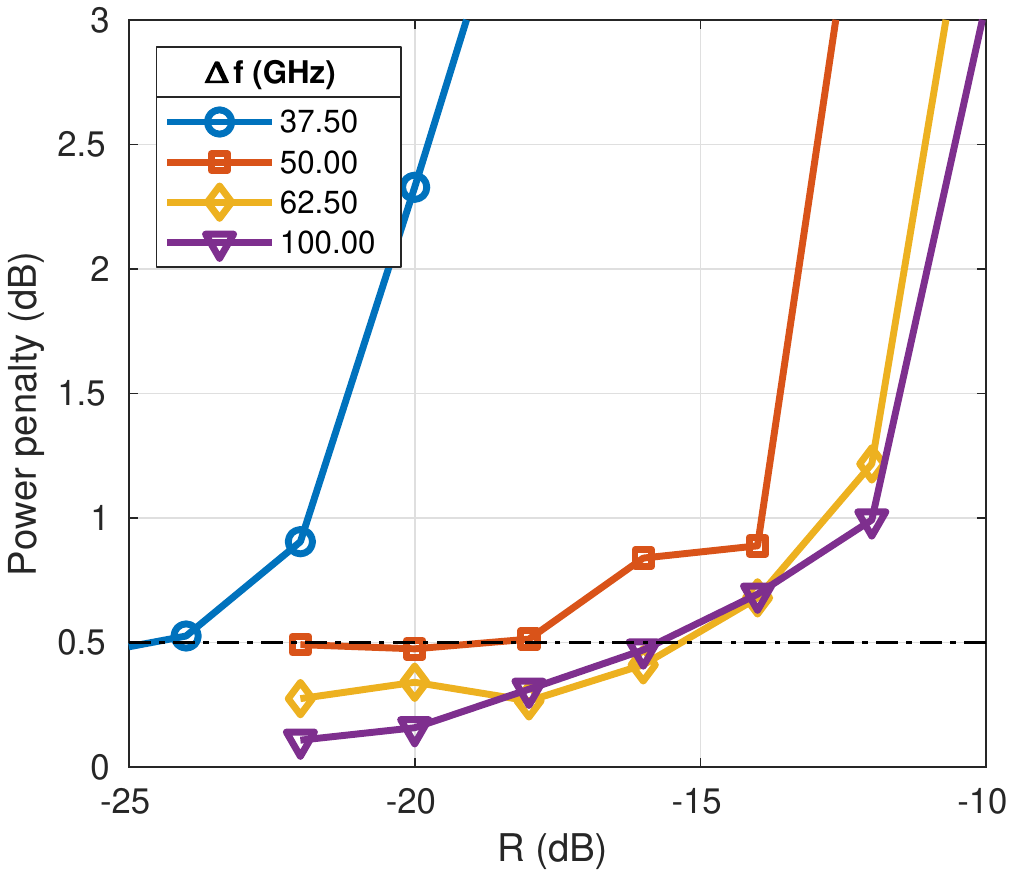}
		\caption{53 GBaud}\label{subfig:singleref53}
	\end{subfigure}
	\caption{Experimental result with a single reflection for different symbol rates.}\label{fig:pamsinglereflector}
\end{figure}
\begin{figure}
	\centering
	\includegraphics[width=0.5\textwidth]{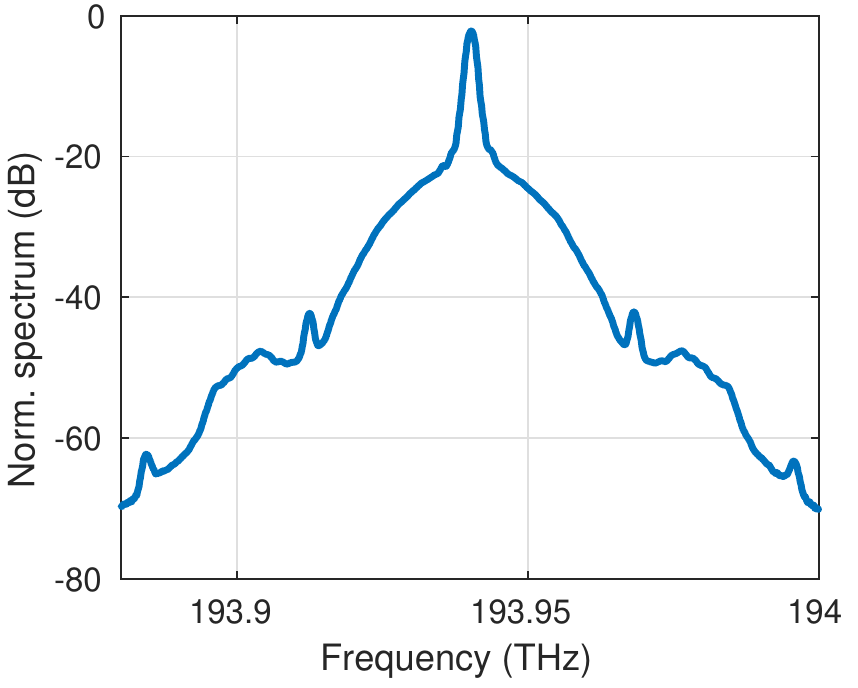}
	\caption{Measured optical spectrum of 28 GBaud PAM-4.}\label{fig:pamspectrum}
\end{figure}
First, we run results with a single reflector ($R_3$ of Fig. \ref{fig:pamexpsetup}). This is the most important test, since, as previously discussed,
single-reflections are the strongest sources of penalty. Results are shown in Fig. \ref{fig:pamsinglereflector} for two PAM-$4$ symbol rates,
$28$ GBaud and $53$ GBaud. The value of reflectivity $R$ has been normalized to the losses of the splitters and polarization controllers, so that, as
done in previous section, the SIR is the opposite of $R$ \eqref{eq:sirreflection}. Then, by tuning the receiver VOA, we calculated the minimum ROP for each combination of $R$ and $\Delta f$
at the FEC threshold, as done with the simulation. Afterwards, all results have been normalized to the minimum ROP without interference, to obtain the optical power penalty.
Analyzing at the results of Fig. \ref{fig:pamsinglereflector}, the conclusions obtained in Sec. \ref{sec:3:simres} are confirmed. 
At both symbol rates, a laser spacing greater than the symbol rate is sufficient to tolerate realistic back-reflections.

The transmitter had strong bandwidth limitations, even at 28 GBaud, as shown in the optical spectrum of Fig. \ref{fig:pamspectrum}.
Therefore, the receiver had to use an adaptive equalizer. With this receiver, the hypothesis of the analytical derivation performed in Sec. \ref{sec:3:theomodelderiv} are not valid anymore,
and its predictions cannot be compared with the experiment. Therefore, cross-validation with the experimental results was performed with the numerical
simulator of Fig. \ref{fig:pamsimsetup}, where we included transmitter bandwidth limitations and adaptive equalizer. Results of the simulations
are shown as dashed lines in Fig. \ref{subfig:singleref28}, where there is good agreement with the experimental results.

\subsection{Three reflections}
\begin{figure}
	\begin{subfigure}[b]{0.48\textwidth}
		\centering \includegraphics[width=\textwidth]{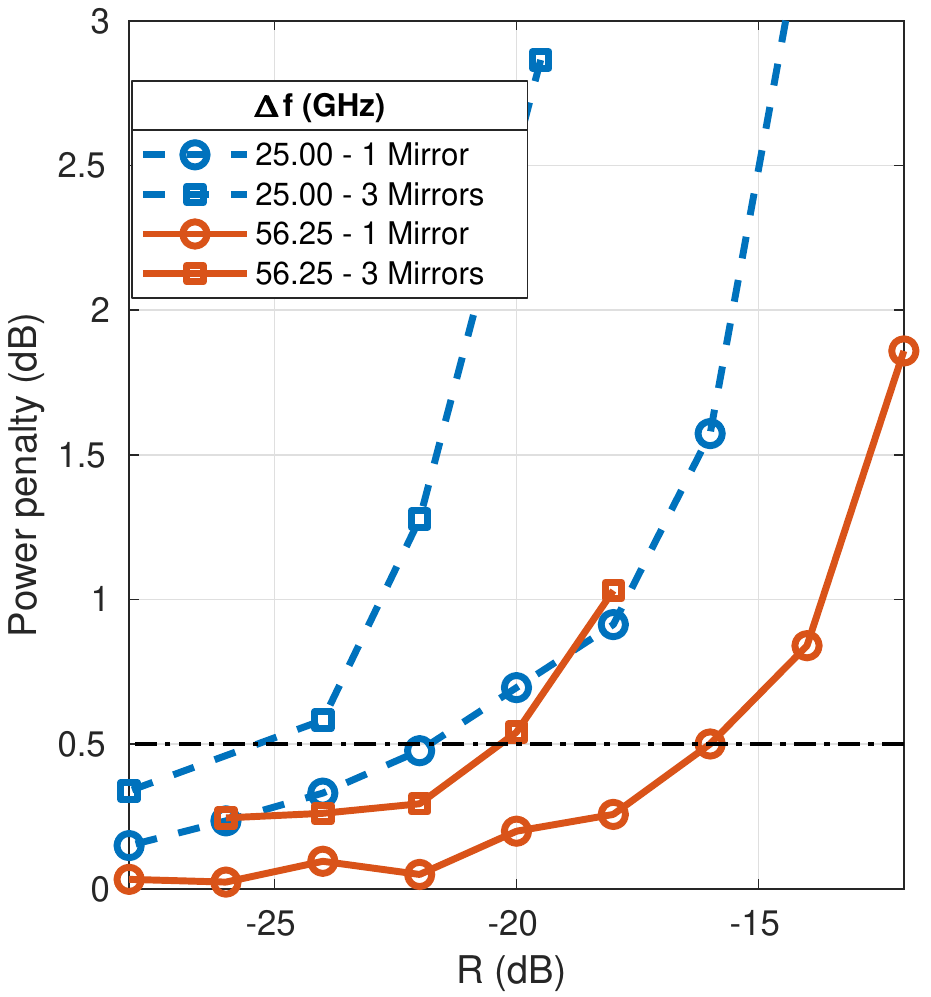}
		\caption{}\label{subfig:threeref}
	\end{subfigure}
	\begin{subfigure}[b]{0.48\textwidth}
		\centering \includegraphics[width=\textwidth]{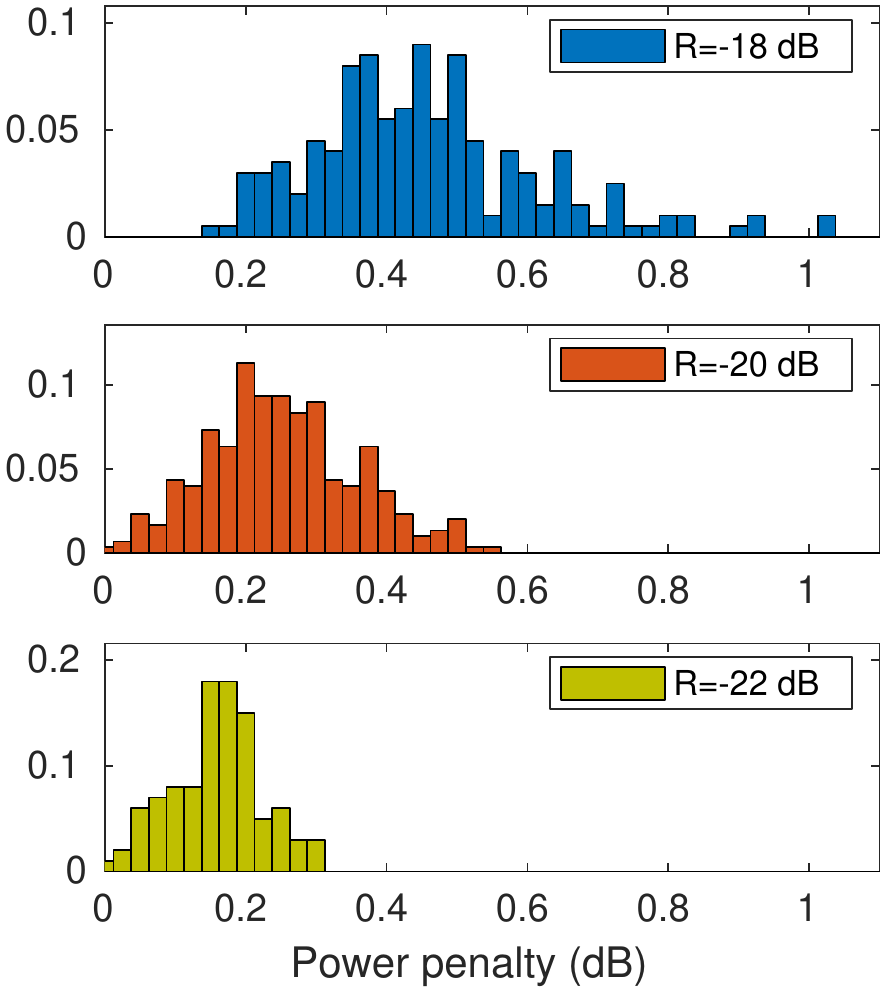}
		\caption{}\label{subfig:threerefhist}
	\end{subfigure}
	\caption{Worst-case over 100 measurements of power penalty for three equal reflections for 28 GBaud PAM-4 (a) and histograms of power penalty
		across different measurements at $\Delta f=56.25$ GHz (b).}\label{fig:pamthreereflections}
\end{figure}
While the main source of penalty are single-reflections, multiple (three) reflections can be detrimental, since crosstalk becomes similar to a Gaussian disturbance \cite{Attard:05}. 
To test this condition, we used the three mirrors $R_1$, $R_2$ and $R_3$ of Fig. \ref{fig:pamexpsetup}. The reflection of each mirror was tuned such that their reflected power 
is equal at the receiver. Results are shown in Fig. \ref{subfig:threeref}, as a function of the total reflectivity $R=R_1+R_2+R_3$, for two different laser spacings, $\Delta f=56.25$ GHz 
(solid red lines) and $\Delta f=25$ GHz (dashed blue lines). Results are compared with single-reflections (circles) at the same value of $R$,
to show the additional penalty of multiple reflections.

In the presence of multiple reflections, the power of the interfering signal is not deterministic anymore, since it is the sum of three contributions with 
a randomly changing phase (due to laser phase noise). Therefore, the penalty is not deterministic as well. Consequently, for each measurement, we captured $100$ waveforms on the
oscilloscope, and Fig. \ref{subfig:threeref} shows the \emph{worst result}. The variation of penalty over different waveforms 
is shown in the histograms of Fig. \ref{subfig:threerefhist} for three values of reflection ($-18$, $-20$ and $-22$ dB) and $\Delta f=56.25$ GHz.

At the $0.5$ dB power penalty line, three reflections have an additional $\sim4$ dB penalty compared to a single reflection. 
This result is expected, since the sum of multiple reflections makes interference close to a Gaussian distribution, which is more detrimental than a single PAM
interferer \cite{Attard:05}. Consequently, 
these results with multiple reflections require an increase of the minimum $\Delta f$ to have acceptable back-reflection penalties. 
While, with a single reflection, it was sufficient a $\Delta f$ larger than symbol rate, in presence of multiple reflections $\Delta f$ should be larger than twice the symbol rate. 
Nevertheless, this value is still within the tolerances of Intra-DC WDM grids.

\section{Conclusion and outlook}
In this chapter, it was proposed a novel spatial-multiplexing architecture, which is able to double the per-laser capacity by using a standard duplex optical cable in both directions. 
The main issue of the proposed scheme are back-reflections, which generate crosstalk between channels transmitted in both directions. 
We found that, by applying a small laser frequency shift, in the order of twice the symbol rate, in one transceiver, the impact of back-reflections is reduced and becomes tolerable. However, it requires a tighter control of laser wavelengths.

The biggest limitation of this experimental evaluation is the absence of a real-time BER measurement, which is necessary to definitively prove the feasibility of the system, and 
will be performed in future research.

\chapter{Single-Sideband Modulation for Inter-DC}\label{ch:ssb}

    \graphicspath{{Chapter4/}}

This Chapter is focused on inter data-center connections, introduced in Sec. \ref{sec:2:inter} of
Chapter \ref{ch:dcarchitect}. First, the Discrete Multitone (DMT) modulation format will be presented as a potential alternative
to PAM for IM/DD systems. 
Then, the single-sideband transmission (and reception) scheme will be presented as a hybrid solution between coherent and IM/DD for
dispersion-uncompensated links. Then, a detailed comparison between IM/DD and SSB, using the DMT modulation format, will be performed
over an $80$-km link. 

Content presented in this Chapter is based on \cite{Randel:15,Pilori:masterthesis,Pilori_JLT:16,Pilori:ICTON2018}.

\section{Discrete multitone modulation}
The goal of Intra-DC connections is transmission of high data-rates over distances in the order of $2-120$ km. 
This operation needs to be done at a cost that is substantially lower than long-haul coherent links. A common option to reduce 
cost is the use of limited-bandwidth components, at the expense of an additional OSNR penalty. 
Since, at those distances, OSNR values are high, this is a feasible solution. 
For this reason, modulation formats that are resilient to bandwidth limitations can be potentially beneficial to Intra-DC links. DMT is, in fact, a modulation format
specifically designed to deal with limited-bandwidth channels.
The goal of this section is briefly describing DMT, focusing only on its main characteristics and differences compared to other modulation formats.

\subsection{Multi-channel transmission}\label{subsec:multichannel}
\begin{figure}
	\centering
	\includegraphics[width=0.8\textwidth]{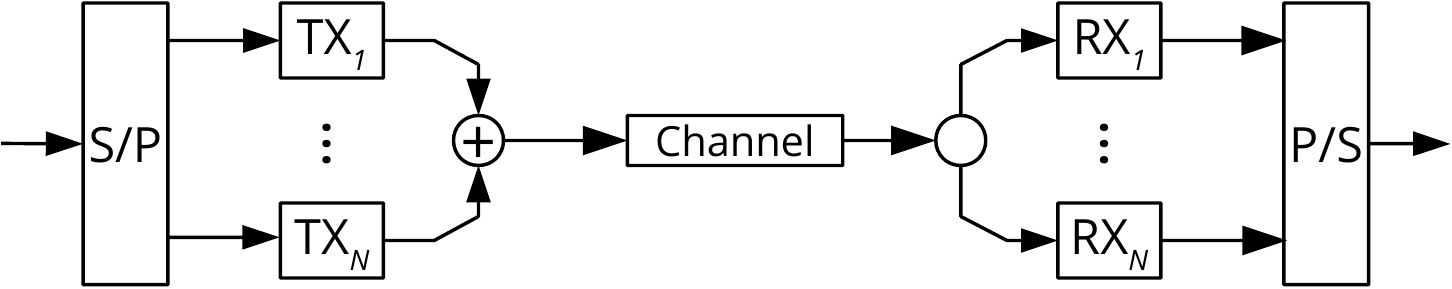}
	\caption{General block diagram of a multi-channel communication system (S/P: serial-to-parallel converter, P/S: parallel-to-serial converter).}\label{fig:multichannelgeneric}
\end{figure}
DMT is a \emph{multi-channel} modulation format. Fig. \ref{fig:multichannelgeneric} shows a block diagram of a generic multi-channel communication system. At the transmitter, information
bits are divided, using a serial-to-parallel converter, into $N$ transmitters. Then, the modulated signals are added and transmitted over the channel. At the receiver, $N$ receivers
detect, respectively, the $N$ transmitted signals. At the end, a parallel-to-serial converter recovers the original stream of bits. These $N$ parallel channels are often called
\emph{subcarriers}.

\begin{figure}
	\centering
	\begin{subfigure}[b]{0.65\textwidth}
		\centering \includegraphics[width=\textwidth]{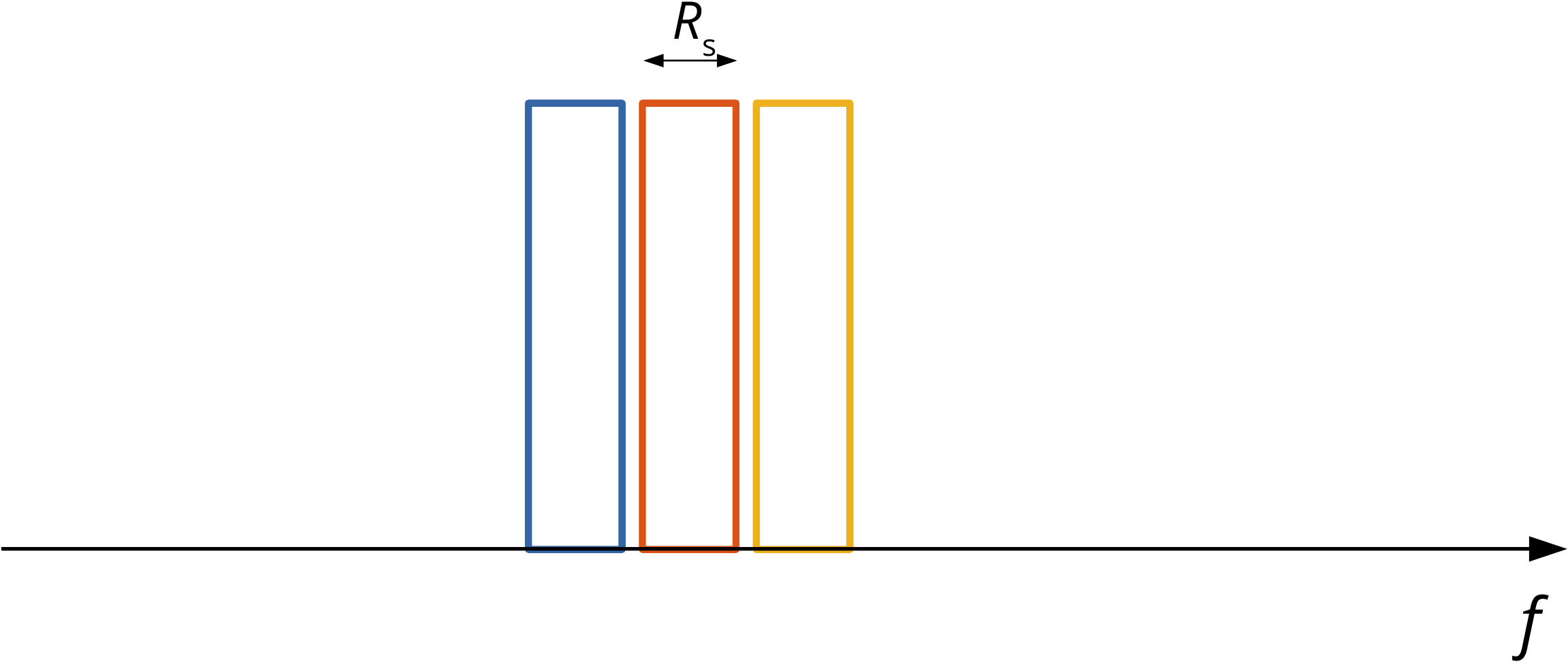}
		\caption{SCM}\label{subfig:scmspectrum}
	\end{subfigure}
	\hfill
	\begin{subfigure}[b]{0.65\textwidth}
		\centering \includegraphics[width=\textwidth]{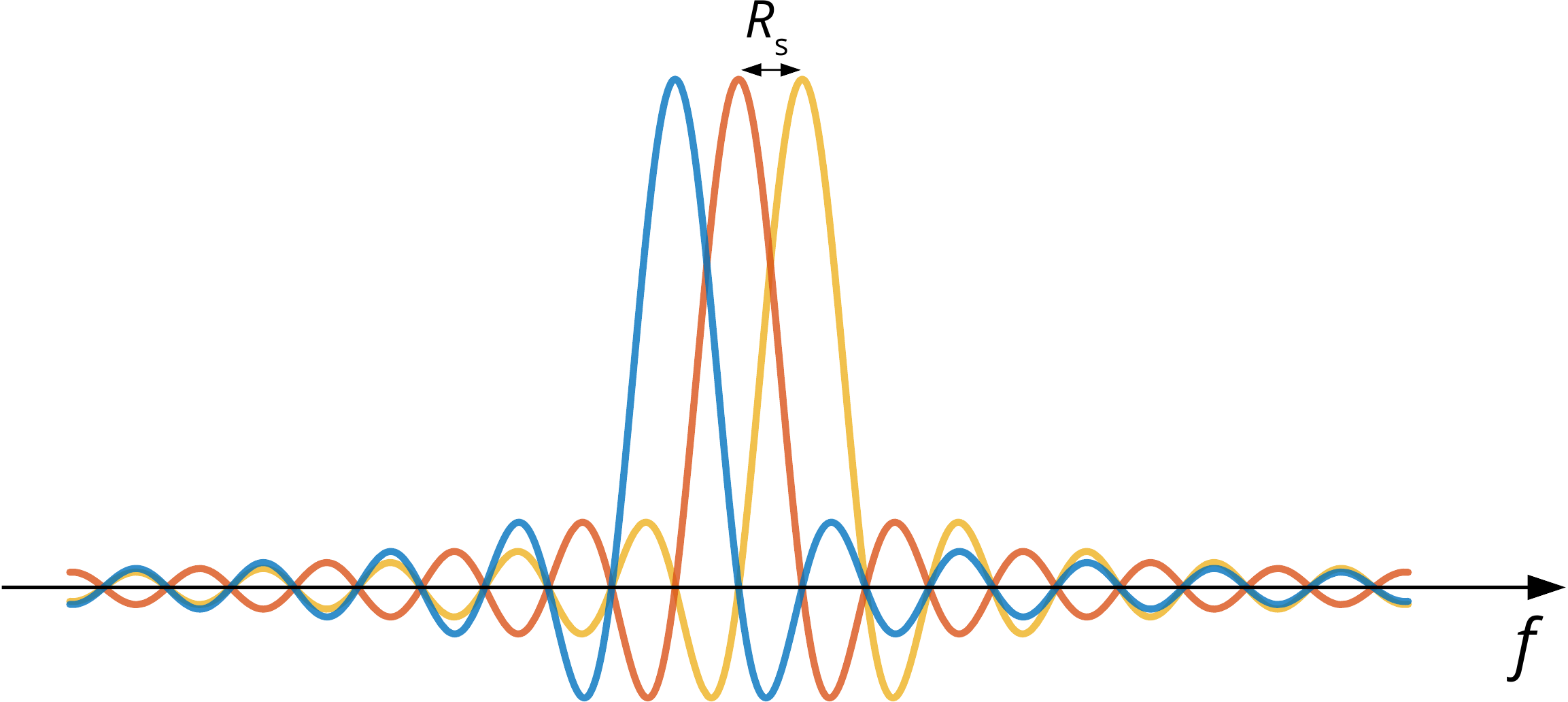}
		\caption{OFDM}\label{subfig:ofdmspectrum}
	\end{subfigure}
	\caption{Spectrum of a three-channel SCM (a) and OFDM (b) signal.}\label{fig:ofdmvsscm}
\end{figure}

It is obvious that the $N$ parallel channels must be separated by a physical dimension to allow detection without interference. 
Usually, the channels are separated in frequency; there are two different methods to achieve frequency separation: Sub-Carrier Multiplexing (SCM), also called Frequency Division Multiplexing (FDM), and Orthogonal FDM (OFDM). In order to highlight the main differences between SCM and OFDM, Fig. \ref{fig:ofdmvsscm} shows the spectra of SCM and OFDM, 
assuming transmission of three channels. 

\paragraph{SCM} Sub-carrier multiplexing fully separates (in frequency) the channels, adding a small guard band between them. 
Channels are usually shaped with a steep (small roll-off) Root-Raised-Cosine (RRC) filter.
SCM requires long (i.e. with many taps) filters at the transmitter and receiver, but it is more robust to various impairments, such as DAC/ADC bandwidth limitations \cite{Bosco:2010}.
Moreover, guard-bands reduce the overall spectral efficiency, making SCM feasible only with a limited number of subcarriers.

\paragraph{OFDM} With OFDM, as opposed to SCM, channels are spaced by exactly the symbol rate, without any guard band. This maximizes spectral efficiency, allowing the use
of large numbers of subcarriers. The receiver is still able to separate, without interference, the channels because they are all \emph{orthogonal} with respect to each other \cite{Shieh:2008}.
However, to preserve orthogonality, the receiver must be tightly synchronized with the transmitter. Moreover, in presence of memory in the channel (i.e. low-pass filtering), 
a cyclic prefix needs to be inserted at each OFDM symbol, reducing spectral efficiency.

\subsubsection{Application to optical communications}
Using coherent transmission and reception, OFDM and SCM can be directly applied to the optical channel, analogously to wireless transmission. 
In the past, these two formats were deeply studied and compared \cite{Shieh:2008,Bosco:2010,Bosco:2011,Shieh:2011}, and it was found that SCM is more suitable than OFDM to the coherent
long-haul transmission scenario. In fact, (at least) one major vendor has successfully implemented SCM in their commercial transponder \cite{Infinera:whitepaper}.

For short-reach applications, single-channel transmission is still widespread. However, with the strong data-rate increase that will be required in the future \cite{Winzer:2017},
 multi-channel modulation can be a potential solution. For this fact, this section will use DMT modulation as an example modulation format for Intra-DC applications.

\subsection{DMT}\label{sec:4:dmtbasic}
In Sec. \ref{subsec:multichannel}, OFDM and SCM were presented from a theoretical point of view, and applied to the coherent long-haul scenario. As discussed in Chapter
\ref{ch:dcarchitect}, coherent detection is still too complex to be widely deployed for Intra-DC applications. Therefore, a direct-detection application of multi-channel transmission
is preferable, and will be discussed in this section.

\subsubsection{Hermitian symmetry}
Discrete Multitone is a variant of OFDM that generates a real-valued (an OFDM signal is complex-valued) signal, which is suitable to be received with a direct-detection
receiver. It was first developed at the beginning of the 1990s \cite{Ruiz:1992} for copper-cable transmission, and subsequently adopted for the Digital Subscriber Line (DSL) standards.
\begin{figure}
	\centering
	\includegraphics[width=0.5\textwidth]{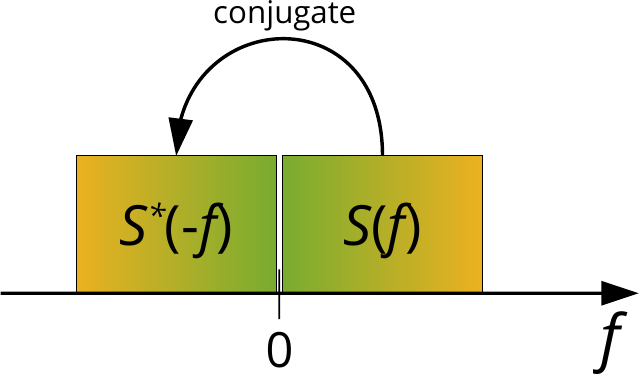}
	\caption{Electrical spectrum of the generation of a DMT signal from an OFDM signal.}\label{fig:dmtfromofdm}
\end{figure}
With DMT, assuming to have $f=0$ at the center of the spectrum, negative-frequency subcarriers are not modulated.
Instead, the negative-frequency spectrum is a mirrored and conjugated copy of the positive-frequency spectrum. This operation is shown 
in Fig. \ref{fig:dmtfromofdm}, where the positive-frequency portion of an OFDM signal $S(f)$ is mirrored and conjugated ($S^*(-f)$) to obtain the negative-portion
of the spectrum. According to signal theory, this property is called ``Hermitian symmetry'', which generates a real-valued time-domain signal. 
Obviously, this operation halves spectral efficiency, which is inevitable due to the loss of one dimension (the imaginary part). 

\subsubsection{Bit and power loading}
DMT is usually employed in strongly frequency-selective channels, such as DSL. In this scenario, transmission quality (i.e. the SNR) is different among different subcarriers, strongly
reducing performances if the same power and modulation format is applied to each subcarrier. Applying the ``water-filling'' power allocation \cite{proakis2007digital}, 
which assigns more power to higher-quality subcarriers, further enlarges the SNR difference between them.

This issue is solved by using \emph{bit loading}, i.e. using different modulation formats on different subcarriers, based on their SNR. The most common bit-loading algorithm is the
Levin-Campello algorithm \cite{Campello:1999,Levin:2001}. A detailed overview of the algorithm, along with a pseudo-code representation, is available on \cite[Chapter 4]{Pilori:masterthesis}.

\subsection{Comparison with PAM}
A comparison between DMT and PAM in a direct-detection scenario is out of the scope of this thesis. Nevertheless, in the past years, several comparisons have been published. Recent experiments \cite{Zhong:2015,Eiselt:2018} found, on average, similar performances if non-linear equalization (e.g. decision-feedback equalization) is applied to PAM.
At the end, the actual choice of modulation format will strongly depend on the scenario, optical components and maximum allowable DSP complexity.

\section{Single side-band}
As discussed in Sec. \ref{sec:2:cohint}, SSB is a potential ``hybrid'' between coherent systems and IM/DD, that is suitable for dispersion-uncompensated links.
In this Chapter, SSB will be compared to direct detection, using the DMT modulation format.  
However, it is important to remark that SSB is modulation-format agnostic, and it can be potentially applied to any format.

\subsection{Basic principle}\label{sec:4:ssbbasic}
\begin{figure}
	\centering
	\includegraphics[width=0.6\textwidth]{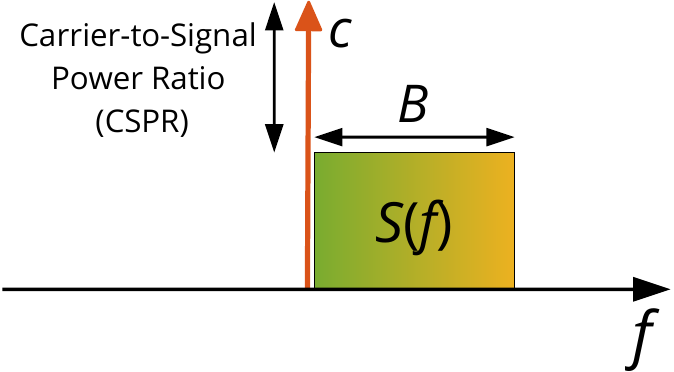}
	\caption{Optical spectrum of a generic SSB signal.}\label{fig:ssbgenericspectrum}
\end{figure}
SSB is a technique that, in general, allows detection of the full optical field (magnitude and phase) with a direct detection receiver. 
The optical spectrum of an SSB signal is shown in Fig. \ref{fig:ssbgenericspectrum}. In the Figure, $S(f)$ is a generic complex-valued coherent signal (e.g. OFDM, QAM, \dots), and
$c$ is a carrier (e.g. laser light), with higher power than the signal. The ratio between the power of $c$ and signal power is called
Carrier-to-Signal Power Ratio (CSPR). Application of SSB to real-valued signals, such as DMT, will be detailed in the next sections.

Assuming a complex-envelope representation around a central frequency set to the middle of the useful signal spectrum,
the time-domain electric field of the signal in Fig. \ref{fig:ssbgenericspectrum} can be written as:
\begin{equation}
E(t) = s(t) + c e^{-j2\pi t B/2}
\label{eq:ssbgeneric}
\end{equation}
where $B$ is the bandwidth of $s(t)$. After direct-detection with an ideal photodiode, the photocurrent $i(t)$ becomes
\begin{equation}
i(t) = |s(t)|^2 + c^2 + 2c\cdot\Re\left\{ s(t) e^{j2\pi t B/2} \right\}
\label{eq:ssbgenericpd}
\end{equation}
This equation has three terms:
\begin{itemize}
\item $c^2$ is a DC component, that can be easily suppressed at the receiver.
\item $2c\cdot\Re\left\{ s(t) e^{j2\pi t B/2} \right\}$ is the complex up-conversion of $s(t)$ at frequency $B/2$, allowing reconstruction of the complex-valued $s(t)$ signal in the DSP. This comes at the expense of a larger receiver bandwidth.
\item $|s(t)|^2$ is an interference term, called Signal-Signal Beating Interference (SSBI), and it must be suppressed at the receiver.
\end{itemize}

In conclusion, SSB allow detection of a single-polarization coherent signal with a direct-detection receiver. This operation increases spectral efficiency, becoming equal to
a single-polarization homodyne coherent system. 
Moreover, it allows electronic dispersion compensation. The main drawbacks are two: increased receiver analog bandwidth (double with
respect to an equivalent coherent system), and SSBI.

\subsection{SSBI compensation}
There are serveral methods to compensate (or avoid generation) of SSBI. The simplest method is based on the observation that $|s(t)|^2$ is a low-pass signal. Then, by introducing an appropriate
frequency gap between $s(t)$ and $c$, SSBI impact can be reduced \cite{Peng:2009}. However, this solution requires an even-larger receiver bandwidth, making it
unsustainable for high data-rate systems. Another possibility is the use of balanced photodiodes to optically remove SSBI \cite{Ma:2013}. However, 
this operation requires a complete optical separation between signal and carrier. This needs steep optical filters and a frequency gap between signal and carrier.
The most promising solution is digital SSBI compensation. In literature, several methods have been proposed \cite{Peng:2009,Randel:15,Li:2016} with different complexity and effectiveness.
Another option to avoid SSBI is the use of a Kramers-Kronig receiver \cite{Mecozzi:2016}, which is a DSP reception technique that, at the expense of higher DSP complexity, is able
to fully suppress SSBI.

A comparison between SSBI compensation techniques is out of the scope of this thesis; the interested reader can refer e.g. to \cite{Li:PhDthesis}. For this thesis, it was adopted
the method presented in \cite{Randel:15,Randel:patent}, which offers good performances with a very low DSP complexity.

\subsection{Generation of SSB signals}
\begin{figure}
	\centering
	\begin{subfigure}[b]{0.45\textwidth}
		\centering \includegraphics[width=0.5\textwidth]{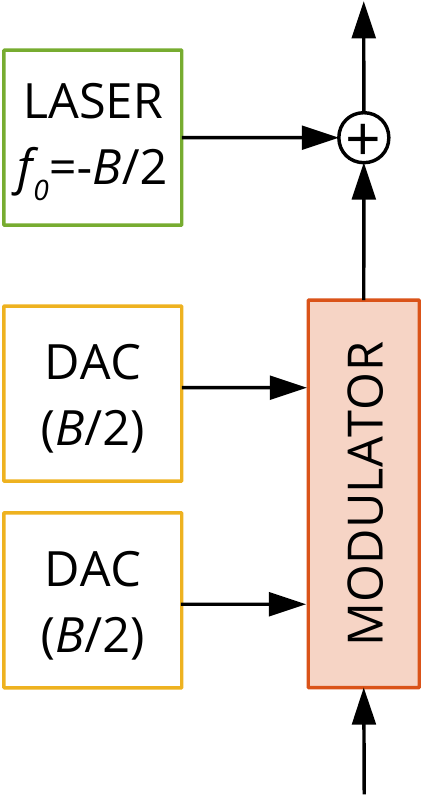}
		\caption{}\label{subfig:ssbgen1}
	\end{subfigure}
	\begin{subfigure}[b]{0.45\textwidth}
		\centering \includegraphics[width=0.5\textwidth]{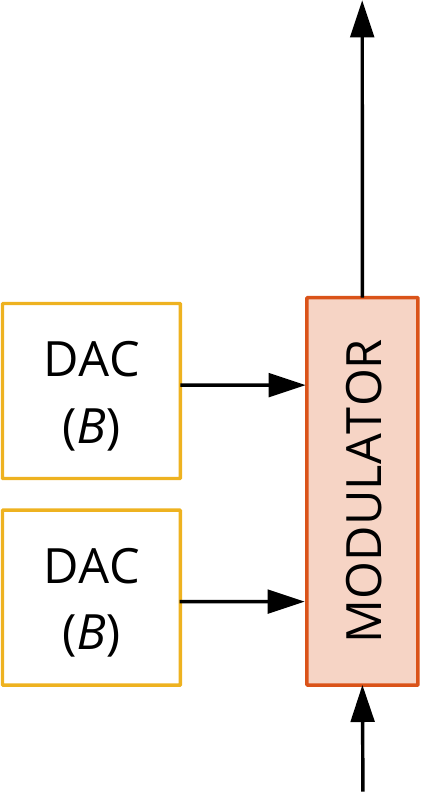}
		\caption{}\label{subfig:ssbgen2}
	\end{subfigure}
	\caption{Possible methods to generate an SSB signal.}\label{fig:ssbgeneration}
\end{figure}
The transmitter is a critical part in the design of a SSB-based transmission system, since it requires the addition of a strong carrier at one edge of the signal spectrum. The two main
methods to achieve this are summarized in Fig. \ref{fig:ssbgeneration}. With the simplest method, shown in Fig. \ref{subfig:ssbgen1}, a second laser, with a $B/2$ frequency difference
between the transmit laser, adds the carrier after I/Q modulation. In this case, the signal is generated by two DACs with bandwidth $B/2$, like a coherent transmitter. However, this
solution requires two lasers, and a tight control of their frequency difference.

An alternative method is shown in Fig. \ref{subfig:ssbgen2}. In this case, the two DACs require twice the bandwidth ($B$) to generate an SSB signal. Then, the carrier is added by 
intentionally leaking un-modulated light from the transmit laser. This is usually achieved by biasing the modulator to an appropriate point \cite{Pilori:ICTON2018}.
Other than bandwidth requirements, this solution increases modulator-induced non-linear distortions, but removes the requirement of a second laser.

In conclusion, the optimal transmitter structure for SSB strongly depends on the choice (and cost) of each component of the system. For the rest of this Chapter, it will be assumed
an ideal transmitter.

\subsection{SSB and DMT}
In Sec. \ref{sec:4:ssbbasic}, SSB was presented from a general point of view. In particular, \eqref{eq:ssbgeneric} defined a SSB electric field $E(t)$ using a \emph{complex}-valued
signal $s(t)$. On the other end, Sec. \ref{sec:4:dmtbasic} defined DMT as a real-valued signal. Apparently, this is a contradiction that makes DMT not compatible with SSB.
With a slight abuse of notation, in an SSB-DMT system, only the \emph{electrical} signal \emph{after} the photodiode (neglecting DC and SSBI) is a DMT signal.
In fact, the term SSB-DMT is not even universally adopted in the literature (e.g. \cite{Schmidt:2008}).

For example, let us consider a general real-valued signal $x(t)$ (e.g. DMT). The I/Q modulator generates the following SSB optical signal:
\begin{equation}
E_\textup{TX}^\textup{(ssb)}(t) = \frac{1}{\sqrt{2}} \left[ x(t) + j\hilb\left\{x(t)\right\} \right] + c
\label{eq:ssbtx}
\end{equation}
In this equation, $\hilb(.)$ denotes \emph{Hilbert transform}, defined as the convolution with a linear filter whose frequency response is
\begin{equation}
H_\textup{hilb}(f) =
\begin{cases}
j & f>0\\
0 & f=0\\
-j & f<0\\
\end{cases}
\end{equation}
The addition, on the imaginary axis, of the Hilbert transform of $x(t)$ eliminates the negative frequencies of $x(t)$. Since the negative spectrum of any real-valued signal
is a copy of the positive spectrum, this operation does not lose any information. The spectrum of the signal is identical to the spectrum of an SSB signal 
\eqref{eq:ssbgeneric}, except from a $B/2$ frequency shift. Detecting this signal with an ideal photodiode generates a photocurrent:
\begin{equation}
i(t) = 2c \cdot x(t) + c^2 + \left[x^2(t) + \hilb^2\left\{x(t)\right\}\right]
\label{eq:ssbcurrentnonoise}
\end{equation}
This equation is equivalent to \eqref{eq:ssbgenericpd}, since it contains the original signal $x(t)$, the DC component and SSBI.

This notation allows the application of SSB to any real-valued modulation format (such as DMT or PAM). It is different than the ``traditional''
SSB notation introduced in Sec. \ref{sec:4:ssbbasic}, but equivalent to it. Since the purpose of this Chapter
is the comparison with IM/DD systems, which use real-valued signals, this is the notation that will be adopted.

\section{Back-to-back comparison}
After introducing DMT and SSB, this Section will be devoted to a detailed comparison between intensity modulation and SSB in a back-to-back scenario. In this section, 
another transmission technique, called Vestigial Side-Band (VSB), will be introduced, as a hybrid between IM and SSB. To highlight the differences between IM and VSB/SSB, Intensity
Modulation will be called Dual Side-Band (DSB). The comparison will be first performed using analytical evaluations, which will be then validated
 using time-domain DMT numerical simulations. 

\subsection{Methodology}
Comparison of different transmission techniques (IM, VSB and SSB) is, in general, not a trivial operation. 
Therefore, some assumptions have to be made in order to perform a fair comparison.

\subsubsection{Modulation}
Simulations will adopt the DMT modulation format. In general, evaluation of system performances with DMT requires transmission of two signals. First, a test DMT signal, using a fixed
modulation format (e.g. 16-QAM) on all subcarriers, is transmitted. The receiver uses this test signal to calculate the per-subcarrier SNR, which is then fed to the bit and power loading
algorithms. These algorithms generate a table which, for each subcarrier, gives the optimal modulation format and relative transmit power. This table is then given to the transmitter, which
generates the ``real'' transmit signal. Then, the receiver measures the BER, comparing it with the FEC threshold. 

It is obvious that this operation is quite onerous in terms of complexity. Therefore, for this comparison, we adopted a simplified approach, based on the ``equivalent SNR'' principle. According
to theory \cite{Cioffi:book}, an ideal multi-channel modulation achieves the same performance as a single-channel transmission with an equivalent SNR which is the \emph{geometric mean}
of the per-subcarrier SNR:
\begin{equation}
\mathrm{SNR}_\textup{eq} = \left( \prod_{i=1}^{N} \mathrm{SNR}_i \right)^{1/N}
\end{equation}
This value is easy to evaluate if the SNRs are expressed in decibels; in this case, the equivalent SNR is simply the arithmetic mean of the per-subcarrier SNRs.

\begin{table}
	\centering
	\begin{tabular}{c c}
		\toprule
		Parameter & Value \\
		\midrule
		Modulation & $16$-QAM \\
		DAC sampling rate & $R_s=64$ Gs/s \\
		FFT size & $512$ \\
		Modulated subcarriers & $255$ \\
		BER threshold & $10^{-3}$ \\
		SNR threshold & $16.54$ dB\\
		\bottomrule
	\end{tabular}
	\caption{Simulation parameters for DMT modulation.}\label{tab:dmtsimparams}
\end{table}
Therefore, in this comparison we transmitted only a test signal, whose parameters are summarized in Table \ref{tab:dmtsimparams}. The receiver evaluates the equivalent SNR by direct comparison with
the transmitted signal, which is then used as performance indicator.

\subsubsection{Simulation setup}
\begin{figure}
	\centering
	\includegraphics[width=0.9\textwidth]{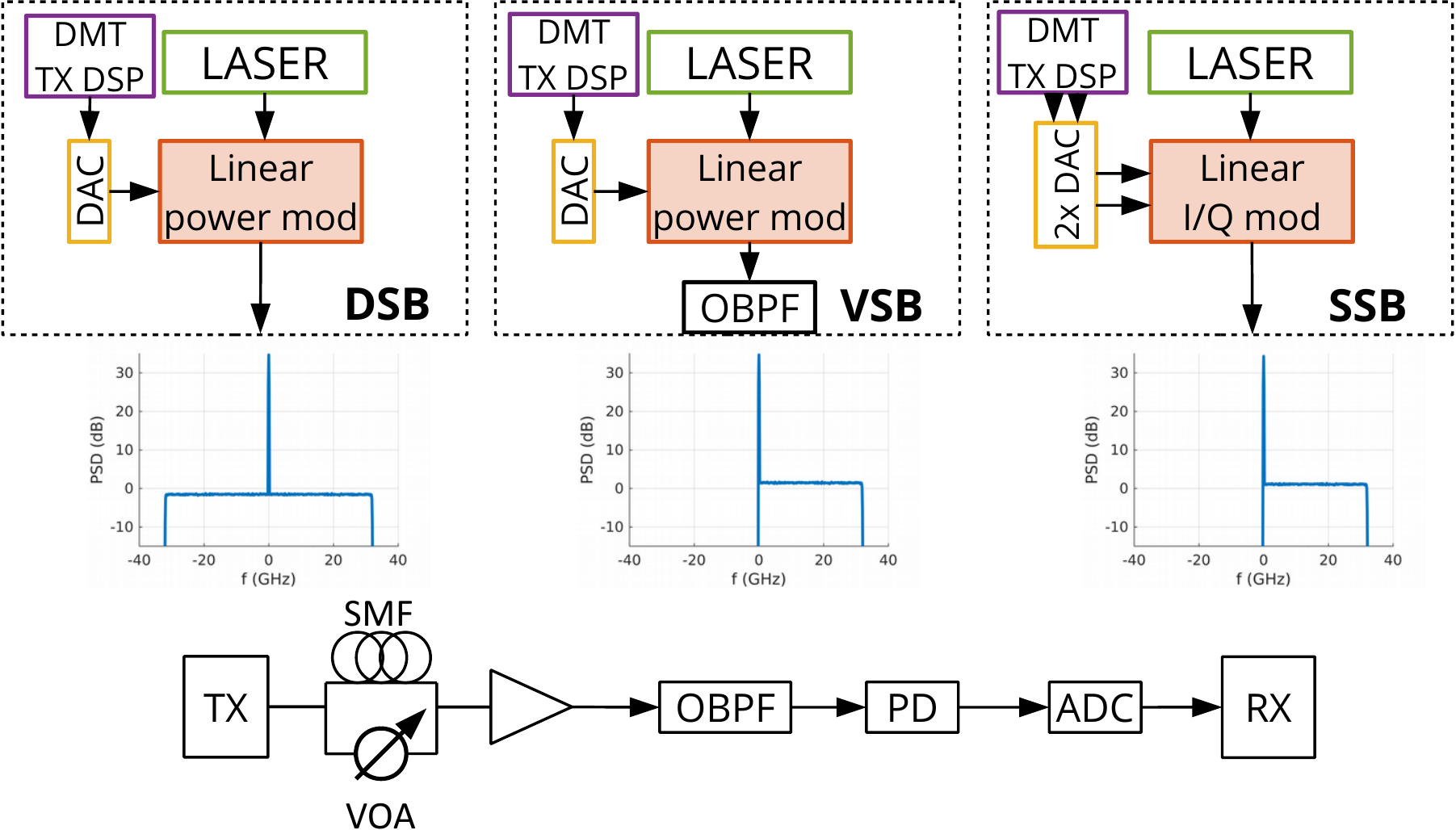}
	\caption{Spectrum of a generic SSB signal.}\label{fig:ssbsimsetup}
\end{figure}
The general schematic of the simulations performed in this Chapter is shown in Fig. \ref{fig:ssbsimsetup}. On the top, there are the three transmitter structures of the three
transmission methods: DSB, VSB and SSB. DSB (synonym for intensity modulation), is the simplest structure, since it requires one DAC and a linear power modulator, which linearly
converts the voltage from the DAC to an optical power. VSB, with respect to DSB, adds an optical band-pass filter (OBPF) to suppress one side-band, emulating a SSB signal.
On the other end, SSB requires (at least) two DACs (same as Fig. \ref{subfig:ssbgen2}) and a linear I/Q modulator, which converts DAC voltages to an electric field.
Examples of optical spectra after the transmitters are shown in the middle of Fig. \ref{fig:ssbsimsetup}.

Transmission line (bottom of Fig. \ref{fig:ssbsimsetup}) is very simple. After the transmitter, signal is transmitted over a certain length of optical fiber, amplified with an EDFA, 
band-pass filtered with an ideal OBPF and detected with a photodiode. For the back-to-back comparisons, fiber is replaced by a VOA.
 The presence of an EDFA adds ASE noise, which will be taken into account in the OSNR. Since, for a DMT signal, it is not
trivial to define a symbol rate, the OSNR will be evaluated on a $0.1$ nm bandwidth, corresponding to $B_o\simeq12.5$ GHz in the C-band.

\subsection{SSB performance}\label{sec:4:ssbb2b}
Let us consider transmission of an SSB signal $E_\textup{TX}^\textup{(ssb)}(t)$, defined in \eqref{eq:ssbtx}. The most important parameter in an SSB transmission
is the CSPR, as shown in Fig. \ref{fig:ssbgenericspectrum}.
Since the Hilbert transform is an all-pass filter, power of the real and imaginary parts are the same and equal to $\sigma^2_x/2$, where $\sigma^2_x$ is the variance of $x(t)$.
Therefore, the CSPR of an SSB system can be expressed as
\begin{equation}
\xi^\textup{(ssb)} = \frac{c^2}{\sigma^2_x}
\label{eq:ssbcspr}
\end{equation}

\subsubsection{Direct detection}
This signal is transmitted in the back-to-back setup shown in Fig. \ref{fig:ssbsimsetup}. ASE noise $n(t)$ is added by the EDFA on both polatizations.
Then, the signal is ideally band-pass filtered to remove out-of-band noise and detected by the photodiode. Assuming, for simplicity, unit responsivity, the photocurrent is
\begin{equation}
i^\textup{(ssb)}(t) = \left\lvert E_\textup{TX}^\textup{(ssb)}(t) + n_\textup{X}(t) \right\rvert^2 + |n_\textup{Y}(t)|^2
\end{equation}
Due to the OBPF, noise-noise beating can be neglected. Removing this contribution, and the DC tone $c^2$, the photocurrent can be expressed as
\begin{equation}
i^\textup{(ssb)}_0(t) \approx \frac{1}{2} \left[ x^2(t) + \hilb^2\{x(t)\} \right] + 2c\left[ \frac{1}{\sqrt{2}} x(t) + \Re\{n_\textup{X}(t)\} \right]
\label{eq:ssbphotocurrent}
\end{equation}
This equation contains the usual two terms: $x(t)$ and SSBI. The main difference with respect to \eqref{eq:ssbcurrentnonoise} is the presence
of ASE noise, which allows evaluation of the equivalent SNR $E_s/N_0$ from the OSNR. Assuming that the SSBI compensation algorithm is able
to completely remove SSBI, the SNR can be expressed as
\begin{equation}\label{eq:ssbosnr}
E_s/N_0 = \mathrm{OSNR}\frac{4B_o}{R_s(1+\xi^\textup{(ssb)})}
\end{equation}

\subsubsection{DMT simulations}
\begin{figure}
	\centering
	\includegraphics[width=0.5\textwidth]{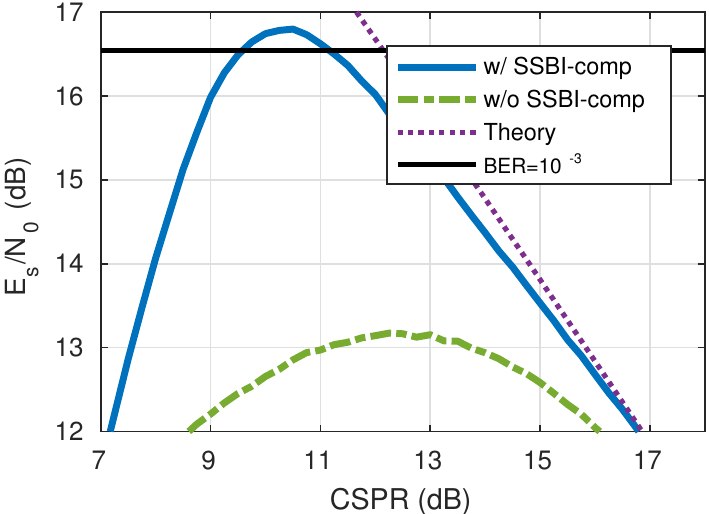}
	\caption{Equivalent SNR as a function of the CSPR of a SSB-DMT signal transmitted in a back-to-back configuration. $\mathrm{OSNR}_{0.1~\mathrm{nm}}=30$ dB.}\label{fig:ssbb2bcomp}
\end{figure}
As previously discussed, the analytical result in \eqref{eq:ssbosnr} will be validated using the DMT modulation, whose
parameters were summarized in Table \ref{tab:dmtsimparams}. We remark again that the choice of DMT modulation
is just an example, since SSB can be potentially combined with any modulation format.

Results are shown in Fig. \ref{fig:ssbb2bcomp}, where the equivalent SNR is shown as a function of the CSPR for a fixed OSNR, equal to $30$ dB measured over a $0.1$-nm bandwidth.
Fig. \ref{fig:ssbb2bcomp} contains the analytical result \eqref{eq:ssbosnr} (dotted line) and DMT simulations,
without (dashed-dotted line) and with SSBI compensation \cite{Randel:15} (solid line). 

At low values of CSPR, main noise source is SSBI. This is explained the $c$-factor that multiplies $x(t)$ in \eqref{eq:ssbphotocurrent}. 
Since \eqref{eq:ssbosnr} does not take into account SSBI, the formula is not accurate at low CSPRs. For high values of CSPR, instead, performance is dominated by ASE
noise, and \eqref{eq:ssbosnr} becomes an accurate upper-bound of performances. Therefore, there exist an optimal CSPR, which represents a balance between SSBI and ASE noise.
Moreover, Fig. \ref{fig:ssbb2bcomp} clearly shows the effectiveness of SSBI compensation. Even if the implemented algorithm \cite{Randel:15} is one of the simplest, it
brings a $\sim3.5$ dB SNR gain at the optimal CSPR. This significant advantage makes SSBI compensation a strict requirement for SSB systems.

\subsection{DSB performance}
An intensity-modulated, also called DSB, signal can be expressed as
\begin{equation}
E_\textup{TX}^\textup{(dsb)}(t) = \sqrt{ \bar{P} \left\{   1+\clip\left[\frac{x(t)}{c'}\right] \right\}  }
\label{eq:dsbtxsignal}
\end{equation}
In this equation, $\bar{P}$ is the average optical power and $\clip(.)$ is a clipping function between $-1$ and $1$, defined as:
\begin{equation}
\clip(\alpha) = \begin{cases}
\alpha & |\alpha|\leq 1 \\
\sign(\alpha) & |\alpha|> 1 \\
\end{cases}
\end{equation}

A first comparison between DSB and SSB can be done by comparing the transmitted signals \eqref{eq:dsbtxsignal} and \eqref{eq:ssbtx}. 
DSB requires a square-root operation, which needs clipping to avoid zero-crossings. This means that, while generation of an SSB signal is performed
using only linear operations (Hilbert transfrom), transmission of a DSB signal requires \emph{two} non-linear operations (clipping and square root). These
operations, as it will be shown later, require several approximations in order to derive analytical results.

The parameter $c'$ is used to tune the amount of clipping. It is customary to define a \emph{clipping ratio}
\begin{equation}
R_\textup{cl} = \frac{c'^2}{\sigma^2_x}
\end{equation}
which means that the signal $x(t)$ is clipped between $\pm\sqrt{R_\textup{cl}\sigma^2_x}$. The clipping ratio is an important design parameter of a DSB-DMT system
\cite{Nadal:2014}, since a DMT signal has a higher Peak-to-Average Power Ratio (PAPR) compared to single-channel modulation formats. 

\subsubsection{Clipping ratio and CSPR}
The definition of clipping ratio is surprisingly similar to the CSPR for SSB \eqref{eq:ssbcspr}. In fact, they both refer to a ratio between carrier and modulated power.
Therefore, in order to perform a fair comparison between SSB and DSB, it is important to find a relation between these two parameters. For this derivation, we
will assume that $x(t)$ is a DMT signal, which means that $x(t)$ is assumed to be Gaussian-distributed. 

For a generic signal $E(t)$, the CSPR can be defined as the ratio between the un-modulated and modulated optical power; in formula,
\begin{equation}
\xi = \frac{\expt^2\left\{E(t)\right\}}{\var\left\{E(t)\right\}}
\label{eq:csprdefinition}
\end{equation}
where $\expt(.)$ defines expectation and $\var(.)$ variance.
Substituting the definition of DSB signal \eqref{eq:dsbtxsignal} into \eqref{eq:csprdefinition} allows obtaining an expression of the CSPR for a DSB signal. 
However, \eqref{eq:dsbtxsignal} contains two non-linear operations, square root and clipping. Therefore, some approximations needs to be performed.

First, the square root can be expanded with a Taylor's series expansion, assuming $R_\textup{cl}\gg1$:
\begin{equation}
E_\textup{TX}^\textup{(dsb)}(t) \approx \sqrt{\bar{P}} \left[ 1 + \frac{y(t)}{2} - \frac{y^2(t)}{8} + \frac{y^3(t)}{16} - \frac{5y^4(t)}{128} \right]
\label{eq:dsbtaylor}
\end{equation}
where
\begin{equation}
y(t) = \clip\left[\frac{x(t)}{c'}\right]
\label{eq:dsbclipdef}
\end{equation}

If $R_\textup{cl}\gg1$, then clipping effects can be neglected, and $y(t)$ can be assumed a Gaussian random variable (like $x(t)$) with zero
mean and variance $\sigma^2_y$. Substituting \eqref{eq:dsbtaylor} into \eqref{eq:csprdefinition} allows obtaining:
\begin{equation}
\xi^\textup{(dsb)} \approx \frac{\expt^2\left\{1 + \frac{y(t)}{2} - \frac{y^2(t)}{8} + \frac{y^3(t)}{16} - \frac{5y^4(t)}{128}\right\}}
{\var\left\{1 + \frac{y(t)}{2} - \frac{y^2(t)}{8} + \frac{y^3(t)}{16} - \frac{5y^4(t)}{128}\right\}}
\end{equation}
After simple mathematical operations, the final result is
\begin{equation}
\xi^\textup{(dsb)} \approx \frac{\left(1-\frac{1}{8}\sigma^2_y - \frac{15}{128}\sigma^4_y \right)^2}{1-\left(1-\frac{1}{8}\sigma^2_y - \frac{15}{128}\sigma^4_y \right)^2}
\end{equation}

\begin{figure}
	\centering
	\includegraphics[width=0.5\textwidth]{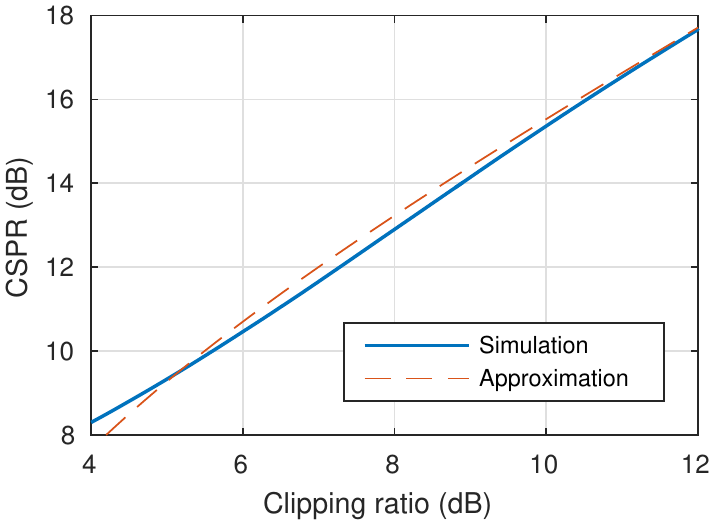}
	\caption{Relation between CSPR and clipping ratio for a DMT signal.}\label{fig:csprclipping}
\end{figure}

Then, neglecting clipping, the variance of $y(t)$ is
\begin{equation}
\sigma^2_y \approx \frac{\sigma^2_x}{c'^2} = \frac{1}{R_\textup{cl}}
\end{equation}
This expression allows a fully-analytical relation between clipping ratio and CSPR. Fig. \ref{fig:csprclipping} shows an example, where the analytical expression
is compared with a numerical simulation, assuming DMT modulation. For high values of CSPR, the analytical expression is very accurate, making it suitable
for comparing DSB and SSB modulations at the same values of CSPR.

\subsubsection{Direct detection}
With the same assumptions made in Sec. \ref{sec:4:ssbb2b}, the phtocurrent of a DSB signal can be expressed as
\begin{equation}
i^\textup{(dsb)}_0(t) \approx \bar{P}\clip\left[\frac{x(t)}{c'}\right] + 
2\Re\{n_\textup{X}(t)\} \cdot \sqrt{\bar{P} \left\{ 1+ \clip\left[\frac{x(t)}{c'}\right] \right\} }
\label{eq:dsbphotocurrent}
\end{equation}

From this expression, neglecting the effects of clipping, the equivalent SNR can be expressed as:
\begin{equation}\label{eq:dsbeqsnr}
E_s/N_0 \approx \mathrm{OSNR}\frac{B_o}{R_\textup{cl}R_s}
\end{equation}
Using the relations derived before, it is possible to convert the clipping ratio to the CSPR, allowing a direct comparison between SSB and DSB. A back-to-back result with DSB-DMT
modulation will be shown in the next section, where it will be compared with SSB and VSB modulation.

\subsection{VSB performance}\label{sec:4:vsbdef}
VSB is a ``hybrid'' transmission scheme between DSB and SSB. It uses the simple DSB transmitter structure, but it adds a steep optical filter that suppresses negative frequencies.
While, at a first sight, the signal may seem identical to SSB, it is not. The main difference is the presence of the two non-linear
operations (square root and clipping), not present in SSB. As it will be shown later, these operations decrease the effectiveness of VSB transmission.

Assuming an ideal filter, a VSB signal is expressed as a function of the DSB transmit signal \eqref{eq:dsbtxsignal}:
\begin{equation}
E_\textup{TX}^\textup{(vsb)}(t) = \frac{1}{\sqrt{2}}\left[ E_\textup{TX}^\textup{(dsb)}(t) + j\hilb\left\{E_\textup{TX}^\textup{(dsb)}(t)\right\} \right]
\label{eq:vsbtxsignal}
\end{equation}

\subsubsection{Direct detection}
Applying the same approximations as previous sections, the photocurrent can be expressed as
\begin{align}
i^\textup{(vsb)}_0(t) &\approx \frac{\bar{P}}{2}\clip\left[\frac{x(t)}{c'}\right] + 
\Re\{n_\textup{X}(t)\} \cdot \sqrt{2\bar{P} \left\{ 1+ \clip\left[\frac{x(t)}{c'}\right] \right\} }+
 \nonumber \\
&+ \frac{\bar{P}}{2} \hilb^2\left\{ \sqrt{\bar{P} \left\{ 1+ \clip\left[\frac{x(t)}{c'}\right] \right\} }  \right\}\label{eq:vsbphotocurrent}
\end{align}

Let us compare this equation with DSB photocurrent \eqref{eq:dsbphotocurrent}. The first two terms, apart from scaling factors, are identical. 
However, there is a third quadratic term, which represents interference, and it is similar to SSBI in SSB systems. For simplicity, this term
will be called SSBI as well.

A very simple compensation algorithm can be designed, similarly to \cite{Randel:15}. The VSB ``SSBI'' can be estimated with
\begin{equation}
\hat{x}_\textup{SSBI}(t) = \gamma \hilb^2\left\{ i^\textup{(vsb)}_0(t) - \frac{1}{4}\left[i^\textup{(vsb)}_0(t)\right]^2 \right\}
\end{equation}
where $\gamma$ is a free optimization parameter.

\subsubsection{DMT simulations}
\begin{figure}
	\centering
	\includegraphics[width=0.5\textwidth]{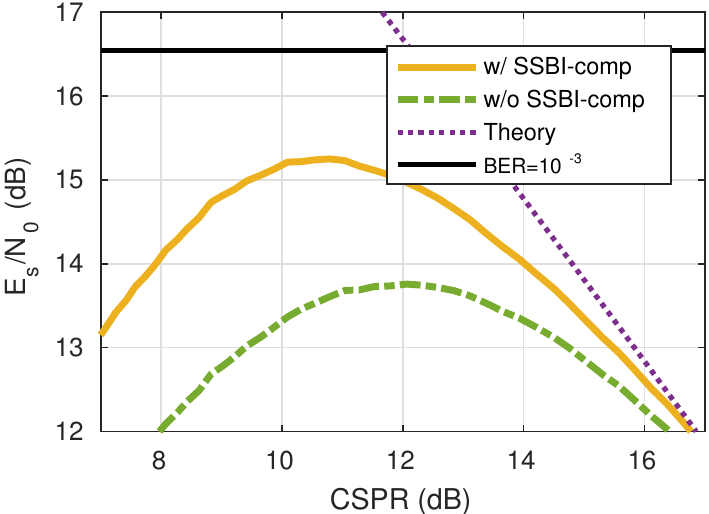}
	\caption{Equivalent SNR as a function of the CSPR of a VSB-DMT signal transmitted in a back-to-back configuration. $\mathrm{OSNR}_{0.1~\mathrm{nm}}=30$ dB.}\label{fig:vsbb2bcomp}
\end{figure}
Fig. \ref{fig:vsbb2bcomp} shows the equivalent SNR of a VSB-DMT system, in back-to-back at a fixed $\mathrm{OSNR}_{0.1~\mathrm{nm}}=30$ dB.
The CSPR have been calculated from the clipping ratio using the relations obtained for DSB. 
These results are similar to SSB (Fig. \ref{fig:ssbb2bcomp}): at low CSPR, system is mainly impaired by SSBI. At high CSPR, system is impaired by ASE noise. SSBI
compensation is somehow less effective than SSBI, but gives still a significant performance improvement.

\subsection{Final comparison}
\begin{figure}
	\centering
	\begin{subfigure}[b]{0.49\textwidth}
		\centering \includegraphics[width=\textwidth]{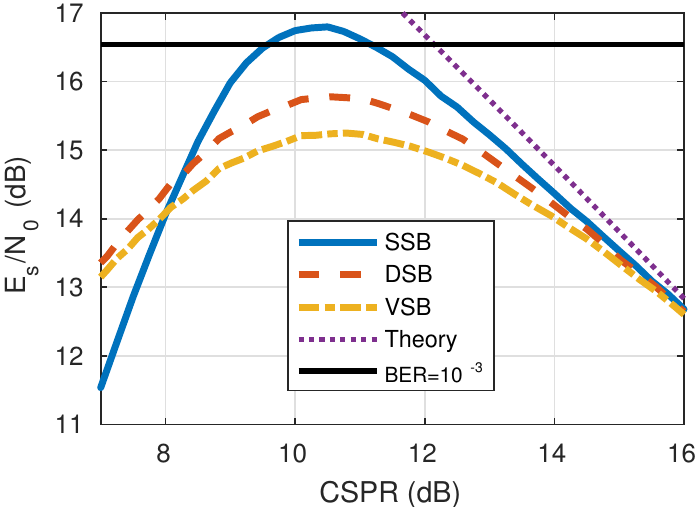}
		\caption{}\label{subfig:b2bfixosnr}
	\end{subfigure}
	\begin{subfigure}[b]{0.49\textwidth}
		\centering \includegraphics[width=\textwidth]{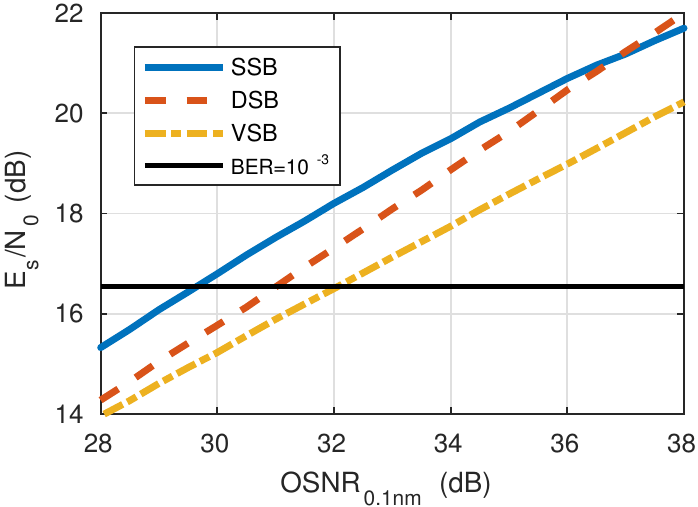}
		\caption{}\label{subfig:b2bsweep}
	\end{subfigure}
	\caption{Back-to-back comparison of SSB, DSB and VSB DMT with the same data-rate. (a): comparison at fixed $\mathrm{OSNR}_{0.1~\mathrm{nm}}=30$ dB. (b):
	comparison at optimal CSPR for different OSNR. }\label{fig:fullb2bcomp}
\end{figure}
With the relation between CSPR and clipping ratio, DSB, VSB and SSB can be directly compared at fixed OSNR, fixed modulation format and CSPR. Results
are shown in Fig. \ref{subfig:b2bfixosnr}. For asymptotically high values of CSPR, where performance is limited by ASE noise, the three variants obtain
approximately the same SNR. This allows to conclude that, even if spectral occupation is different, the three variants have the same performance with ASE noise.
Performance differ, instead, at medium/low CSPRs. At very low CSPR, SSBI with SSB modulation is very strong, making DSB/VSB more effective. However, at the optimal 
CSPR, SSB is more effective. This is explained both by the effectiveness of the SSBI mitigation technique, but also by the absence of non-linear operations in signal generation.

Fig. \ref{subfig:b2bsweep} shows instead the performance, at the optimal CSPR, for different values of OSNR. Close to the FEC threshold, SSB performs better. For
increasing values of OSNR, SSB loses some performance with respect to DSB. In both results, VSB has the worst performance, which is explained by the combination
of non-linear operations and SSBI.

\subsubsection{Final remarks}
From the results presented in this sections, some conclusions can be already made:
\begin{itemize}
\item There is a direct relation between CSPR and clipping ratio, allowing a direct comparison between intensity-modulation (DSB-VSB) and
field modulation (SSB).
\item Performance with respect to ASE noise is identical to all the three methods.
\item At the optimal CSPR, in the considered scenario, SSB is the optimal transmission technique.
\item VSB suffers from both transmitter non-linearities and SSBI, making it the lowest-performing modulation format in back-to-back.
\end{itemize}

\section{Impact of chromatic dispersion}
After comparing the three formats in back-to-back, the goal of this section is the comparison over $80$ km dispersion-uncompensated SMF, which is a typical distance
for Inter-DC links. Given the limited reach, Kerr effect and PMD can be easily neglected, and the only impairment is chromatic dispersion. Therefore,
this section will just focus on its impact on the three modulation formats.

It is well-known that a SSB/VSB signal is resistant to chromatic dispersion \cite{Yonenaga:1993}. On the other end, DSB signals
are severely impaired by chromatic dispersion. The goal of this section is to show the effect of CD on DSB signals, and provide a full comparison between
the three techniques with different amounts of CD.

\subsection{Effect on DSB}\label{sec:4:dsbeffect}
Chromatic dispersion is modeled as an all-pass filter $h_\textup{CD}(t)$, whose Fourier transform is
\begin{equation}
H_\textup{CD}(f) = e^{-j2\pi^2\beta_2f^2z}
\end{equation}
In this equation, $z$ is the distance, $f$ is the frequency in complex baseband, while $\beta_2$ is the chromatic dispersion parameter. For SMF, usually $\beta_2=-21.27~\mathrm{ps}/(\mathrm{THz}\cdot\mathrm{km})$.

With this definition, the received electric field of a DSB signal, before the photodiode, can be expressed as (in one polarization):
\begin{equation}
E_\textup{RX}^\textup{(dsb)}(t) = E_\textup{TX}^\textup{(dsb)}(t)\otimes h_\textup{CD}(t)+n_X(t)
\end{equation}
where $\otimes$ denotes convolution. Since $h_\textup{CD}(t)$ is an all-pass filter, it does not affect noise. To obtain an analytical formulation, 
the square-root in the DSB signal \eqref{eq:dsbtxsignal} needs to be expanded with a Taylor's series expansion, up to the second order.
The result is
\begin{equation}
E_\textup{TX}^\textup{(dsb)}(t) \approx \sqrt{\bar{P}}\left[ 1 + \frac{1}{2} \left(y(t)-\frac{y^2(t)}{4}\right)\otimes h_\textup{CD}(t) \right]
\end{equation}
where $y(t)$ is defined in \eqref{eq:dsbclipdef}. After ideal photodetection, neglecting noise-noise beating and DC components, the photocurrent is
\begin{align}
i^\textup{(dsb)}_0(t) &\approx \bar{P}\Bigg[\left( y(t)-\frac{y^2(t)}{4} \right)\otimes \Re\left\{h_\textup{CD}(t)\right\} + \nonumber \\
&+ \frac{1}{4} \left| \left( y(t)-\frac{y^2(t)}{4} \right)\otimes h_\textup{CD}(t) \right|^2 \Bigg] + 2\sqrt{\bar{P}}\cdot\Re\left\{n_X(t)\right\}
\label{eq:dsbcdphotodiode}
\end{align}

There are two main differences between this equation and the back-to-back result in \eqref{eq:dsbphotocurrent}. First, the signal is impaired
by $\Re\left\{h_\textup{CD}(t)\right\}$, which is \emph{not} all-pass. This filter is the well-known small-signal transfer function of a direct-detection
system \cite{Wang:1992}. In the frequency domain, this transfer functions has deep notches in the frequency, that adds a strong penalty to the system.
A multi-level transmission format, such as DMT, is able to partially work around this issue using bit and power loading. 
In fact, DSB-DMT has been adopted in \cite{Nadal:2014} in dispersion-uncompensated links for DMT's ability to reduce the effect of this transfer function.
Apart from ASE noise, there is a second term, which was not present in back-to-back \eqref{eq:dsbphotocurrent}. This term is a quadratic term, surprisingly
similar to SSBI. 

\subsubsection{Example}
\begin{figure}
	\centering
	\includegraphics[width=0.6\textwidth]{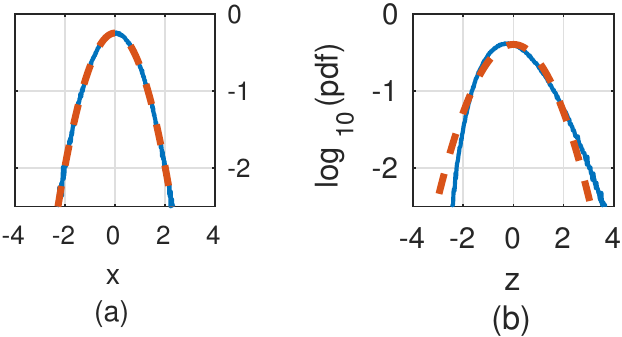}
	\caption{Histograms of a DSB-DMT signal before modulation (a) and after 80-km SMF and photodiode (b). Red dashed lines represent a Gaussian fit.}\label{fig:dsbcdhist}
\end{figure}
The effect of the quadratic term can be seen with a simple numerical example. A time-domain DMT signal is approximately
 Gaussian-distributed, as shown in Fig. \ref{fig:dsbcdhist}a, where
it is shown the histogram (in log-scale) of a time-domain DMT signal (blue solid line) and its Gaussian fit (red dashed line). Gaussian-distributed
signals have the property to keep their 
Probability Densitity Function (PDF) if they
pass through linear systems. Therefore, that signal was power-modulated, propagated over 80-km of linear SMF and detected with an ideal photodiode. The CSPR
was set to $10$ dB, and no ASE noise was added. Fig. \ref{fig:dsbcdhist}b shows the same histogram after photodetection. It can be seen that the PDF is quite different than a Gaussian PDF, which means that the impact of the quadratic term in \eqref{eq:dsbcdphotodiode} is strong enough to change the PDF of the received signal.

This result has some important implications. The small-signal approximation \cite{Wang:1992} that has been widely applied to model the impact of CD in DSB signals do not
apply here. This is because the optimal CSPR, for this scenario, is ``too low'' to assume a small signal. This also means that other small-signal approximations (e.g.
\cite{Bosco:1999} for Kerr effect) cannot be applied to this scenario.

\subsubsection{A pre-distortion proposal}
The quadratic term in \eqref{eq:dsbcdphotodiode} can be mitigated with a method similar to SSBI compensation for SSB/VSB. Here, we propose a combination of
transmitter pre-distortion and receiver compensation, which we found to be more effective than simple receiver mitigation.

Pre-distortion modifies \eqref{eq:dsbtxsignal} into
\begin{equation}
E_\textup{TX}^\textup{(dsb)}(t) = \sqrt{ \bar{P} \left\{1+2\clip\left[\frac{x(t)}{c'}\right] + \clip^2\left[\frac{x(t)}{c'}\right] \right\}  }
\end{equation}
Notice the addition of a quadratic term inside the square root. This pre-distortion allow simplifying the square root to
\begin{equation}
E_\textup{TX}^\textup{(dsb)}(t) = \sqrt{ \bar{P} } \left\{1+\clip\left[\frac{x(t)}{c'}\right] \right\}
\end{equation}
This operation removes one non-linear operation (the square root) in the electric field, allowing a simplification of the equations at the receiver.
Apart from clipping, this signal is similar to a field-modulated SSB signal \eqref{eq:ssbtx}. 

The pre-distorted signal after photodetection becomes
\begin{align}
	i^\textup{(dsb)}_0(t) &\approx 2\bar{P}\clip\left[\frac{x(t)}{c'}\right] \otimes \Re\left\{h_\textup{CD}(t)\right\} + \nonumber \\
	&+\bar{P}\left| \clip\left[\frac{x(t)}{c'}\right]\otimes h_\textup{CD}(t) \right|^2 + 2\sqrt{\bar{P}}\cdot\Re\left\{n_X(t)\right\}
	\label{eq:dsbcdphotodiodepredist}
\end{align}
Comparing to \eqref{eq:dsbcdphotodiode}, the quadratic term is now similar to SSBI. Therefore, it can be mitigated with a similar method to SSB \cite{Randel:15}.

A potential drawback of this scheme is the spectral broadening that is caused by the addition of the quadratic term. This will increase the penalties
added by a realistic DAC, potentially reducing the benefits of this scheme. Nevertheless, a detailed analysis of the pratical implementation penalties
of this scheme is out of the scope of this thesis, and may be a topic of future work.

\subsection{DMT simulations}
After introducing the impact of CD over DSB, this section will perform the final comparison between DSB, VSB and SSB over different scenarions. As with
back-to-back results, the comparison will be performed using DMT modulation.

For these results, the Levin-Campello algorithm was applied to the per-subcarrier SNR, measured using the ``test'' 16-QAM DMT signal described in Table
\ref{tab:dmtsimparams}. The target gross bit rate of the algorithm is 120 Gbit/s which, assuming a 20\% overhead taking into account FEC and protocol headers, 
corresponds to a net bit rate of 100 Gbit/s.

\subsubsection{Impact of quadratic terms in DSB}
\begin{figure}
	\centering
	\includegraphics[width=0.5\textwidth]{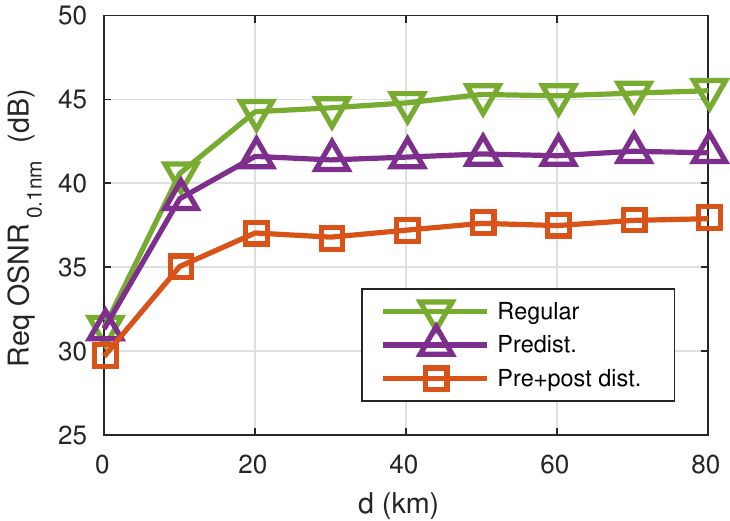}
	\caption{Impact of quadratic terms in DSB transmission and advantage of compensation techniques.}\label{fig:dsbquadcomp}
\end{figure}
In Sec. \ref{sec:4:dsbeffect} it was shown that the impact of CD on a DSB signal is made by linear (frequency-selective transfer function) and non-linear (quadratic terms)
impairments. While linear distortions cannot be compensated, quadratic terms can be (at least partially) compensated with transmitter pre-distortion and/or receiver
compensation. Fig. \ref{fig:dsbquadcomp} shows the required OSNR of the test DSB-DMT signal with different values of cumulated dispersion (up to $80$-km SMF).
At the maximum distance, the required OSNR for standard DSB-DMT is quite high ($45.5$ dB). Pre-distortion reduces it by $3.7$ dB, while pre-distortion followed
by receiver compensation further reduces it by $4.4$ dB. Overall, compensation of the quadratic effect gives a significant gain ($8.1$ dB) in terms of required OSNR, which
makes it fundamental in dispersion-uncompensated DSB links. Looking at the variation over distance, after $10$ km performance rapidly degrades due
to the effect of $\Re\left\{h_\textup{CD}(t)\right\}$, which makes the signal strongly frequency selective.

\subsubsection{VSB optical filter}
\begin{figure}
	\centering
	\includegraphics[width=0.5\textwidth]{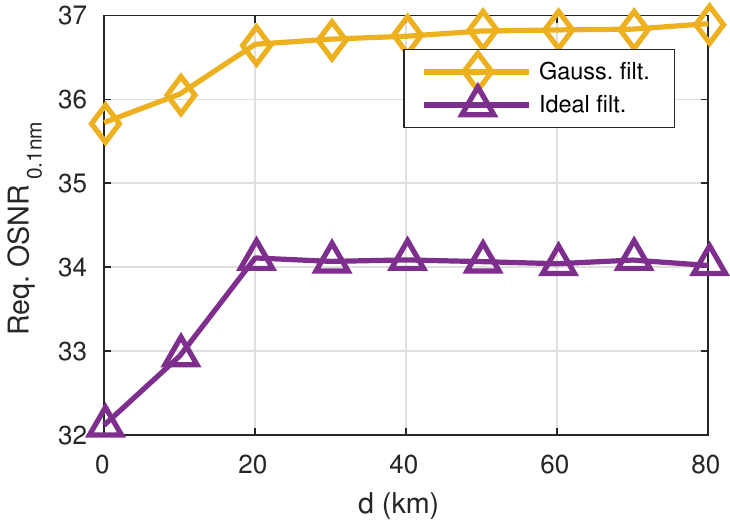}
	\caption{Performance difference between an ideal and a super-Gaussian filter for VSB.}\label{fig:vsbgaussfilt}
\end{figure}
In Sec. \ref{sec:4:vsbdef} VSB was introduced as a DSB transmitter followed by a steep OBPF. Results presented in Sec. \ref{sec:4:vsbdef} assumed an ideal (rectangular) OBPF.
Since an ideal OBPF is not realistic, Fig. \ref{fig:vsbgaussfilt} shows the impact of CD on a VSB signal assuming an ideal filter, and a realistic fourth-order
super-Gaussian filter with a $35$-GHz $3$-dB bandwidth. Parameters of this filter were chosen to emulate the shape of a typical $50$-GHz DWDM demultiplexer.
As shown in the Figure, the use of a realistic filter adds an additional $\sim2.88$ dB penalty at $80$-km. This filter also adds a penalty in back-to-back due to the linear
distortions added by the filter.

\begin{figure}
	\centering
	\includegraphics[width=0.5\textwidth]{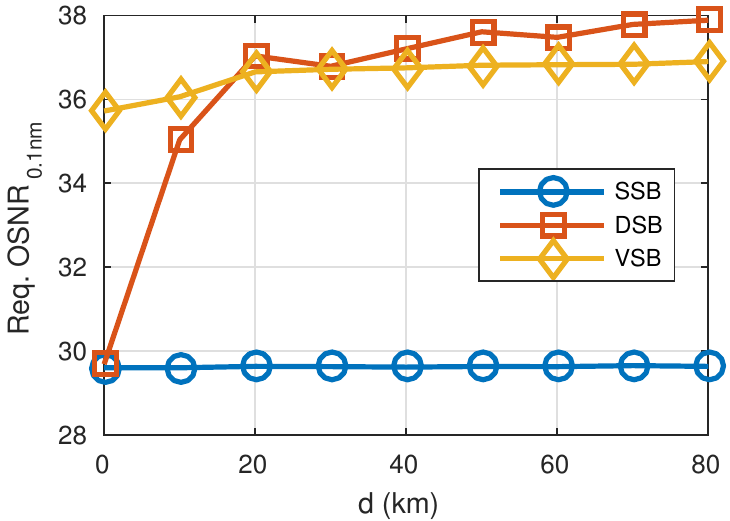}
	\caption{Final comparison DSB/VSB/SSB in presence of chromatic dispersion. VSB has been obtained with a super-Gaussian filter.}\label{fig:dsbvsbssbfinalcomp}
\end{figure}
Finally, the comparison between DSB, VSB and SSB DMT with dispersion is shown in Fig. \ref{fig:dsbvsbssbfinalcomp}. As expected, SSB shows a negligible penalty with distance, since CD
is electrically compensated at the receiver. VSB and DSB instead show similar penalties at $80$ km. The surprisingly good performance of DSB, compared to VSB (generated using
a realistic super-Gaussian filter), is due to DMT's bit and power loading algorithms, which are able to partially compensate for channel frequency-selectivity. 

\section{Conclusions and outlook}
This Chapter proposed SSB transmission for dispersion-uncompensated Inter-DC links, as a hybrid solution between IM and coherent systems. While dispersion
compensation is still widely adopted in legacy links, removal of DCF will ease the inevitable future transition to coherent transmission for Inter-DC.
Therefore, SSB can be a viable choice for the transition between intensity modulation and coherent transmission. 

In this comparison, we found that SSB is not only highly resistant to chromatic dispersion (at least up to $80$ km), 
but it also has an advantage in back-to-back with respect to IM. This comes at the expense of a higher transmitter complexity, which can be reduced with a careful choice of the optical and electrical components. Obviously, this analytical derivation neglected several impairments (e.g CD-induced phase noise \cite{Gatto:2017}, fiber non-linear effects,\dots) which
have to be carefully taken into account to fully prove the feasibility of this solution.

VSB, which have been proposed as a low-cost alternative to SSB, was found to be performing almost equal to DSB. Therefore, it is not a feasible
choice for this particular scenario.

\part{Coherent Systems}\label{part:coh}

\chapter{Introduction to Coherent Systems}\label{ch:coherent}

    \graphicspath{{Chapter5/}}

In Part \ref{part:dd}, direct-detection systems were introduced for short-distance links.
It was shown that advanced digital signal processing is able to give substantial improvements for Intra-DC links (Chapter \ref{ch:bidir}) and
Inter-DC links (Chapter \ref{ch:ssb}). The goal of this part is, instead, the application of advanced DSP algorithms to long-haul links, where
coherent modulation and detection is adopted. 

This Chapter is divided in two sections. First, the coherent long-distance optical channel is briefly introduced.
Then, the generic structure of a coherent transceiver is described, with a specific focus on the receiver DSP algorithms.

\section{Long-haul links}
In Chapter \ref{ch:dcarchitect}, a simplified channel model for Intra- and Inter-DC links was introduced. Likewise, this section will introduce
a model for long-haul links. As opposed to short-reach DC communications, long-haul links are rarely point-to-point connections, but are part of large networks,
optically routed at the wavelength level \cite{Saleh:2012}. Fortunately, the study of DSP at the transceiver level does not need a realistic optical network model. In fact,
transceivers \cite{Raybon:2018} and propagation effects \cite{Saavedra:2017,Cantono:2018} are often characterized in point-to-point links. 
This operation greatly simplifies simulations and experiments. On the other end, it does not allow to fully evaluate some networking-related
aspects of physical layer impairments. Since a study of the optical networking impact of physical-layer impairments is out of the scope of this thesis, point-to-point links
for long-haul communication will be considered in the following. For more details on networking-related aspects, the reader can refer to \cite{Cantono:PhDThesis} (and references therein).

Therefore, in this Section optical point-to-point long-haul links will be described, starting from a system-level overview, followed by details on some specific sub-systems.

\subsection{Link model}
\begin{figure}
	\centering
	\includegraphics[width=0.9\textwidth]{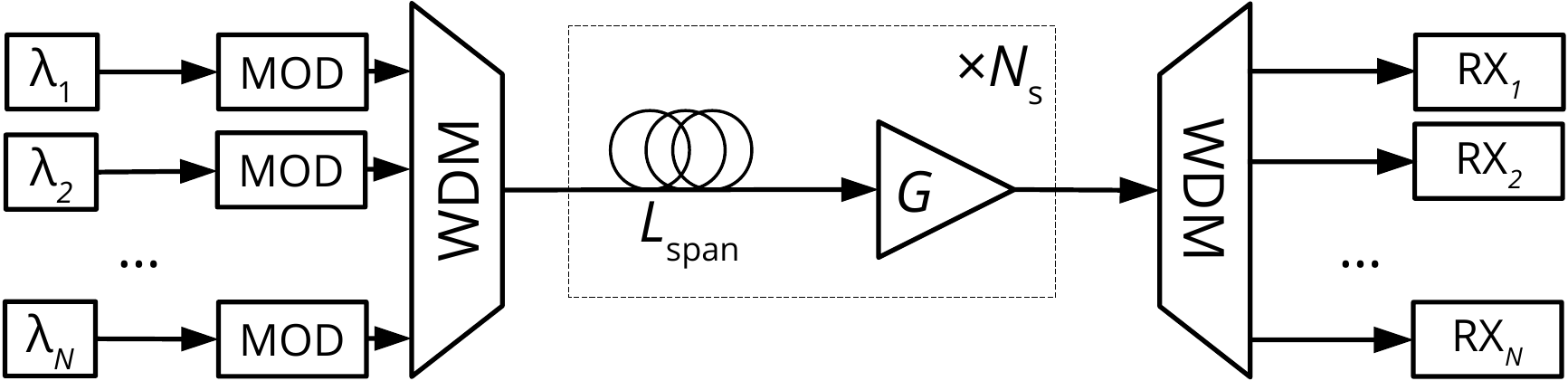}
	\caption{Simplified model of a long-haul uncompensated link.}\label{fig:longhaulmodel}
\end{figure}
As opposed to short-reach systems, long-haul links need to be periodically amplified. This increases complexity of the link, but it also gives more freedom of choice to the link designer. A simplified model of a dispersion-uncompensated point-to-point link is shown in Fig. \ref{fig:longhaulmodel}.
In the model, $N$ transmitters generate $N$ WDM channels in the DWDM grid \cite{std:dwdm}. These channels are first multiplexed, then transmitted over $N_s$ identical spans. Each
span is made of a certain length ($L_\textup{span}$) of optical fiber, followed by an amplifier that fully recovers span loss. Afterwards, a WDM demultiplexer separates the WDM channels, which
are independently received. While this model may seem overly simplified for a terrestrial link, in which spans can be very different from each other, it is widely
adopted for optical transmission simulations and experiments \cite{Raybon:2018,Saavedra:2017,Cantono:2018}.

To test the overall performances of a transceiver, the performance of one (or more) channel-under test (CUT) is measured over different system conditions. Usually, for a given
system setup, two parameters are scanned: the per-channel transmit optical power ($P_\textup{ch}$) and the number of spans ($N_\textup{s}$). 
A metric commonly used to evaluate the performance of the system, which can be derived from those results, is the \emph{maximum reach} \cite{Curri:2013}. It is defined as
the maximum number of spans that keep the system on-service, at the optimal launch power. This condenses in a ``single number'' the performance of a specific link, 
allowing the development of optimization strategies based on its maximization \cite{Curri:2015}.

\subsection{ASE and phase noise}
The main noise source of an optical link is Amplified Spontaneous Emission (ASE) noise that is emitted by optical amplifier. Generation of ASE
depend on amplifier technology (EDFA or Raman) \cite{Giles:1991,Bromage:2004}. A simple model for EDFA amplifiers relates ASE power spectral density $N_0$, measured
after the amplifier, with its gain and its spontaneous emission factor (or population-inversion factor) $n_\textup{sp}$ \cite[Eq. (4)]{Yariv:1990}
\begin{equation}
N_0 = h f_0 (G-1) n_\textup{sp}
\end{equation}
In this equation, $h$ is Planck's constant, $f_0$ the central frequency of the signal, $G$ is the amplifier gain and $F$ is the EDFA noise figure. It is customary
to use the amplifier \emph{noise figure} instead of its spontaneous emission factor, which is defined as the SNR degration that it is induced by the amplifier itself.
Assuming a large gain ($G\gg1$), the noise factor $F$ can be evaluated as \cite[Eq. (11)]{Yariv:1990}
\begin{equation}
F \approx 2n_\textup{sp}
\end{equation}

ASE noise is added incoherently in each span, so that the OSNR (evaluated over the symbol rate) is equal to
\begin{equation}
\mathrm{OSNR}_{R_s} = \frac{P_\textup{ch}}{N_\textup{span} N_0 R_\textup{s}}
\end{equation}

An additional noise source is phase noise that is generated by the phase difference between the lasers at the transmitter and at the receiver. In Chapter \ref{ch:bidir}, it was shown
the impact of phase noise in an IM/DD system. Over a coherent system, this phase noise must be compensated at the reciever using a proper algorithm, which will be discussed later
in this Chapter. A common discrete-time model for laser phase noise that is affecting a digital transmission is the first-order random walk, also called Wiener process
\begin{equation}
\phi(n) = \phi(n-1) + \nu(n)
\end{equation}
where $\nu(n)$ is an white Gaussian process with variance equal to $2\pi \Delta\nu T$, $\Delta\nu$ is the laser \emph{linewidth}, $T$ is the symbol duration and $n$ the discrete-time
index ($T$-spaced). More details on phase noise will be given in Chapter \ref{ch:phnoise}.

\subsection{Fiber non-linear effects}
Glass is a non-linear medium. Therefore, accurate modeling on the effects of fiber propagation of a modulated signal is not an easy task. In this section (and in the following),
the most important fiber non-linear effects are shown, along with their impact on a coherent modulated signal. 
More details on fiber non-linear effects can be found in \cite{Agrawal:nonlinear}.

\subsubsection{Kerr effect}
Kerr effects originate in the third-order $\chi^{(3)}$ susceptibility of glass. This introduces
a linear dependence between the refractive index and the instantaneous optical power.
Propagation of an optical modulated signal inside a fiber with Kerr effect is governed by a set of $2$ (one per polarization)
coupled non-linear differential equations. 
These equations take the generic name of Dual-Polarization Non-Linear Schr\"{o}dinger Equation (DP-NLSE), and,
for a coherent modulated signal, they cannot be solved in closed form. Therefore, evaluation of propagation of a signal inside a fiber requires a numerical integration of the 
DP-NLSE. A common solving technique is called Split-Step Fourier Method (SSFM). 

For results presented in this thesis, we used an internally-developed SSFM routine, called Fast Fiber Simulator Software (FFSS) \cite{Pilori:FFSS2017}. More details
on it are shown in appendix \ref{ch:ffss}.

\subsubsection{Stimulated Raman and Brillouin scattering}
Stimulated Raman Scattering (SRS) and Stimulated
Brillouin Scattering (SBS) are two effects which are generated by stimulated inelastic scattering of photons inside glass. While SBS is generally not present in the coherent long-haul scenario, SRS
can be detrimental for wide-band links. In simple words, SRS causes a power transfer from high-frequency channels to low-frequency ones, creating a ``tilted'' received
power profile of the WDM channels. More details on the impact of SRS in a coherent link are shown in appendix \ref{ch:ffss}.

\subsection{Impact on a communication system}\label{sec:5:commkerimpact}
Modeling the impact of these non-linear effects on a coherent communications link is not an easy task. In the past, several analytical models have been developed
\cite{Secondini:2012,Mecozzi:2012,Bononi:2012,Dar:2013,Poggiolini:2014,Carena:2014,Golani:2016} to solve this issue, each with a different
degree of complexity and accuracy. 

\subsubsection{Gaussian-noise approximation}
A crude approximation considers Kerr effect as an AWGN source, called Non-Linear Interference (NLI), 
which is statistically independent from ASE noise and transmit data. In formulas,
\begin{equation}
y[k] = a[k] + n_\textup{ASE}[k] + n_\textup{NLI}[k]
\end{equation}
where $k$ is the time instant, $y$ and $a$ the received and transmitted symbol (respectively) and $n$ is an AWGN process. 
Surprisingly, this approximation was found to be very accurate, especially for long-distance links \cite{Splett:1993,Carena:2010}. 
In particular, it was found that the power of NLI is proportional to the cube of signal power (per WDM channel):
\begin{equation}
P_\textup{NLI} = \eta P_\textup{ch}^3
\end{equation}
Therefore, with this assumption, the SNR, measured at the receiver, can be expressed as:
\begin{equation}
\mathrm{SNR} = \frac{P_\textup{ch}}{P_\textup{ASE}+\eta P_\textup{ch}^3}
\label{eq:cohsnrexample}
\end{equation}
\begin{figure}
\centering
\includegraphics[width=0.6\textwidth]{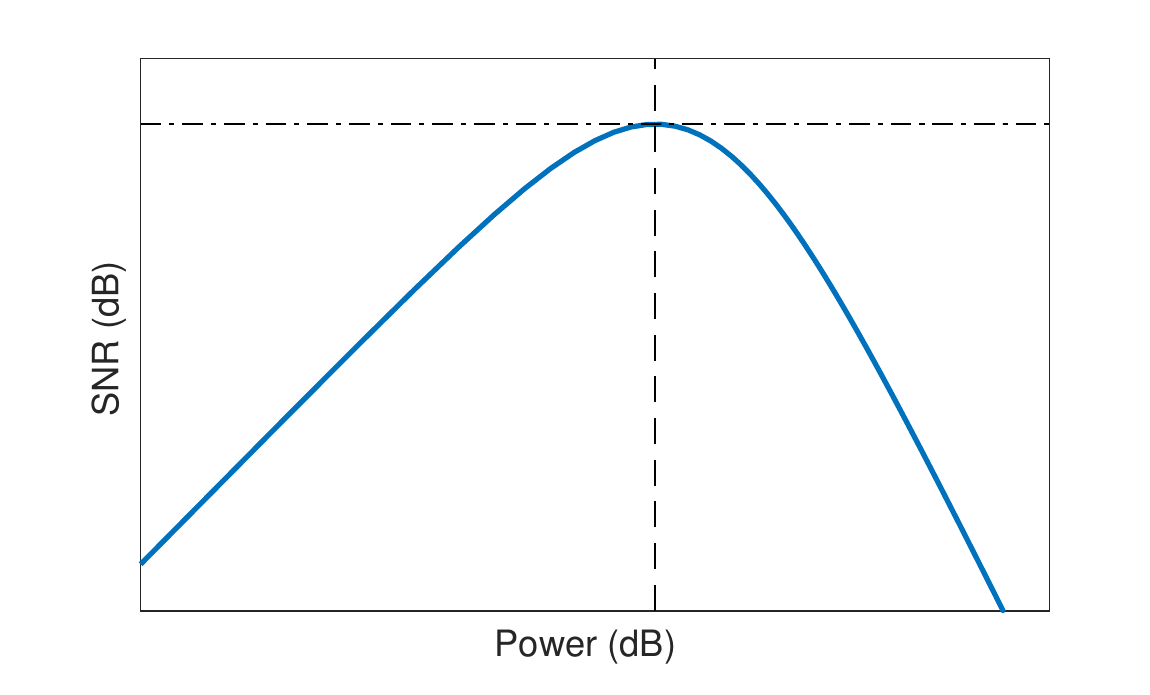}
\caption{Example of SNR dependence on power using \eqref{eq:cohsnrexample}. Optimal power is indicated with a dashed line, while
maximum SNR is indicated with a dashed-dotted line.}\label{fig:optpchexample}
\end{figure}
An example of SNR evaluation using \eqref{eq:cohsnrexample} is shown in Fig. \ref{fig:optpchexample}. Given a value of $\eta$, which depends
on link parameters, and ASE power $P_\textup{ASE}$, there always exists an optimal power $P_\textup{ch}^\textup{(opt)}$ which gives
the maximum signal-to-noise ratio $\mathrm{SNR}_\textup{max}$. Moreover, it was found that, with good accuracy, $P_\textup{NLI}$ accumulates
incoherently for each span. This allows optimizing independently each fiber span (so-called LOGO strategy \cite{Poggiolini:2014}), which makes
this model very useful for network design and analysis (e.g. \cite{Curri:2017}).

\subsubsection{Non-linear phase noise}
Unfortunately, there are situations where the Gaussian-noise approximation is not accurate. 
For instance, in \cite[Fig. 1]{Dar:2013}, the authors showed that cross-channel NLI (i.e. NLI generated by the interaction between the CUT and another WDM channel)
has a phase-noise component (called non-linear phase noise, NLPN), which is significant where the interfering channel is 16-QAM instead of QPSK. This effect is
explained by the theory presented in \cite{Dar:2013,Dar:2016}, where cross-channel NLI is seen as a time-varying intersymbol interference process. This effect was not seen in the
initial investigations on the Gaussian approximation \cite{Carena:2010}, since it is strongly dependent on the constellation shape. While it vanishes for phase-only modulation (like QPSK), it
becomes stronger on higher-cardinality constellations. In details, NLPN strength is linked to the statistical parameter \emph{kurtosis} of the constellation
\cite{Dar:2013,Carena:2014}:
\begin{equation}
\kurt(a) = \frac{\expt[|x|^4]}{\expt^2[|x|^2]}
\label{eq:kurtosis}
\end{equation}
This parameter also affects the overall strength of NLI \cite{Dar:2014,Carena:2014}, which is stronger for higher value of kurtosis. Both these effects
were found to vanish for long links \cite{Dar:2016}, making them negligible for long-distance (e.g. submarine) links, but significant for high-capacity short-distance 
(e.g. metro) links.

A critical parameter of the NLPN random process is its memory (or, equivalently, the bandwidth). It was found that its variation is usually slower than the symbol rate of the CUT
, which means that NLPN may be compensated by standard phase-recovery algorithms \cite{Dar:2017,Poggiolini:2017}. In particular, it was found that
full NLPN compensation (using a ``genie-aided'' algorithm) \cite{Nespola:2016} is able to obtain approximately the same amount of NLI that is generated by a QPSK system.

More details on this aspect will be given in Chapters \ref{ch:shaping} and \ref{ch:phnoise}.

\section{Coherent transceivers}
\begin{figure}
	\centering
	\includegraphics[width=0.8\textwidth]{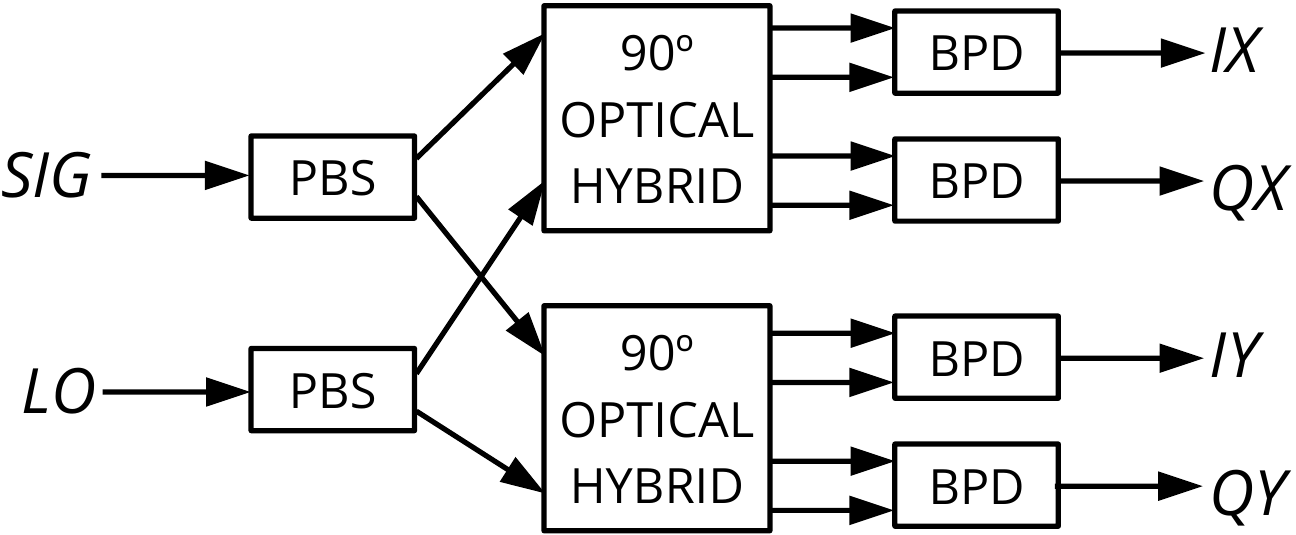}
	\caption{Block scheme of an intradyne coherent receiver \cite{Kikuchi:2016}.}\label{fig:cohrxgeneric}
\end{figure}
A single-polarization coherent transmitter is identical to the SSB transmitter described in Chapter \ref{ch:ssb}. 
Extension to dual-polarization is trivial. However, the receiver
is quite different, since it needs to detect a full dual-polarization optical field. A generic schematic of such receiver is shown in Fig. \ref{fig:cohrxgeneric}
\cite{Kikuchi:2016}. The incoming signal (\emph{SIG} in Fig. \ref{fig:cohrxgeneric}) is divided by a Polarization Beam Splitter (PBS) into two $90^\circ$ optical hybrids. These
hybrids mix signal light with a laser, approximately tuned at the central frequency of the channel (called \emph{intradyne} reception), called local oscillator (LO), in both polarizations.
The outputs of the hybrids are then sent to four balanced photodiodes (BPDs), which generate four electrical signals (IX, QX, IY and QY). These electrical signals
define the full optical field in complex baseband, centered around LO frequency. The interested reader can refer to \cite{Kikuchi:2016} (and references therein) for
details on coherent reception.

\subsection{DSP chain}\label{ch:5:dspchain}
\begin{figure}
	\centering
	\includegraphics[width=0.3\textwidth]{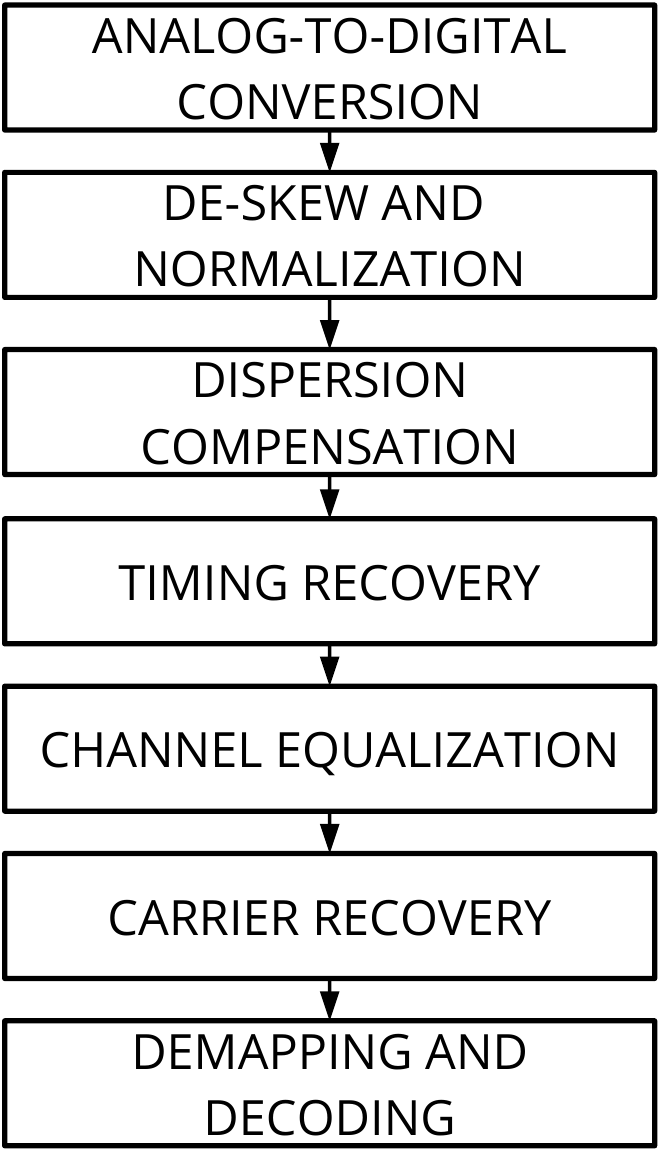}
	\caption{DSP chain of a coherent receiver.}\label{fig:cohdspchain}
\end{figure}
After coherent reception, the received signal needs to be processed by some DSP algorithms to recover the transmitted signal. This is a strict requirement, since both fiber
propagation and coherent reception add impairments that must be digitally reversed. The common DSP stack of a coherent receiver is explained in
detail in several references, e.g. \cite{Savory:2010}.

The basic algorithms, which are present in any coherent receiver, are summarized in Fig. \ref{fig:cohdspchain}. After analog-to-digital conversion at
two samples per symbol, the receiver compensates for analog impairments in the electrical paths, such as skew (i.e. delay between different signals) and gain. 
Afterwards, chromatic dispersion is estimated (e.g. using \cite{Malouin:2012}) and compensated with an all-pass Finite-Impulse-Response (FIR) filter. 
Then, the signal is interpolated to recover the transmitter clock (timing recovery). Common algorithms employed in optical communications
are Godard's tone clock \cite{Godard:1978}, Gardner's scheme \cite{Gardner:1986} and Mueller-M\"{u}ller \cite{Mueller:1976}. 
Signal is then equalized using a fractionally-spaced $2\times2$ MIMO adaptive equalizer, 
which is able to compensate for several linear impairments (PMD, birifrengence, bandwidth limitations) and automatically converges
to the matched filter \cite{Barry:2004,proakis2007digital}. Equalizer error can be either computed with data-aided methods (e.g. with pilot symbols) using the 
LMS method \cite{Barry:2004}, or using fully blind methods, such as the Constant Modulus Algorithm (CMA) \cite{Godard:1980}. 
For the LMS algorithm to converge, a simple phase recovery is required inside the error computation circuit \cite{Randel:2013}.
After adaptive equalization, a frequency and phase recovery circuit recovers frequency and phase of transmitter carrier (i.e. the laser at the transmitter). 
The phase recovery is also called Carrier Phase Estimator (CPE). Due to the high symbol rates
commonly employed in optical communications, feed-forward algorithms are preferred 
compared to Phase-Locked Loops (PLLs). Commonly adopted algorithms are Viterbi-Viterbi for QPSK \cite{Viterbi:1983} and Blind Phase Search (BPS) \cite{Pfau:2009}.
At the end, the received symbols are detected and decoded.

\subsection{Demapping and decoding}\label{sec:5:decoding}
\begin{figure}
	\centering
	\includegraphics[width=\textwidth]{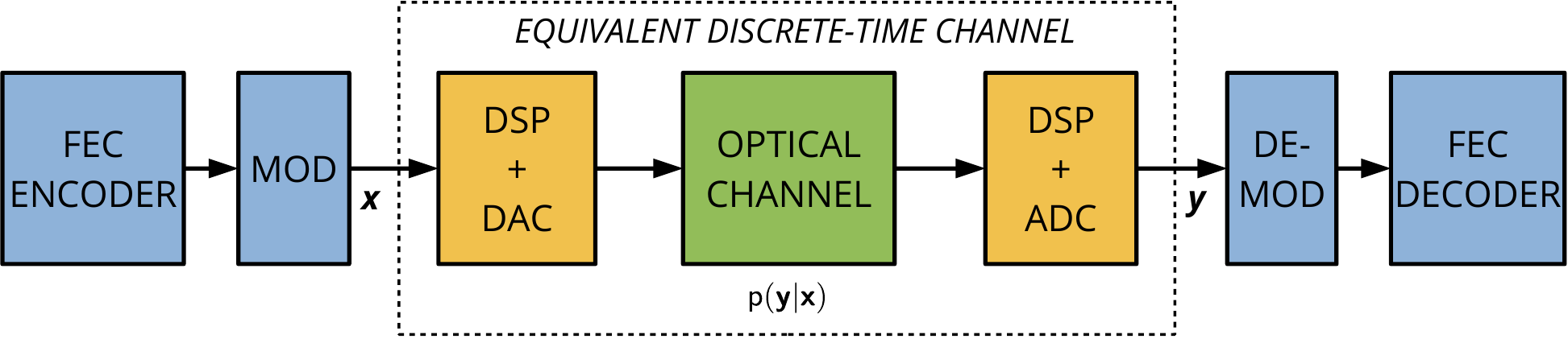}
	\caption{Information-theoretical channel model of the coherent long-haul optical channel \cite{Agrell:2018}.}\label{fig:infothchannel}
\end{figure}
While short-reach communications (described in Part \ref{part:dd}) and legacy long-haul communications employed very simple FEC schemes, modern long-distance
communications employ powerful FEC schemes, which closely operate within the Shannon limit. 

From an information-theoretical perspective, the optical channel can be modeled as shown in Fig. \ref{fig:infothchannel}. Both the optical channel (Fig. \ref{fig:longhaulmodel})
and the full transmitter/receiver DSP chain can be modeled as a black-box, called ``equivalent discrete-time channel'' in Fig. \ref{fig:infothchannel}. This ``box''
takes as an input a complex-valued random vector $\mathbf{X}$ and outputs another complex-valued random vector $\mathbf{Y}$. Realizations of these random variables
are indicated with lowercase letters, i.e. $\mathbf{x}$ and $\mathbf{y}$.

Therefore, this channel is fully characterized
by its conditional probability $p(\mathbf{y}|\mathbf{x})$. In general, this channel is multi-dimensional, because it propagates dual-polarization signals and
it has memory. However, this makes it very difficult to be handled from an information-theory perspective \cite{Alvarado:2018,Agrell:2018}. Therefore, we will
assume that transmitter and receiver DSP are able to remove all channel memory, and the two polarizations are fully orthogonal. With these assumptions, the channel
can be assumed scalar, with conditional probability $p(y|x)$.

\subsubsection{Achievable rates}
According to information theory \cite{Cover:2005}, the maximum information rate that can be transmitted over such channel with an arbitrary low error rate is given by a quantity that is called
\emph{Mutual Information} (MI)
\begin{equation}
\mathcal{I}(X;Y) = \expt\left[\log_2\frac{p(y|x)}{p(y)}\right]
\label{eq:midef}
\end{equation}
In other words, this quantity represents the maximum number of information bits that can be transmitted for each channel use (i.e. transmitted variable $x$) and received
with an arbitrarily low error probability \cite{Cover:2005}. Note that this quantity depends on $p(x)$, which can be arbitrarily chosen by the transmitter.
Maximization of $\mathcal{I}(X;Y)$ over all the possibile $p(x)$ gives the celebrated \emph{channel capacity} $C$. It is important to note that the capacity
represents the maximum data-rate that can be transmitted (without errors), given $p(y|x)$. This limits the validity of the ``maximum-rate'' statement
to the knowledge of $p(y|x)$, which is still not fully known for the optical channel.
Therefore, computation of the true capacity of an optical fiber transmission system is, therefore, still an open research question 
\cite{Secondini:2017}.

The MI is not a convenient metric for a realistic communication system, since it assumes perfect knowledge
of $p(y|x)$ at the receiver. A more practical metric is the so-called Achievable Information Rate (AIR) \cite{Agrell:2018,Merhav:1994}, which is (in general)
defined as
\begin{equation}
\mathcal{I}_q(X;Y) = \expt\left[\log_2\frac{q(y|x)}{q(y)}\right]
\label{eq:airdef}
\end{equation}, where $q(y|x)$ is the channel law that is used by the receiver to perform decoding. 
It can be shown that $\mathcal{I}_q(X;Y)\leq\mathcal{I}(X;Y)$, and equality 
holds when $q(y|x)=p(y|x)$. In fact, the AIR can be also seen as a generalization of the MI where the receiver does not have full knowledge of the channel.
In other words, the MI is an AIR if the receiver has such knowledge. 
As a final remark, the expectation in \eqref{eq:airdef} 
can be evaluated with Monte-Carlo integration, without any knowledge of the true $p(y|x)$. This makes the AIR a convenient performance metric for optical experiments
\cite{Alvarado:2018}.

\subsubsection{Bit Interleaved Coded Modulation}\label{sec:5:bicm}
Modern optical long-haul communication systems use binary soft-decision FEC. There are different methods to apply a binary code to a multi-level modulation format, such as QAM,
called \emph{coded modulation} schemes \cite{Alvarado:2018}.
The most employed method is the so-called Bit Interleaved Coded Mo\texttt{}dulation (BICM) \cite{Caire:1998}. Even if it is suboptimal, it allows obtaining very good performances
with a small complexity. In a nutshell, a BICM receiver separates the constellation demapping and binary decoding into two independent stages. First, the demapper
calculates bit reliability metrics called Log-Likelihood Ratios (LLRs), computed as
\begin{equation}
l_i = \log\frac{P(b_i=1|y)}{P(b_i=0|y)}
\label{eq:llrdef}
\end{equation}where $b_i$ is the $i$-th bit that is mapped to the transmitted symbol $x$. Assuming that the transmitted symbols are drawn from a constellation
$\chi$ with cardinality $|\chi|=M$, so that each symbol carries $\log_2(M)$ bits, the LLRs are computed as
\begin{equation}
l_i = \log\frac{\sum_{x_1\in\chi_i^1} q(y|x_1) P(x_1) }{\sum_{x_0\in\chi_i^0} q(y|x_0) P(x_0)}
\end{equation}where $\chi_i^b$ is the subset of $\chi$ in which the $i$-th bit is equal to $b$. $q(y|x)$ is the same symbol decoding metric used to evaluate the AIR
in \eqref{eq:airdef}. Note that the evaluation of the LLR does not only depend on the constellation symbol, but also on the mapping function between bit and symbols.
Square QAM constellations allow the use of Gray mapping, which is applied separately on the In-Phase and Quadrature components. As it will be shown later, Gray mapping
is able to give very high performances at medium/high SNRs. For other constellations, a bit mapping needs to be assumed or optimized. For this thesis, unless
explicitly stated, Gray mapping will be assumed for square QAM constellations. The bit mapping of other constellations will be detailed in Appendix \ref{app:qamconst}.

An AIR for the BICM channel is the Generalized Mutual Information (GMI) \cite{Martinez:2009,Szczecinski:2014}, which can be computed from the LLRs using
\begin{equation}\label{eq:gmidef}
\mathcal{G}(X;Y) = \log_2(M) - \min_{s\geq0}\sum_{i=1}^{\log_2(M)}\sum_{b\in\{0,1\}} \frac{1}{2}
\int_{-\infty}^{+\infty}p_{L_i|B_i}(l|b)\log_2\left(1+e^{s(-1)^bl}\right)\,\mathrm{d}l
\end{equation} Similarly, if the receiver has full knowledge of the channel (contained in $p_{L_i|B_i}(l|b)$),
 the GMI is maximized, and it can be proven that it is the sum of the \emph{bit-wise} mutual informations:
\begin{equation}
\mathcal{G}(X;Y) = \sum_{i=1}^{\log_2(M)} \mathcal{I}(B_i;L_i)
\end{equation}

\subsubsection{Predicting FEC performance}
AIRs, even with a mismatched channel metric, represent a rate that can be obtained only with ideal (i.e. with infinite block length) FEC. 
Realistic FEC codes have a penalty with respect to ideal FEC, and operate at a data-rate that is inferior to the AIR. 

Compared to Intra-DC links (Sec. \ref{sec:2:intra}), the complexity constraint is less strict. 
Therefore, codes for this application use large block sizes and large overhead \cite{Alvarado:2018}. 
Moreover, the required post-FEC BER for these applications is $10^{-15}$ (or even smaller), which is difficult to be measured with direct error counting.
Consequently, a pre-FEC metric to infer post-FEC performances is required.
While for hard-decision FECs (Chapter \ref{ch:dcarchitect}) such metric is simply the pre-FEC BER, for soft-decision FECs this metric is not appropriate \cite{Alvarado:2016}.
It was found \cite{Alvarado:2016} that the AIRs are indeed good metrics for this purpose. In particular, the MI was found to be a good metric for symbol-wise FECs, while the GMI
is a good metric for bit-wise FECs. An example of GMI thresholds for soft-decision DVB-S.2 LDPC codes 
concatenated with a $6.25\%$ hard-decision staircase code are shown in \cite[Table I]{Alvarado:2018}.

\chapter{Constellation Shaping}\label{ch:shaping}
    \graphicspath{{Chapter6/}}

This Chapter focuses on constellation shaping techniques for coherent optical communications. Since, as discussed in Chapter \ref{ch:coherent}, the coherent optical channel
is very well approximated by the AWGN channel, this channel will be first studied in detail. Afterwards, two different kinds of shaping will be presented: probabilistic
and geometrical shaping. After an introduction on the different probabilistic shaping techniques, special focus will be devoted on the Probabilistic Amplitude
Shaping architecture. An experimental test of this technique will be presented. At the end, geometric shaping will be briefly presented, using the Amplitude-Phase
Shift Keying (APSK) constellation family as an example.

Results presented in this chapter are based on \cite{Bosco:2017,Pilori:2017,Pilori:IPC2017,Bertignono:2017,Pilori:JLT2018}.

\section{Additive White Gaussian Noise Channel}
As illustrated in Sec. \ref{sec:5:commkerimpact}, the long-haul coherent optical channel is well approximated by an AWGN channel, i.e.
a channel that just adds a white Gaussian random process to the wanted signal. In fact, unless explicitly written otherwise, an AWGN channel will always be assumed in this thesis.
To understand the benefits of constellation shaping, some information-theoretical
details on the AWGN channel need to be given. More details can be found in specialized references, such as \cite{Cover:2005,Barry:2004,proakis2007digital}.

\subsection{Capacity of the Gaussian channel}
In this Chapter, the Gaussian channel (or AWGN channel) is defined as a channel that adds a random white process to the signal. Both input ($x$) and output ($y$) signals are 
complex-valued. Noise is distributed according to the circularly-symmetric Gaussian distribution, i.e. real and imaginary part are statistically independent and have
the same variance $\sigma^2_n/2$. With this hypothesis, channel conditional probability is
\begin{equation}
p(y|x) = \frac{1}{\pi\sigma^2_n}e^{-\frac{|y-x|^2}{\sigma^2_n}}
\end{equation}

\begin{figure}
	\begin{subfigure}[b]{0.48\textwidth}
		\centering 	\includegraphics[width=\textwidth]{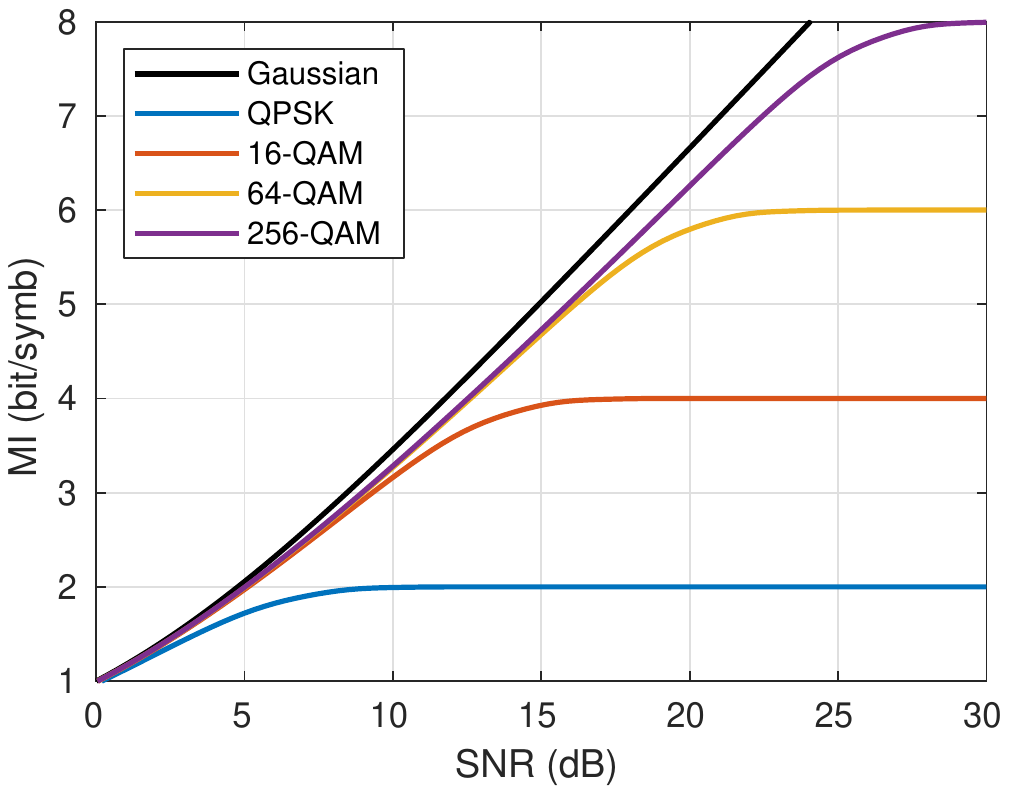}
		\caption{}
	\end{subfigure}
	\begin{subfigure}[b]{0.48\textwidth}
		\centering \includegraphics[width=\textwidth]{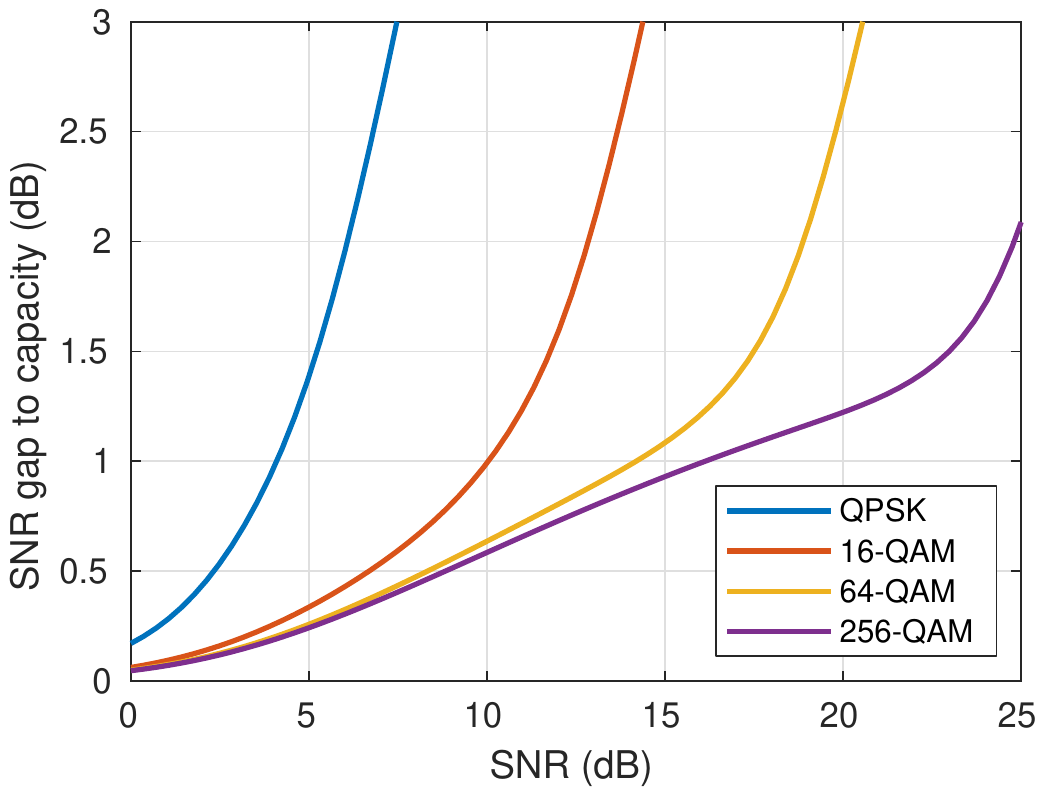}
		\caption{}
	\end{subfigure}
	\caption{Mutual information of QAM constellations over an AWGN channel (a). To ease comparison between constellations, (b) shows
		their SNR gap with respect to the Gaussian constellation.}\label{fig:qamthmi}
\end{figure}

Substituting this equation in \eqref{eq:midef} allows obtaining the MI of a Gaussian channel for an arbitrary input distribution $p(x)$. It can be shown
\cite{Cover:2005} that the MI is maximized (obtaining the capacity) when the input distribution is Gaussian too. In this case, the capacity is expressed by the famous
formula:
\begin{equation}
C = \log_2\left(1+\frac{\sigma^2_x}{\sigma^2_n}\right)
\label{eq:awgncapacity}
\end{equation}
However, the Gaussian constellation has an infinite number of points, which makes it not implementable in a communication system.
Real-world constellations have a finite number of points, such as $M$-QAM. 
The MIs that can be achieved by QAM constellations are shown in Fig. \ref{fig:qamthmi}, where they are compared with the Gaussian constellation.
As seen in the Figure, at low SNR the performance of QAM constellations is close to the Gaussian constellation. 
Then, each constellation saturates at its entropy, i.e. the maximum data-rate in bit/symbol that can be transmitted by it:
\begin{equation}
H(\chi) = -\sum_{x\in\chi} p(x)\log_2 p(x)
\end{equation}
where $\chi$ is the set of transmitted constellation symbols. However, at high values of SNR there is also a fixed gap between QAM
constellations and channel capacity. This gap, called \emph{shaping gap}, is asymptotically equal to $\pi e/6\approx1.53$ dB (e.g. \cite{Forney:1998})
 and is related to the sub-optimality of the 2D square shape. One of the purposes of constellation shaping is trying to close this gap.

\subsubsection{BICM channel}
\begin{figure}
	\begin{subfigure}[b]{0.48\textwidth}
		\centering 	\includegraphics[width=\textwidth]{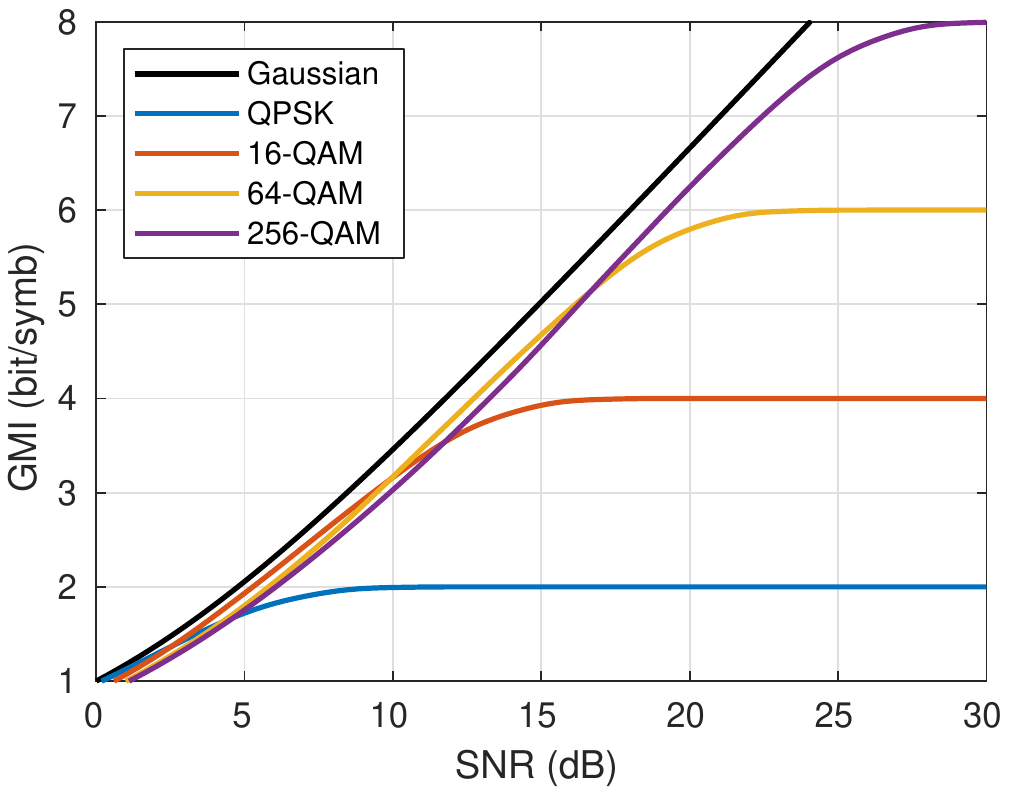}
		\caption{}
	\end{subfigure}
	\begin{subfigure}[b]{0.48\textwidth}
		\centering \includegraphics[width=\textwidth]{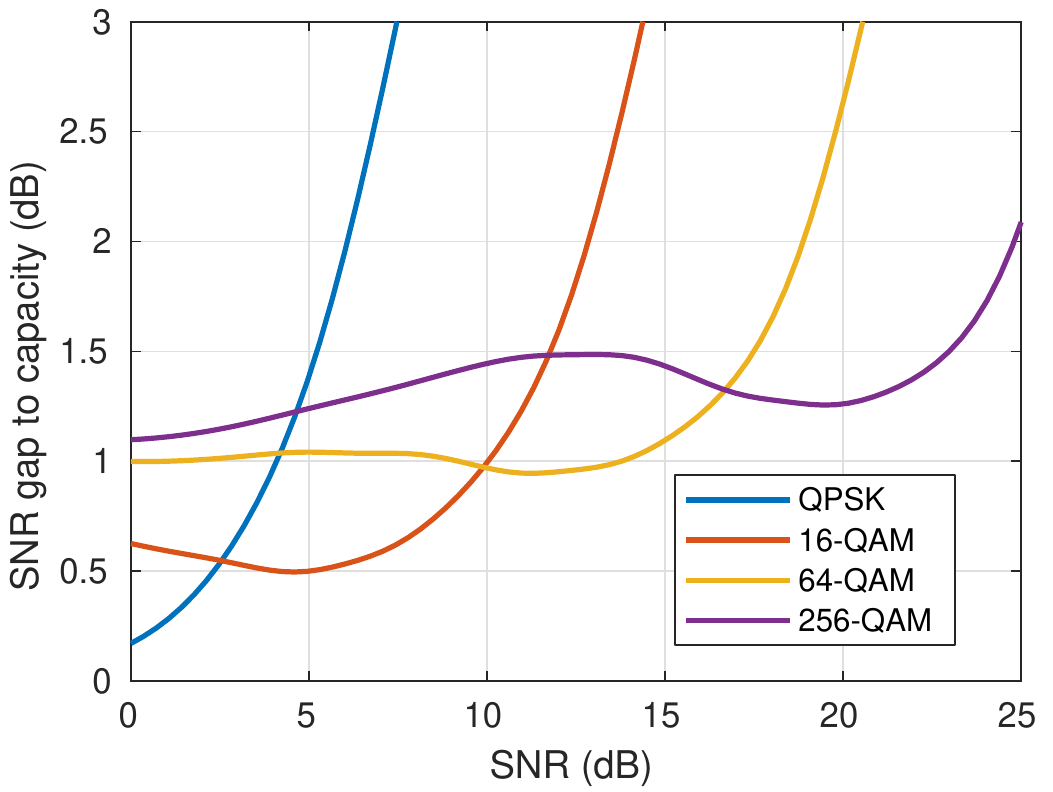}
		\caption{}
	\end{subfigure}
	\caption{Generalized mutual information of QAM constellations over an AWGN channel (a). To ease comparison between constellations, (b) shows
	their SNR gap with respect to the Gaussian constellation.}\label{fig:qamthgmi}
\end{figure}
If BICM is used (Sec. \ref{sec:5:bicm}), then the GMI \eqref{eq:gmidef} is the correct AIR to be used. The GMI of QAM constellations over
an AWGN channel is shown in Fig. \ref{fig:qamthgmi}, where it was assumed a Gray mapping between bits and constellation symbols. Comparing it
with the MI, it can be seen that at medium/high SNR performance is almost identical, decreasing at low SNR. This also means that for every SNR there
is an ``optimal'' QAM constellation which gives the highest GMI, as can be clearly seen in Fig. \ref{fig:qamthgmi}b.
In conclusion, at reasonable values of SNR, BICM does not add any significant penalty with respect to ideal coded modulation schemes.

\subsection{Filling the shaping gap}
As shown in Figs. \ref{fig:qamthmi} and \ref{fig:qamthgmi}, QAM constellation have an asymptotic gap with respect to the optimal Gaussian constellations. Therefore, some
modifications of QAM constellation are necessary to fill this gap. Several techniques that take the generic name of
\emph{constellation shaping} have been proposed to overcome this issue (see e.g. 
\cite[Sec. II]{Bocherer:2015}). In general, two different shaping methods can be applied to a QAM-based communication system:
\begin{itemize}
\item Geometric shaping (GS).
\item Probabilistic shaping (PS).
\end{itemize}
Geometric shaping techniques optimize the position of constellation points to maximize a metric (e.g. the MI), while probabilistic shaping changes the probability
to be transmitted of standard QAM constellations. Both methods are asymptotically optimal, i.e. they fully close the shaping gap for an infinite number of constellation points. 
With the same number of constellation points, it was shown that GS slightly outperforms PS \cite{Qu:2017,Qu:2018}. However, application of PS reduces the data-rate of the constellation,
which means that comparisons between GS and PS are generally performed with a different number of constellation points. In that case, PS has an advantage \cite{Buchali:2016,Zhang:2018}.
In conclusion, for the time being, there is not any clear winner, at least in the long-haul optical communications channel. At the time of writing this thesis, several
major coherent DSP vendors claim to include constellation shaping in their product. Unfortunately, none of them is providing any detail on the implemented technique.

\section{Probabilistic shaping}
\subsection{PAS architecture}
Optimization of the input probability distribution is a widely used information-theoretical technique aiming at maximizing the mutual
information \cite{Shannon:1948}. For instance, the Blahut-Arimoto algorithm \cite{Blahut:1972,Arimoto:1972} 
is a well-known algorithm to optimize symbol probabilities on a given channel. 
However, implementing a practical algorithm to combine PS and FEC (which is required in modern optical communication systems) 
is a hard task \cite{Cho:OFCinvited:2018}.
 In \cite{Bocherer:2015}, the authors developed an ingenious algorithm, called
Probability Amplitude Shaping (PAS). This algorithm allows combining PS and FEC by using a systematic code which inserts redundancy
bits as \emph{sign} of the In-Phase and Quadrature components of the constellation. In this case, FEC can be applied after PS without
modifying the amplitude distribution of constellation points. The key component of the PAS architecture is the Distribution Matcher (DM), which is a block that converts
an equiprobable stream of bits into a multi-level sequence with unequal probability. Finding an efficient hardware-implementable DM is a current topic of active research
(e.g. \cite{Bocherer:2015,Bocherer:2017,Qu:OFC2018,Yoshida:2018}).
In \cite{Bocherer:2015}, the authors present a Constant-Composition Distribution Matcher \cite{Schulte:2016} (CCDM). This algorithm uses
fixed block-sizes, and each output block has the same distribution. These features make this algorithm very easy to be used. 
However, for optimal performances it requires large block sizes, which increase its complexity. 

\subsection{Data-rate of PAS}\label{sec:6:pasdatarate}
\begin{figure}
	\centering
	\includegraphics[width=0.8\textwidth]{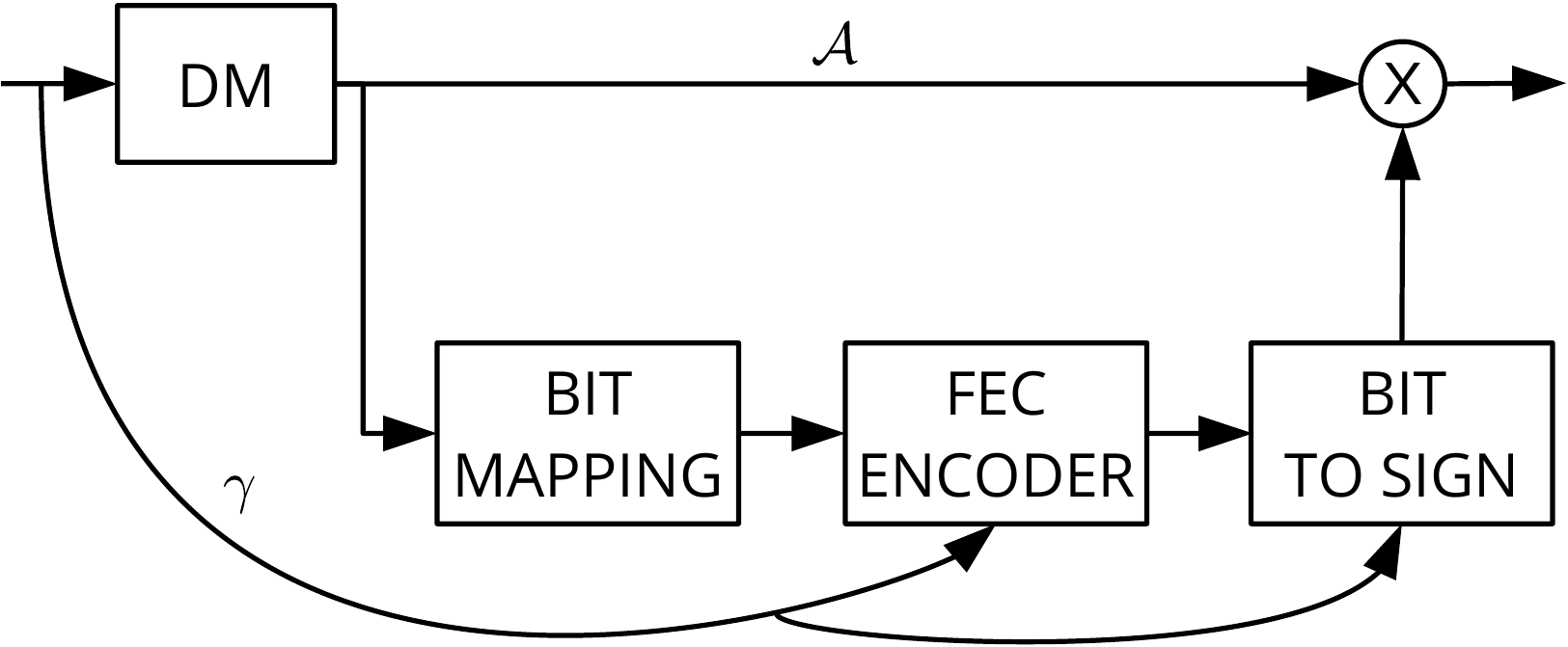}
	\caption{High-level block scheme of the PAS architecture \cite{Bocherer:2015}.}\label{fig:pascheme}
\end{figure}
The net data-rate (i.e. post-FEC) of a multilevel (e.g. QAM) coded system is trivial to evaluate:
\begin{equation}
R_\textup{info} = m\mathcal{R}_\textup{FEC}
\label{eq:uniformrate}
\end{equation}
In this equation, $m$ is the number of bit/symbol carried by each constellation point, and $\mathcal{R}_\textup{FEC}$ is the rate of the FEC code. In optical
communications, the FEC overhead $\mathrm{OH}$ is often used:
\begin{equation}
\mathrm{OH} = \frac{1-\mathcal{R}_\textup{FEC}}{\mathcal{R}_\textup{FEC}}
\end{equation}
Simplicity of these equations makes very simple to evaluate both the pre-FEC and the post-FEC data-rates. 
However, when PAS is used, these equations change significantly, making it difficult to properly evaluate the data-rates in a PAS scheme \cite{Cho:2018}. 

A high-level scheme of PAS is shown in Fig. \ref{fig:pascheme}. A stream of random equiprobable bits is converted into a sequence of un-signed ASK
amplitudes $\mathcal{A}$ by the DM. These amplitudes are then converted to bits with a Gray mapping and encoded with a systematic FEC encoder. The redundancy
bits are then converted to sign and multiplied to the ASK sequence. To allow flexibility in the FEC rate, a portion $\gamma$ of the input bits
is not shaped, and it is directly encoded and added as sign to $\mathcal{A}$. For complex-valued constellations (e.g. QAM), this operation 
is applied separately on each quadrature.

To evaluate the data-rate of PAS, the first step is the evaluation of the DM rate.
Assuming to apply PS to a $2^m$-QAM modulation, each un-signed ASK sequence has $2^{m/2-1}$ possible amplitudes.
Therefore, the DM rate $R_\textup{DM}$ is
\begin{equation}
R_\textup{DM} \leq H(\mathcal{A})
\label{eq:dmrate}
\end{equation}
An ideal DM has a data-rate that is exactly equal to the entropy of the $2^{m/2-1}$-ASK sequence $\mathcal{A}$. Therefore, \eqref{eq:dmrate} becomes an equality for an ideal DM. 
A realistic DM will have a penalty, i.e. a data-rate reduction. If all the amplitudes are equiprobable, i.e. no PS is applied, then the rate is exactly equal to $m/2-1$.

Applying such DM into the PAS architecture, the total data-rate becomes \cite{Bocherer:2015}
\begin{equation}
R_\textup{info,PAS} = 2(R_\textup{DM}+1) + m (\mathcal{R}_\textup{FEC}-1)
\label{eq:pasrate}
\end{equation}
It is self-evident that \eqref{eq:uniformrate} and \eqref{eq:pasrate} are different. While standard coded QAM has a direct proportionality between net data-rate and FEC rate, with
PAS this is not valid. This makes it quite difficult to correctly evaluate data-rates and FEC thresholds in PAS architectures \cite{Cho:2018}.

Since the FEC overhead is inserted as signs of the In-Phase and Quadrature components, there is a constraint on the minimum FEC rate (or maximum overhead) that can be applied to the architecture
\begin{equation}
\mathcal{R}_\textup{FEC} \geq \frac{m-2}{m}
\label{eq:pasminrate}
\end{equation}This constraint is another significant difference with respect to standard coded QAM. An example of maximum code rates (and overheads)  for different
square $2^m$-QAM constellations is shown in Table \ref{tab:pasminrate}.
\begin{table}
	\centering
	\begin{tabular}{c c c}
		\toprule
		Constellation & Min. FEC rate & Max. FEC overhead \\
		\midrule
		16-QAM & 0.50 & 100.0\% \\
		64-QAM & 0.66 & 50.0\% \\
		256-QAM & 0.75 & 33.3\% \\
   		1024-QAM & 0.80 & 25.0\% \\
		\bottomrule
	\end{tabular}
	\caption{Minimum FEC rates and corresponding maximum overheads for $2^m$-QAM with PAS architecture \eqref{eq:pasminrate}.}\label{tab:pasminrate}
\end{table}

\subsubsection{Achievable information rates}
Finding a suitable AIR, or equivalently, a suitable performance metric for PS constellations is still a topic of current research. Obviously, the Mutual Information
\eqref{eq:midef} is always an AIR, but in order to be ``achievable'' it needs a non-binary FEC architecture.

For binary FEC, in Sec. \ref{sec:5:decoding} we presented the GMI as an AIR for the BICM channel. At the time of writing this thesis, the GMI
is the most popular AIR for shaped constellations. However, in \cite{Cho:NGMI:2017} the authors found that the GMI may not be the most appropriate metric to compare
different modulation formats. Indeed, they propose to use the \emph{normalized} GMI (NGMI). It is defined as
\begin{equation}
\mathrm{NGMI}_u = \frac{\mathcal{G}}{m}
\end{equation}
for coded QAM (unshaped) and
\begin{equation}
\mathrm{NGMI}_p = 1-\frac{H(\chi)-\mathcal{G}}{m}
\end{equation}
for PS QAM constellations using the PAS architecture. The NGMI allows both to capture the non-ideality of a FEC code (which operates at a data-rate that is smaller
than the GMI) and the difference between \eqref{eq:uniformrate} and \eqref{eq:pasrate}.

A discussion of the possible AIR for PS constellations is out of the scope of this thesis. However, a performance metric must be chosen for the experiments with PS 
constellations. Since this thesis collects several results, published in different articles, different AIRs will be used.

\subsubsection{Calculation of AIRs}
In this Chapter and the next one, AIRs, introduced in Sec. \ref{sec:5:decoding}, will be used as performance metrics. 
In both situations, the real channel law ($q(y|x)$ in \eqref{eq:airdef}) is unknown. This channel law can be either estimated (e.g. using a histograms), or assumed.
While the first solution gives the highest AIR \cite{Agrell:2018,Alvarado:2018}, it makes difficult the calculation of the LLRs, since it may require very large
look-up tables. Therefore, for this thesis, an analytical channel law will always be assumed. Afterwards, the AIRs will be evaluated
directly from the received samples using Monte-Carlo integration \cite{Alvarado:2018}.
In particular, this Chapter assumes the AWGN channel law. In the next Chapter, a different metric will be introduced.

\subsection{Comparison between PAS and uniform QAM}
Comparing PAS QAM and uniform (i.e. standard coded) QAM transmission is not an easy task, due to the differences between their
data-rates. First of all, any fair comparison must be done at the same \emph{net} data-rate, i.e. 
\eqref{eq:uniformrate} and \eqref{eq:pasrate} must be equal.
Then, there is the choice of FEC overhead. If one wants to use the same constellation, then different FEC rates must be used. Otherwise, by constraining to use
the same FEC rate in both modulations, then the PAS solution must use a larger constellation. In both cases, the comparison is at the same data-rate, but it is not completely
``fair''. Assuming to adopt the second solution (same FEC rate), the rate of the DM is fixed and it is equal to
\begin{equation}
R_\textup{DM} = \frac{m-\mathcal{R}_\textup{FEC}(m-m_\textup{u})}{2}-1
\label{eq:samefeccom}
\end{equation}where the PS $2^m$-QAM constellation is compared to uniform $2^{m_\textup{u}}$-QAM, where $m_\textup{u}<m$. 
Note that this choice, in general, is sub-optimal since the symbol probability (which gives the DM rate) should be matched to the channel to maximize the mutual information
\cite{Pilori:2017}. However, this operation would require different FEC rates for each net data-rate, which could be difficult to implement.

\subsection{Shaping for the AWGN channel}\label{sec:6:shapawgn}
Let us assume that, as discussed in Sec. \ref{sec:5:commkerimpact}, the optical channel is an AWGN channel (the impact of eventual non-linear phase noise
will be discussed in Chapter \ref{ch:phnoise}). While one can use the Blahut-Arimoto algorithm to optimize symbol probabilities for each channel, it was shown that
the optimal distribution for the AWGN channel is the Maxwell-Boltzmann distribution \cite{Kschischang:1993}:
\begin{equation}
P(a) = \frac{e^{-\lambda|a|^2}}{\sum_i e^{-\lambda|a_i|^2}} \qquad \forall a\in\chi
\end{equation}In this equation, $a$ is a QAM symbol and $\lambda>0$ is a scaling parameter which sets the constellation entropy. 
For each SNR, there is an optimal $\lambda$ (or, equivalently, optimal entropy), i.e. the value that maximizes the mutual information \cite{Pilori:2017}.
As said before, the use of the optimal $\lambda$ in the PAS
architecture require the use of FECs with arbitrary code rates.

\subsubsection{Example}
\begin{table}
	\centering
	\begin{tabular}{c c c}
		\toprule
		Constellation & Entropy &  Net data rate \\
		& (bit/symbol) &  (bit/symbol) \\
		\midrule
		16-QAM & $4$ & $10/3\approx3.33$ \\
		PS 64-QAM & $13/3\approx4.33$ & $10/3\approx3.33$ \\
		32-QAM & $5$ & $25/6\approx4.17$ \\
		PS 64-QAM & $31/6\approx5.17$ & $25/6\approx4.17$ \\		\bottomrule
	\end{tabular}
	\caption{Details on the constellations under-test.}\label{tab:psconstellationsundertest}
\end{table}
\begin{figure}
	\begin{subfigure}[b]{0.48\textwidth}
		\centering 	\includegraphics[width=\textwidth]{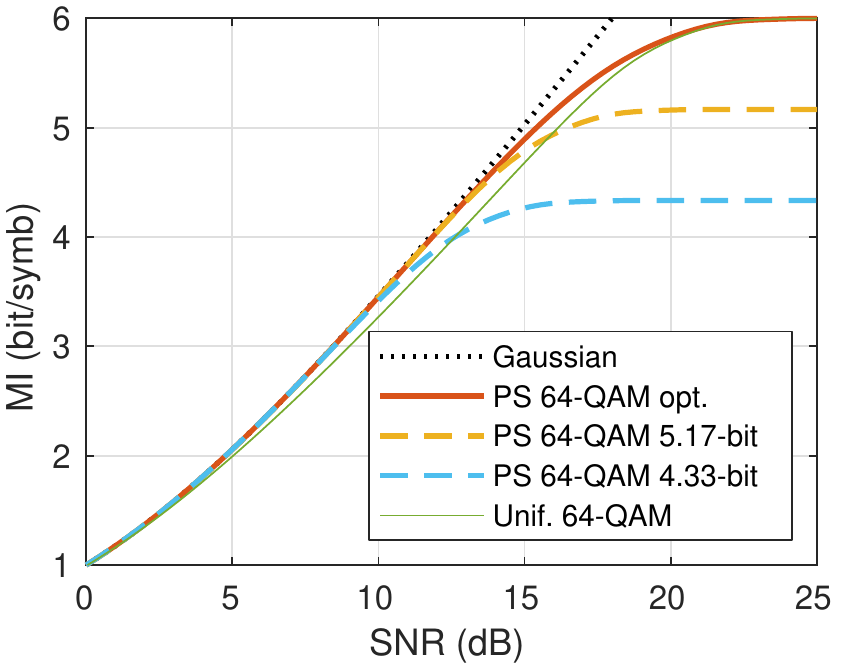}
		\caption{}\label{subfig:psqamexample}
	\end{subfigure}
	\begin{subfigure}[b]{0.48\textwidth}
		\centering \includegraphics[width=\textwidth]{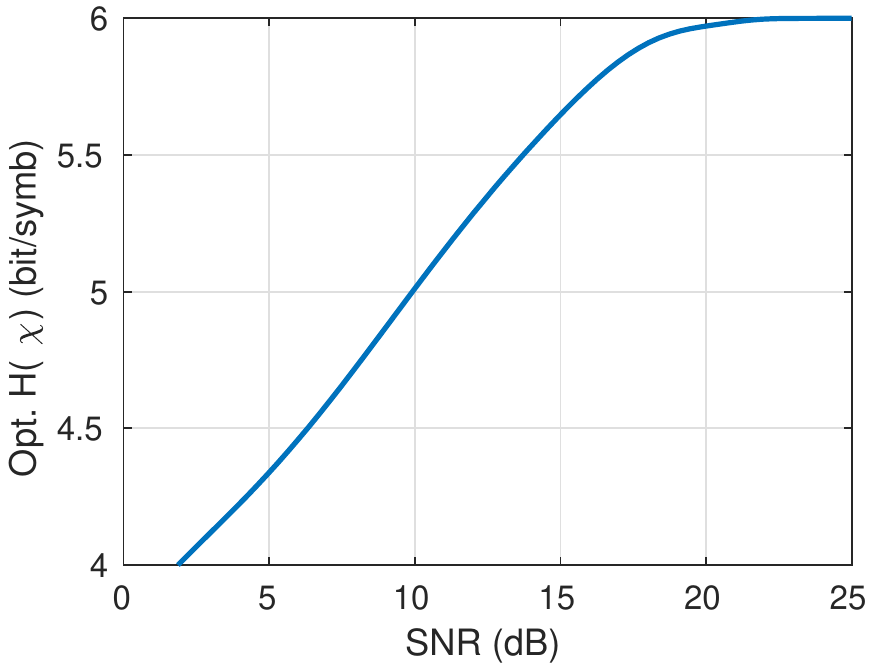}
		\caption{}\label{subfig:psqamoptentr}
	\end{subfigure}
	\caption{(a): Mutual information over an AWGN channel for PS 64-QAM with optimal $\lambda$ (blue solid line), with fixed values of $\lambda$ (dashed lines) and
	without shaping (solid green line). (b): optimal entropy for each SNR to obtain the blue solid line in (a).}\label{fig:psqamexample}
\end{figure}
As an application example, we consider different ``flavors'' of PS 64-QAM with a Maxwell-Boltzmann distribution over an AWGN channel, shown in Fig. \ref{subfig:psqamexample}. 
The solid red line (PS 64-QAM opt.) represents the curve where $\lambda$ has been optimized for every SNR. The corresponing optimal constellation 
entropy $H(\chi)$ is shown in Fig. \ref{subfig:psqamoptentr}. 
Far from the saturation at $6$ bit/symb, performance of the constellation is very close to channel capacity (black dots of Fig. \ref{subfig:psqamexample}). 
Then, we set two fixed values of $\lambda$ to obtain constellation entropies of $31/6\approx 5.17$ bit/symbol and $13/3\approx 4.33$ bit/symbol. These two 
distributions were chosen to perform a comparison with 32-QAM and 16-QAM (respectively) at the same data rate and with a $20\%$ FEC, using \eqref{eq:samefeccom}.
With an ideal FEC, i.e. with MI thresholds equal to $25/6\approx 4.17$ and $10/3\approx 3.33$ bit/symbol for (respectively) 32-QAM and 16-QAM, performance
difference between entropy-optimized constellations and fixed entropy is very small ($<0.1$ dB). The main parameters of these four constellations under test are
summarized in Table \ref{tab:psconstellationsundertest}.

\subsubsection{Exponential distribution}
\begin{figure}
	\begin{subfigure}[b]{0.48\textwidth}
		\centering \includegraphics[width=\textwidth]{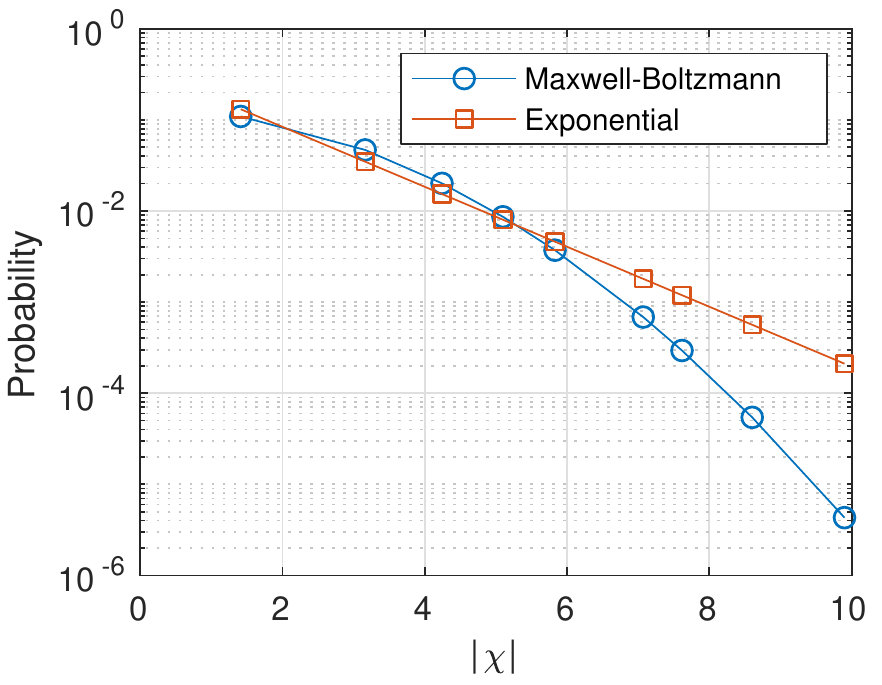}
		\caption{}
	\end{subfigure}
	\begin{subfigure}[b]{0.48\textwidth}
		\centering \includegraphics[width=\textwidth]{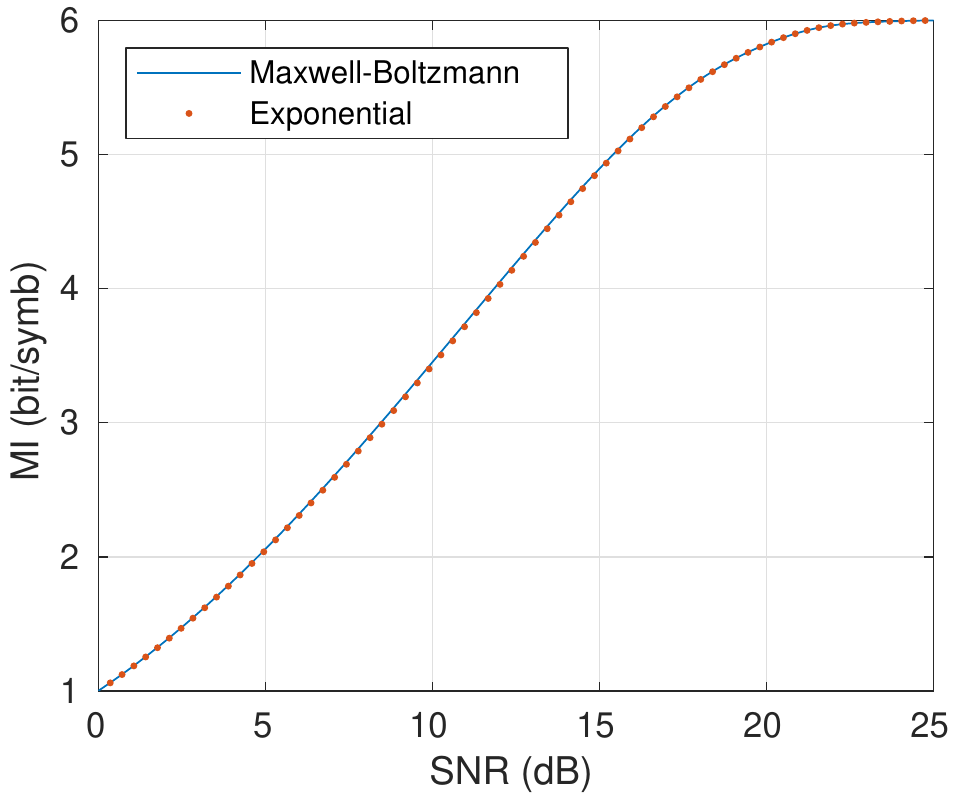}
		\caption{}
	\end{subfigure}
	\caption{Amplitude probability distribution of a PS 64-QAM constellation with entropy $13/3$ bit/symbol (a). 
Mutual information of PS 64-QAM with optimized Maxwell-Bolzmann and exponential distributions (b).}\label{fig:expvsmb}
\end{figure}
\begin{figure}
	\begin{subfigure}[b]{0.48\textwidth}
		\centering \includegraphics[width=0.8\textwidth]{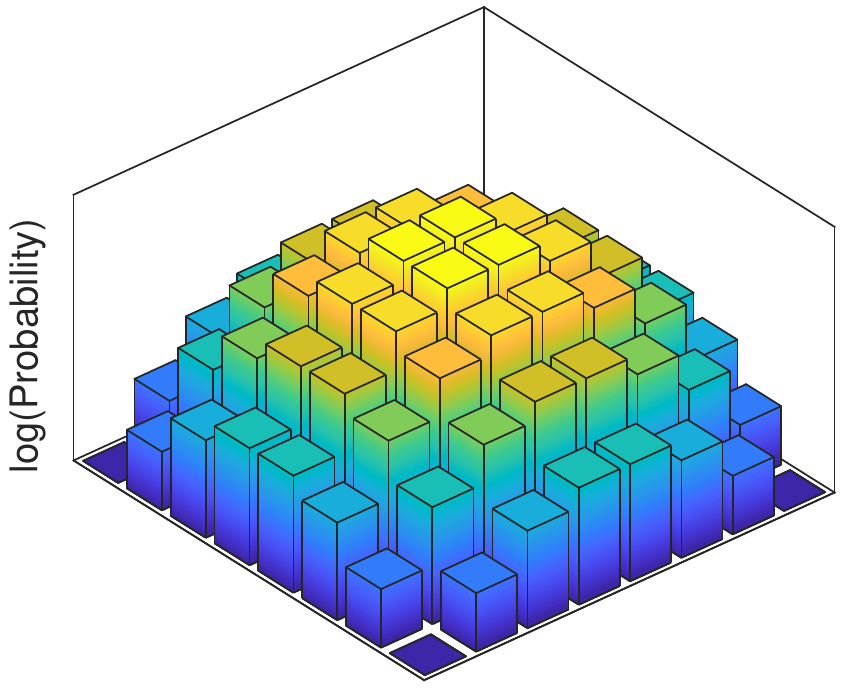}
		\caption{Maxwell-Boltzmann}
	\end{subfigure}
	\begin{subfigure}[b]{0.48\textwidth}
		\centering \includegraphics[width=0.8\textwidth]{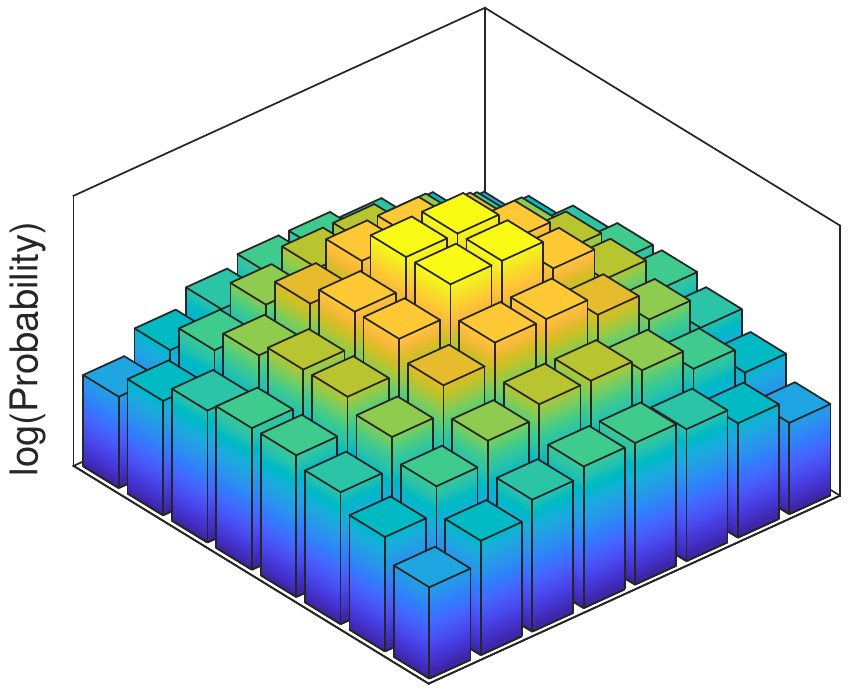}
		\caption{Exponential}
	\end{subfigure}
	\caption{3D plots of the amplitude distribution (logarithmic scale) of PS 64-QAM with entropy equal to $13/3$ bit/symb, using the Maxwell-Boltzmann distribution (a) or the
	exponential distribution (b).}\label{fig:expmb3d}
\end{figure}

While the Maxwell-Boltzmann distribution is optimal, it has some drawbacks. At low entropies, the ratio between the probability of the innermost and outermost
points of the constellation can become significantly large. For instance, assuming PS 64-QAM with entropy $H(\chi)=13/3$ bit/symb, the ratio is approximately
$\sim2.5\times10^4$. This large ratio makes very difficult its implementation, especially with a CCDM. 

The exponential distribution, instead, 
\begin{equation}
P(a) = \frac{e^{-\lambda|a|}}{\sum_i e^{-\lambda|a_i|}}
\end{equation} is more ``uniform'' than the Maxwell-Boltzmann, at the same entropy. With the same example described before (PS 64-QAM with entropy $H(\chi)=13/3$ bit/symb), 
the ratio is only $\sim6.2\times10^2$. This also reduces the PAPR by $\sim0.2$ dB. This example is graphically
shown in Fig. \ref{fig:expvsmb}a, where it is shown the probability of each ``ring'' of a 64-QAM constellation using both distributions with the same entropy. A 3D plot
of the two distributions, in logarithmic scale, is shown in Fig. \ref{fig:expmb3d}.

Obviously, the exponential distribution is not optimal in the AWGN channel. Fig. \ref{fig:expvsmb}b shows its performance, compared to Maxwell-Bolzmann, over
an AWGN channel. Performance difference is very small, which makes the exponential distribution a good compromise between performance and complexity. For instance, it has been
also applied to PAM for data-center communications \cite{Han:2018}.

\subsection{Experimental demonstration}\label{sec:6:expdem}
In recent years, several experimental demonstrations of PS in optical communications 
\cite{Yankov:2014,Buchali:2015,Pan:2016,Buchali:2016,Ghazisaeidi:2017,Idler:2017,Cho:PDP:2018,Olsson:2018} have been published. Among these results, 
ECOC 2015 post-deadline paper \cite{Buchali:2015} (extended in the \emph{Journal of Lightwave Technology} in \cite{Buchali:2016}) was the first
optical demonstration of the PAS architecture, ``kick-starting'' plenty of other research works. It is interesting that, in those early works, the reach gain enabled
by PS that was claimed significantly differ between different works. Claims ranged from $7\%$ \cite{Pan:2016} to $40\%$ \cite{Buchali:2016}. Some works even
claimed $300\%$ reach gains \cite{Zhu:2017}! In fact, the gain of PS largely depends on several factors, such as system scenario, modulation format, FEC rate
and target net data rate. Moreover, as discussed before, difference between PAS data rate \eqref{eq:pasrate} and coded QAM data rate \eqref{eq:uniformrate} makes difficult
to perform fair comparisons.

Therefore, we decided to perform a thorough experimental comparison of PS 64-QAM and uniform 16-QAM and 32-QAM at the same net data rate. 
Assuming a $20\%$ FEC code (rate $\mathcal{R}_\textup{FEC}=5/6)$, the constellation entropies were the same as the constellations described
in Sec. \ref{sec:6:shapawgn}, but we used the exponential distribution. As an AIR, for this work we adopted the MI, which gives an AIR without
making any assumption on the receiver structure. 
These results were presented at OFC 2017 conference \cite{Bertignono:2017}, and then extended, as an invited submission, in the \emph{Journal of Ligthwave Technology} \cite{Pilori:JLT2018}.

\subsubsection{Setup}
\begin{figure}
	\centering
	\includegraphics[width=0.8\textwidth]{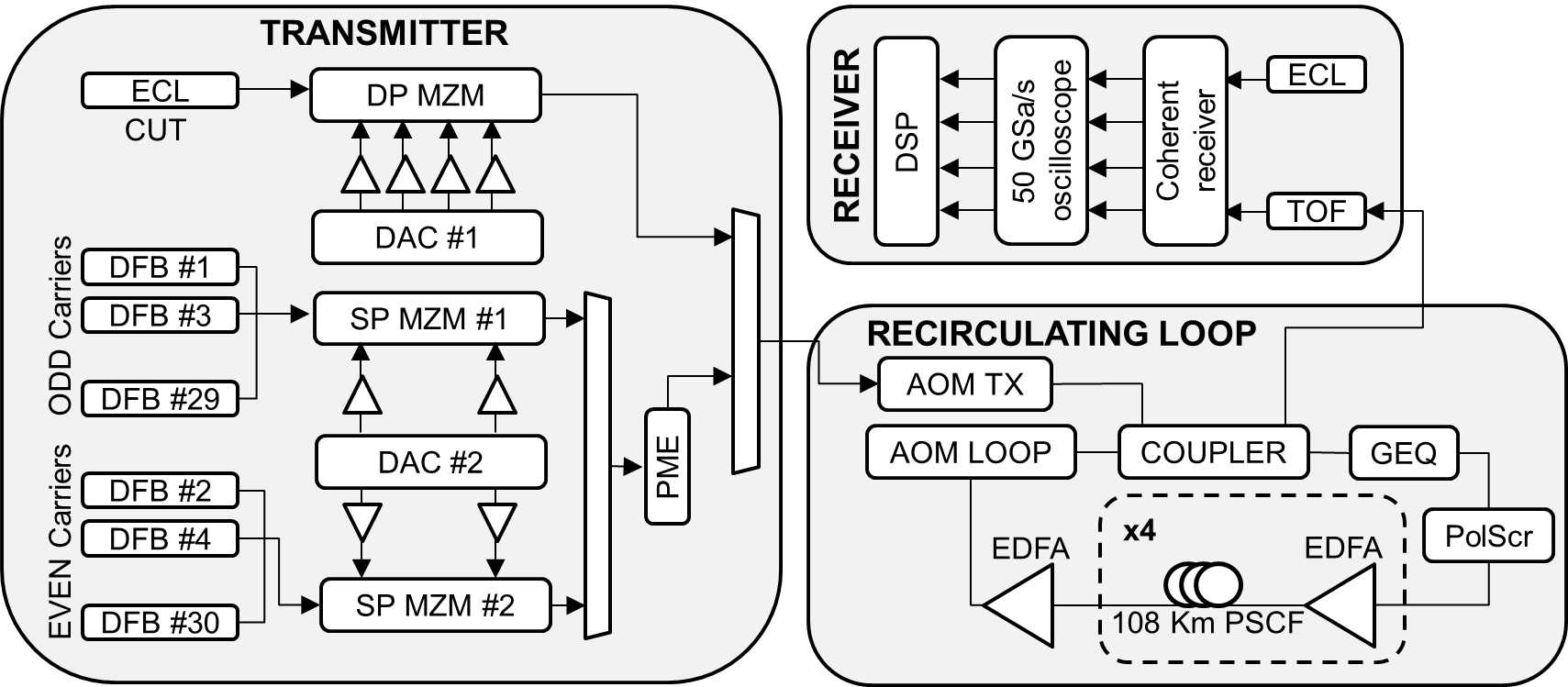}
	\caption{Experimental setup.}\label{fig:expsetuppspscf}
\end{figure}
A scheme of the experimental setup is shown in Fig. \ref{fig:expsetuppspscf}. The transmitter generates $31$ WDM channels at $16$ GBaud each, with a $25$ GHz spacing.
The Channel-under-Test is the central channel, and it is generated with a $<100$-kHz External Cavity Laser (ECL) and modulated with a dual-polarization
lithium-niobate Mach-Zehnder modulator. The other channels were generated using DFBs, modulated by two single-polarization MZMs, followed by a PM-emulator.
All channels are shaped with a $15\%$-roll-off Root Raised Cosine filter. This signal can be either directly connected to the receiver to perform optical
back-to-back measurements, or propagated over a recirculating loop \cite{Bergano:1991}. Each loop is made by four $108$-km spans of Pure Silica Core Fiber (PSCF), followed
by an EDFA with $5.2$ dB noise figure that fully recovers span loss. Fiber had an attenuation of $0.17$ dB/km, chromatic dispersion of $20.17$ ps/nm/km and Kerr coefficient
$0.75$ 1/W/km. Inside each loop, a Gain Equalizer filter (GEQ) equalizes the power of the WDM spectrum, and a loop
synchronous polarization scrambler emulates polarization characteristic of a transmission line \cite{Yu:2003}.

At the receiver, a tunable optical filter selects the CUT, which is then detected by an integrated coherent receiver, and digitized by a four-channel $50$-Gs/s real-time
oscilloscope.

\subsubsection{Receiver phase recovery}\label{sec:6:rxcpe}
Receiver DSP uses the same structure detailed in Sec. \ref{ch:5:dspchain}. The only critical component is the CPE, which compensates for both laser phase noise
and non-linear phase noise (explained in Sec. \ref{sec:5:commkerimpact}). In this work we adopted two different phase recovery strategies.

The first strategy is the ``Ideal CPE'', also called Phase Noise Receiver \cite{Fehenberger:2015}, which removes any phase noise by using a fully data-aided algorithm. This
algorithm is obviously not realistic, but it represents the best performance that an ideal feed-forward phase recovery algorithm can achieve. To avoid over-compensation of noise, we
measured the non-circularity index of the constellation \cite{Poggiolini:2017} and set the memory of the CPE to have the constellation points as close as possible
to circles.

\begin{figure}
	\centering
	\includegraphics[width=0.7\textwidth]{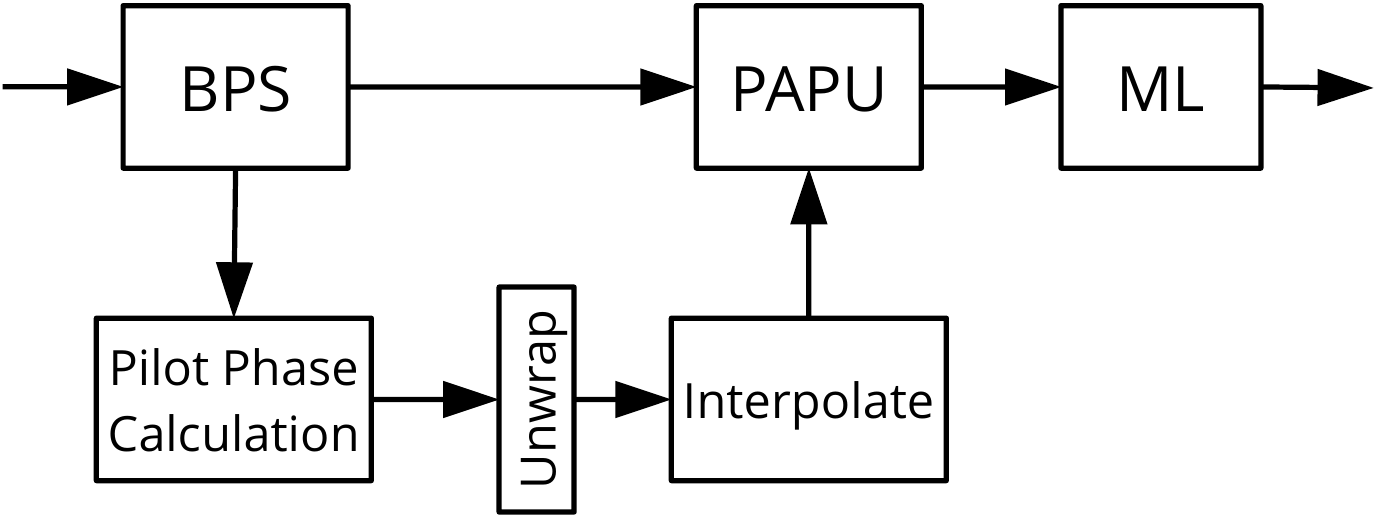}
	\caption{Block scheme of the pilot-aided BPS-ML phase recovery algorithm used in this thesis.}\label{fig:cpescheme}
\end{figure}
The second scheme is more realistic and it is based on a combination of Blind Phase Search \cite{Pfau:2009} and Maximum Likelihood (ML) \cite{Zhou:2010}. While these algorithms
are blind, BPS is only able to estimate phase within a quadrant. Therefore, a proper phase-unwrap algorithm is required. Since blind phase phase-unwrap algorithms work
only for low symbol error rates, we periodically inserted pilot symbols to train the algorithm \cite{Magarini:2012,Nishihara:2014}.
A high-level schematic of this CPE is shown in Fig. \ref{fig:cpescheme}. Pilots are used \emph{only} for phase unwrapping, therefore a large pilot overhead will not
increase performance of the algorithm, but it will only cancel cycle slips. For this work we used BPS with $18$ test angles in the range
$[-\pi/4,\pi/4)$ and a $\sim2\%$ pilot overhead. It is noteworthy to remark that this algorithm is far from being ideal. Indeed, it was shown that BPS
has a penalty when applied to PS \cite{Mello:2018}. Therefore, more sophisticated algorithms (such as \cite{Yankov:2015}) may give better performances.

\subsubsection{Back-to-back results}
\begin{figure}
	\begin{subfigure}[b]{0.48\textwidth}
		\centering 	\includegraphics[width=\textwidth]{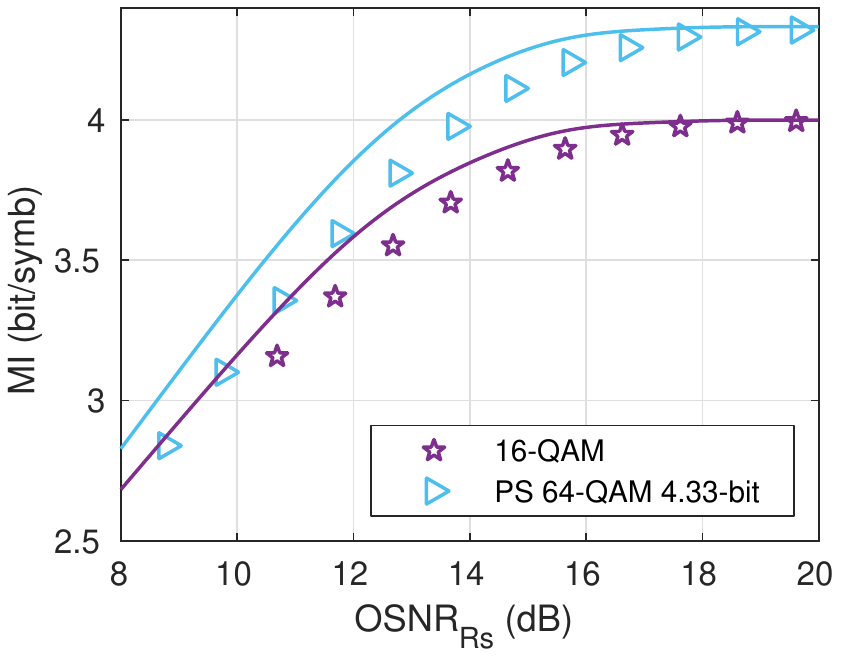}
		\caption{}\label{subfig:4bitb2b}
	\end{subfigure}
	\begin{subfigure}[b]{0.48\textwidth}
		\centering \includegraphics[width=\textwidth]{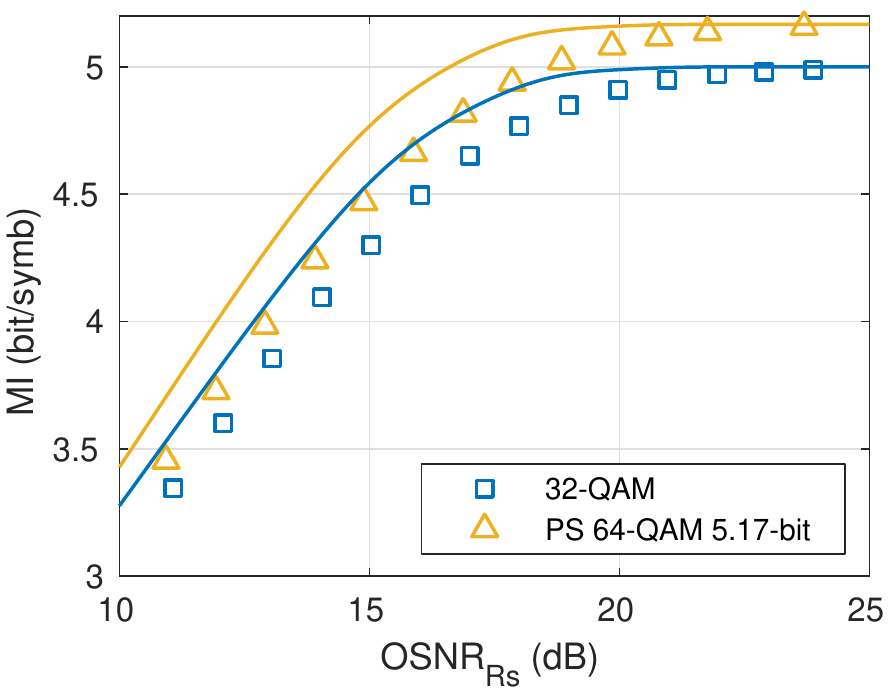}
		\caption{}\label{subfig:5bitb2b}
	\end{subfigure}
	\caption{Back-to-back results at 16 GBaud (markers), alongside the theoretical curves (solid lines). 
		In each figure the constellations bear the same net data rate, assuming a $20\%$ FEC.}\label{fig:psqamb2bres}
\end{figure}
Optical back-to-back performances of the two PS 64-QAM constellations, compared to 16-QAM and 32-QAM, are shown in Fig. \ref{fig:psqamb2bres}. The theoretical
curves were obtained by solving \eqref{eq:midef} using the Gauss-Hermite quadrature method. Uniform constellations
have a $\sim0.9$ dB penalty with respect to theory, while PS 64-QAM have a slightly larger penalty of $\sim1.05$ dB. In fact, quantization penalty
of PS 64-QAM constellations is not expected to be significantly larger than uniform constellations (see Sec. \ref{sec:6:quant} for details).

\subsubsection{Propagation results}
\begin{figure}
	\begin{subfigure}[b]{0.48\textwidth}
		\centering \includegraphics[width=\textwidth]{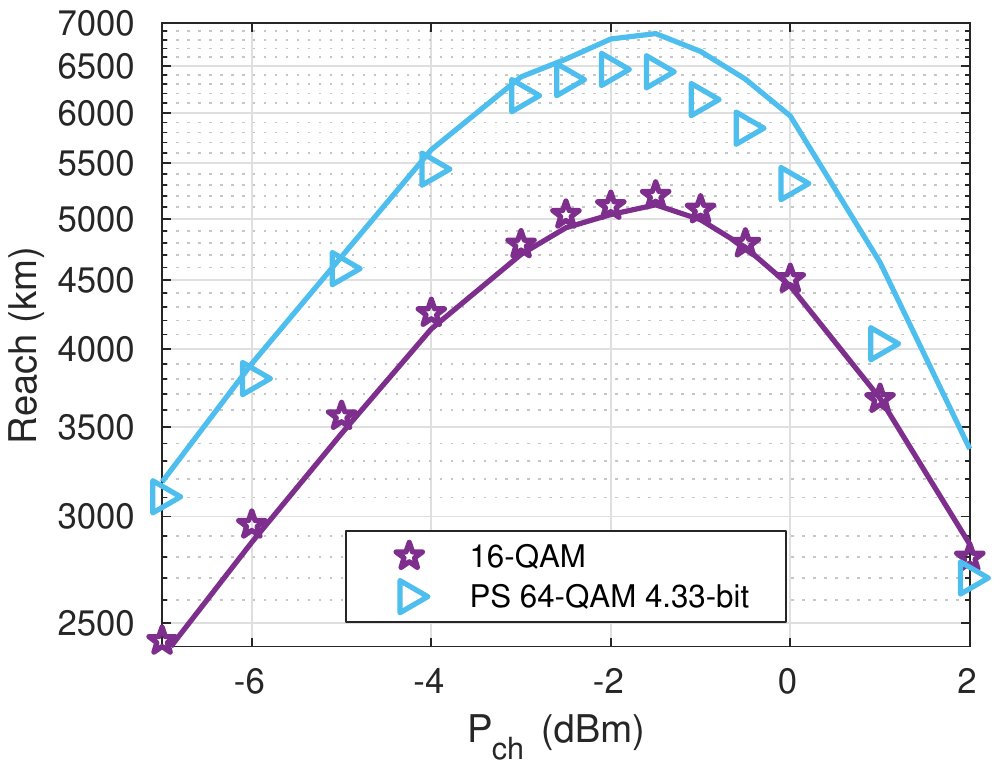}
		\caption{}\label{subfig:4bitpscfpch}
	\end{subfigure}
	\begin{subfigure}[b]{0.48\textwidth}
		\centering \includegraphics[width=\textwidth]{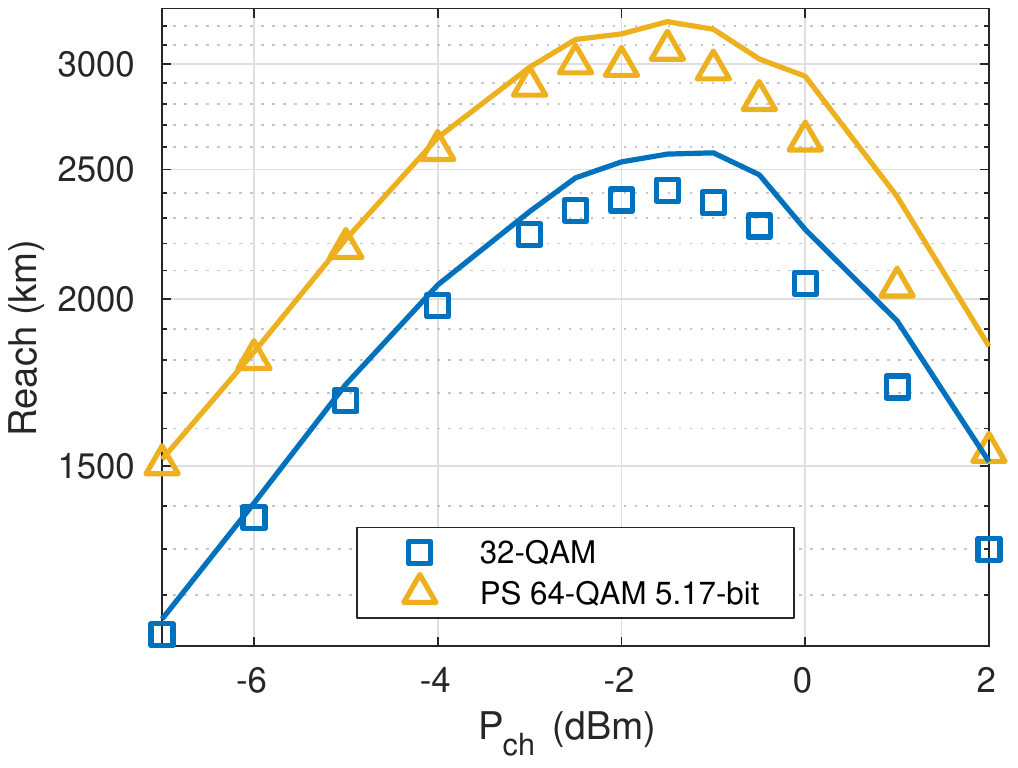}
		\caption{}\label{subfig:5bitpscfpch}
	\end{subfigure}
	\caption{Maximum reach for different per-channel transmit power over $108$-km spans of PSCF. Solid lines were obtained with the Ideal CPE, while markers
	are obtained with BPS-ML CPE.}\label{fig:psqampscfpch}
\end{figure}
Results after propagation over $108$-km spans of PSCF are shown in Fig. \ref{fig:psqampscfpch} as maximum reach (expressed in kilometers) as a function of the
per-channel launch power in dBm. To evaluate the maximum reach, we arbitrarily set the MI thresholds to $3.6$ bit/symbol for 16-QAM/PS 64-QAM 4.33-bit and 
$4.5$ bit/symbol for 32-QAM/PS 64-QAM 5.17-bit. Results were obtained both using an ideal CPE (solid lines) and a more realistic BPS-ML
(markers). At low transmit power, where performance is ASE-limited, both CPEs fully compensate for laser phase noise, obtaining the same values of reach. Approaching
the optimal power, there is a slight penalty with BPS due to the effect of NLPN on all constellations except for 16-QAM, which has
the smallest kurtosis \eqref{eq:kurtosis}. Nevertheless, the difference is very small. Maximum-reach gains using an ideal CPE are $34.4\%$ and $24.9\%$ for 16-QAM
and 32-QAM, respectively. Using instead BPS-ML, gains become $24.4\%$ and $27.5\%$.

\subsubsection{Discussion and conclusion}
Looking at the comparison, both in back-to-back and after propagation, some conclusions can already be drawn. First of all, by comparing PS 64-QAM with lower cardinality
uniform constellations, realistic maximum-reach gains are $\sim30\%$. Higher gains are obtained only if the comparison
is performed at different data-rates (or symbol rates). Then, even if the kurtosis is increased, we did not measure any significant increase in NLI generation. This may be due
the long distance, or the use of large effective area fiber. More details on this will be given in Chapter \ref{ch:phnoise}.

These conclusions have been obtained using the MI performance metric, which assumes an ideal receiver. Using a more ``realistic'' metric (e.g. the NGMI), results
are expected to be different. However, we do not expect significant differences on the impact of Kerr effect. Therefore, we expect
that the main conclusions are still valid. Nevertheless, future work should thoroughly investigate this aspect,
assuming a realistic receiver structure (e.g. PAS architecture with BICM), and using the corresponding AIR.

\subsection{Impact of DAC quantization noise}\label{sec:6:quant}
\begin{figure}
	\centering
	\includegraphics[width=0.6\textwidth]{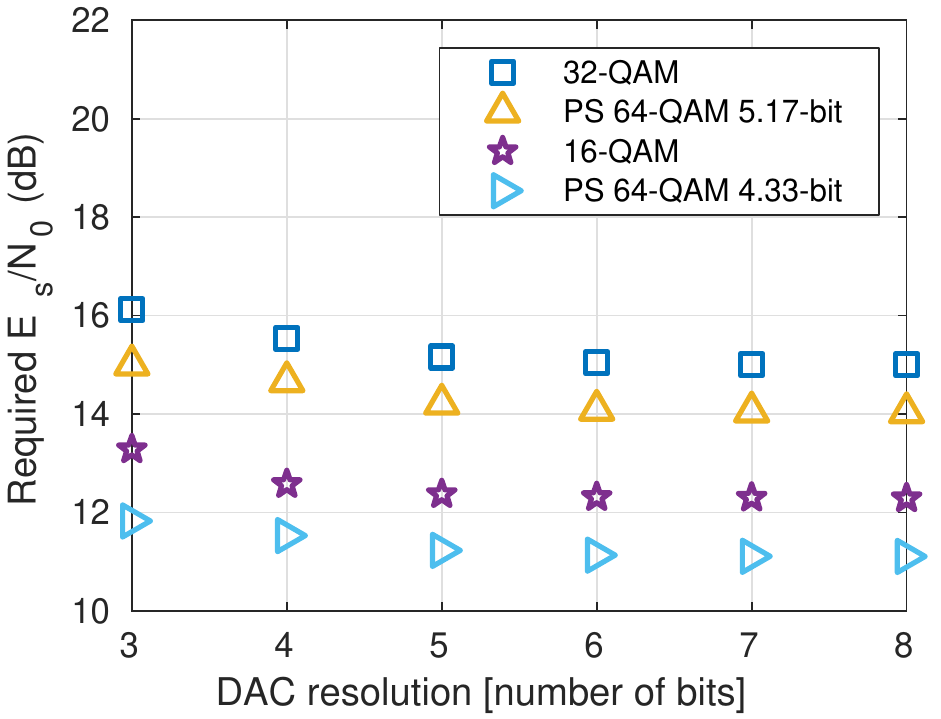}
	\caption{Required SNR of the PS 64-QAM in the presence of DAC quantization noise. Results are compared to 16 and 32-QAM \cite{Pilori:IPC2017}.}\label{fig:dacquantization}
\end{figure}
In Fig. \ref{fig:psqamb2bres}, it was shown that PS 64-QAM does not have a significant back-to-back penalty with respect to uniform QAM
constellations. This can be somehow counter-intuitive, since 64-QAM is expected to be more sensitive to quantization noise. 
The answer lies on the fact that, even if the number of points is different, the target OSNR is the same (or even lower) as lower-cardinality QAM.

To prove this hypothesis, it was set up a simple back-to-back simulation, where the DAC was assumed to have a finite number of resolution bits, constant
across the full bandwidth \cite{Pilori:IPC2017}. All the other components of the simulation were assumed ideal, and signals were shaped using a $15\%$ roll-off RRC filter.  
Results, assuming the same MI threshold as previous Sec. \ref{sec:6:expdem}, are shown in Fig. \ref{fig:dacquantization}. For resolution bits larger than $5$ bits, penalty
is negligible. For lower values of resolution, penalty is only slightly higher for PS 64-QAM. For instance, at $4$-bit resolution, penalty with respect to an ideal DAC
is $0.43$ dB and $0.64$ dB for PS 64-QAM with 4.33 and 5.17 bit/symb entropy, respectively. Penalty for 16-QAM and 32-QAM is instead $0.29$ dB and $0.55$ dB
respectively. 

\section{Geometric shaping}
Compared to PS, implementation of GS is relatively straightforward. Given a channel (i.e. $p(y|x)$), positioning of constellation points is optimized using either
the MI or the GMI performance metric \cite{Zhang:2018}. Optimizing with the GMI metric is more difficult, because bit mapping needs to be optimized
alongside symbol position. The optimization is usually done with iterative algorithms, such as Simulated Annealing (SA) \cite{Kirkpatrick:1983,Kayhan:2014} 
or modified gradient descent \cite{Foschini:1974}.
Afterwards, a GS constellation is used like a standard QAM constellation. With respect to PS, it does not give any data-rate flexibility, but it does not require
special architectures like PAS.

As an example, the APSK constellation will be briefly presented. More sophisticated GS constellations will be presented in Chapter \ref{ch:phnoise}.

\subsubsection{Example: APSK constellation family}
\begin{figure}
	\centering
	\includegraphics[width=0.4\textwidth]{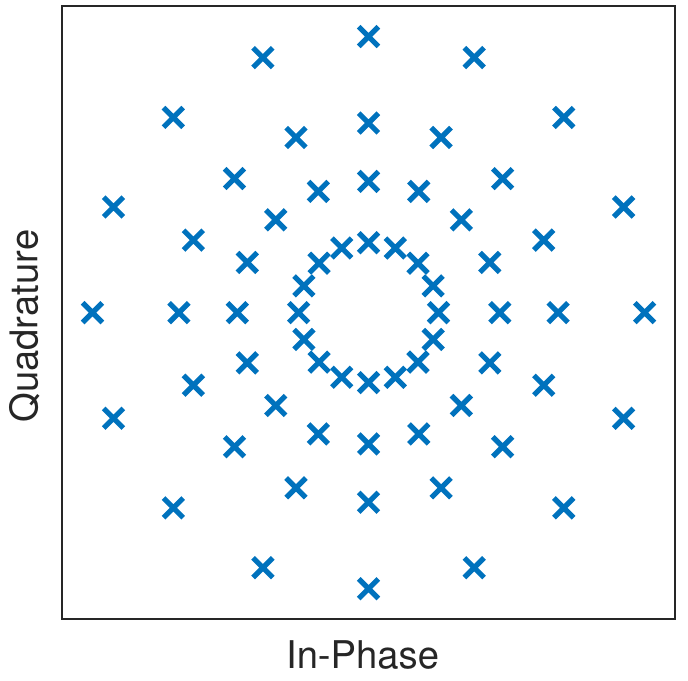}
	\caption{Constellation diagram of a 64-APSK constellation.}\label{fig:64apskexample}
\end{figure}
A popular geometrically-shaped constellation for the AWGN channel is APSK. APSK constellations are made by several concentric PSK rings. 
Number of rings, relative phase shift and amplitude spacing are free parameters that are optimized, based on the received channel. 
In \cite{Liu:2011} the authors describe a method to generate such constellation with a good bit mapping, which allows the use of BICM at the receiver.
An example of such APSK constellation is shown in Fig. \ref{fig:64apskexample}, in which the constellation is made by $4$ concentric $16$-PSK constellation, for a total of $64$ points. 

\begin{figure}
	\centering
	\includegraphics[width=0.7\textwidth]{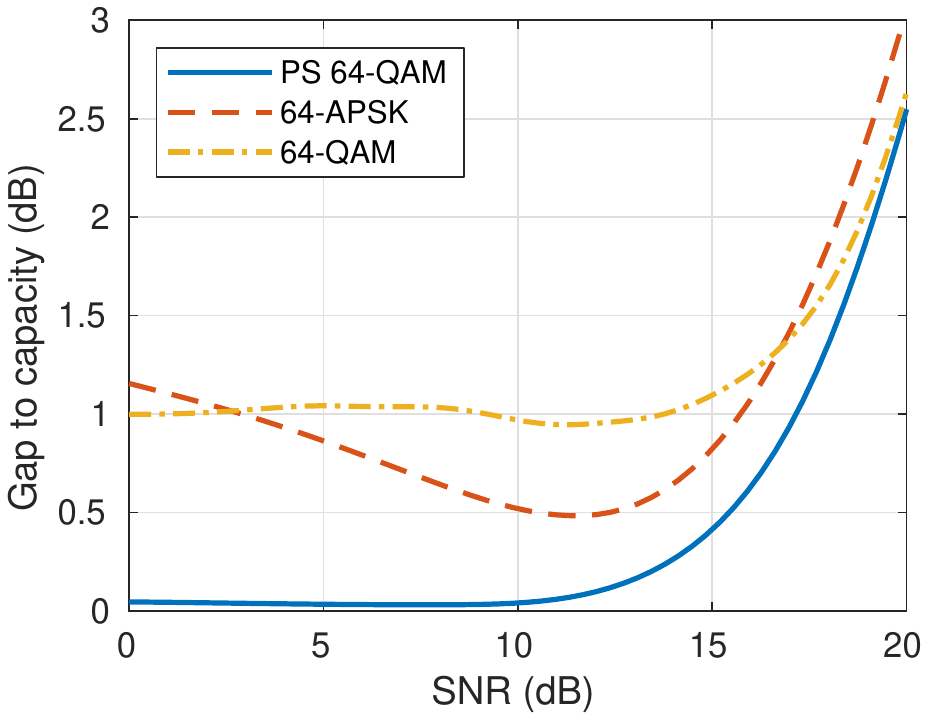}
	\caption{Performance of 64-APSK in an BICM-AWGN channel compared to PS 64-QAM and 64-QAM.}\label{fig:64apskgmi}
\end{figure}
Performance of this constellation over a BICM-AWGN channel is shown in Fig. \ref{fig:64apskgmi}. For 64-QAM, we assumed a standard Gray mapping, while 64-APSK
used the bit mapping described in \cite{Liu:2011}.
To ease comparison between different constellations, instead of the GMI, the SNR gap (in dB) to the AWGN capacity \eqref{eq:awgncapacity} is shown. PS 64-QAM, obtained
with an optimized Maxwell-Boltzmann distribution, achieves the best result. 64-APSK is able to obtain better results than 64-QAM only in a limited SNR region. This pattern
happens quite often in GS constellations, which (in general) do not perform best outside of their ``target SNR'' region.
Nevertheless, the SNR gain ($\sim0.5$ dB in the optimal region) of 64-APSK is still significant, keeping in mind that it can be achieved with standard DSP algorithms. Moreover,
64-APSK can be combined with PS \cite[Appendix I]{Jardel:2018} to further increase performance and allow flexible data-rates.

\chapter{Non-Linear Phase Noise}\label{ch:phnoise}
    \graphicspath{{Chapter7/}}\
In Chapter \ref{ch:coherent} it was shown that the AWGN channel is a good approximation of Kerr effect in a coherent optical channel. However, there
are situations where Kerr effect generates phase-noise, called non-linear phase noise. 
This Chapter collects several results related to the non-AWGN portion of NLI, with a specific focus on systems using constellation shaping.
Afterwards, some possible compensation techniques will be presented.

Results presented in this chapter are based on \cite{Pilori:JLT2018,Pilori:OFC2018,Pilori:OFC2019} and other submitted (or in-progress) works.

\section{Modeling NLPN}
The presence of a phase-noise component of NLI has been known for several years \cite{Secondini:2012,Dar:2013,Dar:2015,Dar:2016,Poggiolini:2017}. The cause of this NLPN
is the non-linear beating between the channel under test and another WDM channel, which is an effect that takes the name of Cross Phase Modulation (XPM).

Generation of NLPN strongly depends on the system characteristics. While long-distance systems ($>1000$ km) with lumped amplification exhibit low (or none) NLPN, metro links
or systems with Raman amplifications exhibit a significant phase-noise component \cite[Tab. I]{Dar:2015}. Unfortunately, the popular GN \cite{Poggiolini:2014} and EGN \cite{Carena:2014}
models approximate all NLI as an AWGN source, which means that they cannot be used to estimate phase-noise generation \cite{Poggiolini:2017}. Consequently, a different theoretical model must be used.

\subsubsection{Pulse-collision theory of NLPN}
A good theoretical framework to study the generation of NLPN is the ``pulse-collision theory'' \cite{Dar:2016}, which is based on the time-domain model developed in
\cite{Mecozzi:2012,Dar:2013}. This model, based on a first-order perturbation approximation, assumes that the received symbol at the $n$-th time instant $a_n$ can be expressed as
\begin{equation}
y_n = a_n + \Delta a_n
\end{equation}
In this equation, $\Delta a_n$ represents the effect of NLI. Without any loss of generality, we consider the $n=0$ time instant, since we assume that NLI is stationary.
According to the model, $\Delta a_n$ caused by XPM with an interfering WDM channel can be expressed as
\begin{equation}
\Delta a_0 = 2j\gamma \sum_{h,k,m}a_h b^*_k b_m \cdot X_{h,k,m}
\label{eq:deltaa0}
\end{equation}
In this equation, $b_n$ are the symbols of the interfering WDM channel, $\gamma$ is the non-linear Kerr parameter and $X_{h,k,m}$ are the NLI coefficients.
Assuming that all symbols are statistically independent from each other, the terms with $h=0$ generate phase-noise,
while terms with $h\neq0$ generate an additive circular noise. 
Therefore, an analysis of the strength of the $X_{h,k,m}$ allows to
distinguish between generation of amplitude and phase noise.

It was shown that these coefficents depend on the ``collision'', during propagation over $L$ meters of fiber, of (up to) four pulses $g(z,t)$:
\begin{align}
X_{h,k,m} &= \int_0^L f(z) \int_{-\infty}^{+\infty} g^*(z,t) g(z,t-hT) \cdot \\
& g(z,t-kT-z\beta_2  \Delta f) g(z,t-mT-z\beta_2 \Delta f) \,\mathrm{d}t \,\mathrm{d}z
\label{eq:pulsecollision}
\end{align}
The waveform $g(z,t)$ is the transmit pulse-shaping filter (e.g. root raised cosine in the frequency domain),
dispersed by chromatic dispersion $g(z,t)=g(0,t)\exp(-0.5j\beta_2\partial^2/\partial t^2)$, and $T$
is the symbol duration. As discussed before, phase-noise is generated by the terms where $h=0$. These terms are distinguished between 
two-pulse collisions ($k=m$) and three-pulse collisions ($k\neq k$). In \cite{Dar:2016}, the authors showed that three-pulse collisions are often very small; therefore, three-pulse
collisions are neglected, focusing on two and four-pulse collisions.
Assuming $h=0$ and $k=m$ in \eqref{eq:deltaa0} and \eqref{eq:pulsecollision}, the perturbation is equal to $y_0 = a_0(1+j2\gamma |b_m|^2 X_{0,m,m})$, where $X_{0,m,m}$ is positive
and real-valued. The $|b_m|^2$ terms means that power of this phase noise depends on the fourth-order moment of $b_m$ \eqref{eq:kurtosis}, which depends on the shape
of the constellation. On the other end, four-pulse collisions depend only on the second-order moment of $b$, therefore they are not constellation-shape dependent.

Relative significance between two-pulse and four-pulse collisions depend on system conditions. It was found that two-pulse collisions prevail in short-distance
systems, links using distributed (e.g. Raman) amplification, low-dispersion fibers, and systems operating at relatively low symbol rates \cite{Dar:2016,Poggiolini:2016,Dar:2017}.
Since the strength of two-pulse collisions depend on the kurtosis of the constellation, signal shaping will inevitably worsen the impact of those collisions. 
In the next sections, some examples will be given.

\section{Examples}\label{sec:7:examples}

\subsection{Low symbol rates}\label{sec:7:lowrs}
A system scenario where NLPN is expected to dominate is subcarrier multiplexing, introduced in Sec. \ref{subsec:multichannel} of Part \ref{part:dd}. In long-haul
systems, in addition to the features illustrated in Part \ref{part:dd}, there is a convex dependence between the symbol rate and overall NLI variance.
In particular, it was found that there is a specific symbol rate that minimizes NLI generation.
This has been shown both theoretically, using the EGN model \cite{Carena:2014,Poggiolini:2015}, and experimentally \cite{Poggiolini:2016}. However, it was shown
\cite{Dar:2017} that, reducing the symbol rate, two-pulse collisions become stronger, i.e. more NLPN is generated. Moreover, given the longer duration
of symbols, NLPN memory is expected to be shorter, which makes it difficult to be compensated by CPE algorithms. 

Therefore, we simulated a communication system with $100$-km spans of SMF (Table \ref{tab:fibparams}) at different symbol rates.
The constellations under test are identical to the ones used in Chapter \ref{ch:shaping}, which are summarized in Table \ref{tab:psconstellationsundertest}.
To perform a fair comparison, we kept constant the total modulated optical bandwidth, equal to $480$ GHz, 
and the relative channel spacing $\delta f = 25/16$, which correspond to 32 GBaud channels in the 50 GHz DWDM grid. 
Laser phase noise was scaled with the symbol rate, such that the linewidth-symbol rate ratio
is constant $\nu/Rs = 2.5$ kHz/GBaud. Receiver used the pilot-aided BPS-ML phase recovery introduced in Sec. \ref{sec:6:rxcpe}.
Performance was measured with the MI metric, with thresholds of 3.6 and 4.5 bit/symb, for, respectively, 16-QAM and 32-QAM (and their
PS 64-QAM counterparts). 

\begin{figure}
	\begin{subfigure}[b]{0.48\textwidth}
		\centering 	\includegraphics[width=\textwidth]{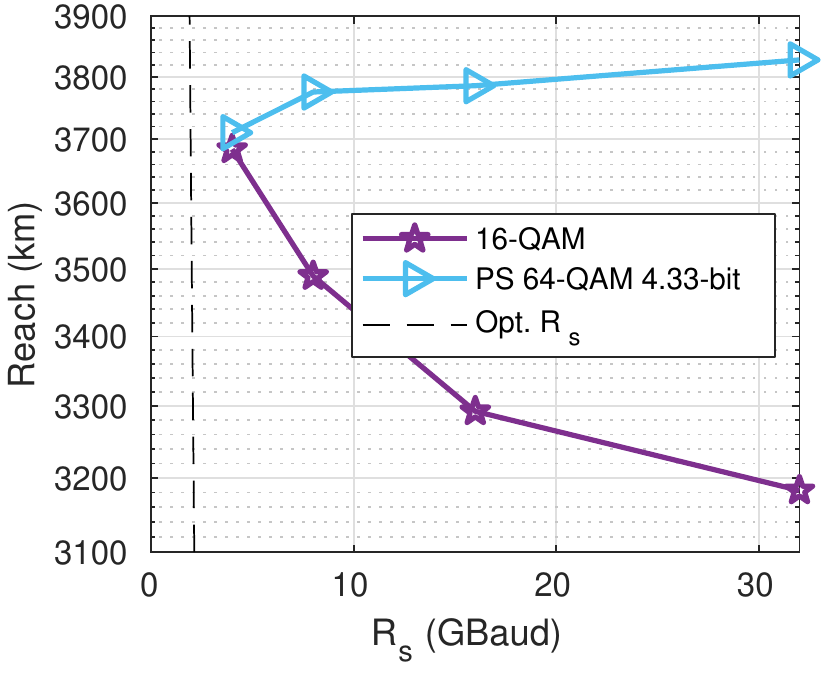}
		\caption{}
	\end{subfigure}
	\begin{subfigure}[b]{0.48\textwidth}
		\centering \includegraphics[width=\textwidth]{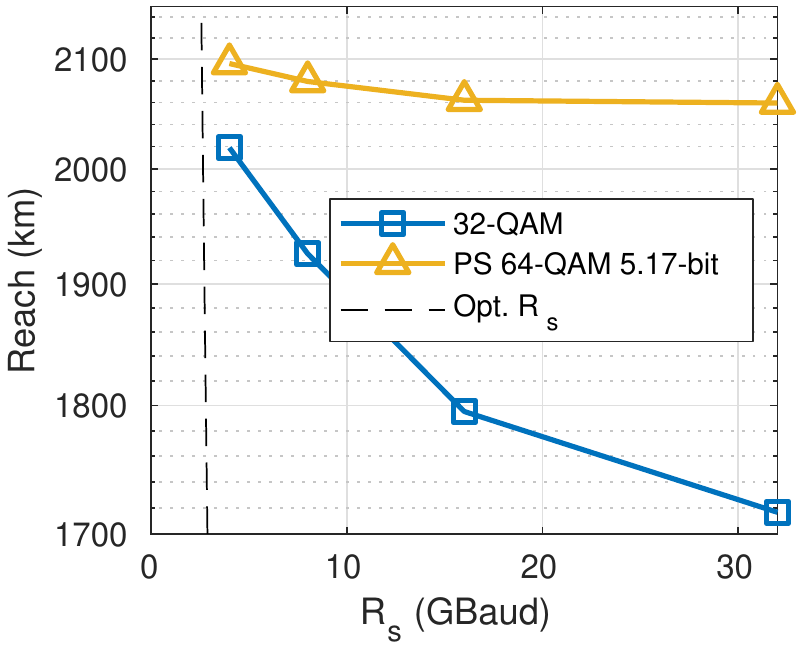}
		\caption{}
	\end{subfigure}
	\caption{Maximum reach of 16-QAM, 32-QAM and PS 64-QAM over $100$-km spans of SMF at different symbol rates with the same modulated optical bandwidth.}\label{fig:srosmf}
\end{figure}
\begin{equation}
R_\textup{opt} = \sqrt{\frac{2}{\pi |\beta_2| L_\textup{span}N_\textup{span} (2\delta f-1) }}
\end{equation}
Considering 16-QAM and 32-QAM, moving from 32 GBaud to 4 GBaud gives a maximum-reach increase of $+19.12\%$ and $+19.09\%$, respectively. This effect
is caused by the reduction of the overall generated NLI. However, 
PS 64-QAM constellations do not experience such gain. Instead, the maximum-reach gains of PS 64-QAM were $-2.62\%$ (reach decrease) and $+0.65\%$ for 4.33-bit and 5.17-bit, respectively.
\begin{figure}
	\begin{subfigure}[b]{0.48\textwidth}
		\centering 	\includegraphics[width=\textwidth]{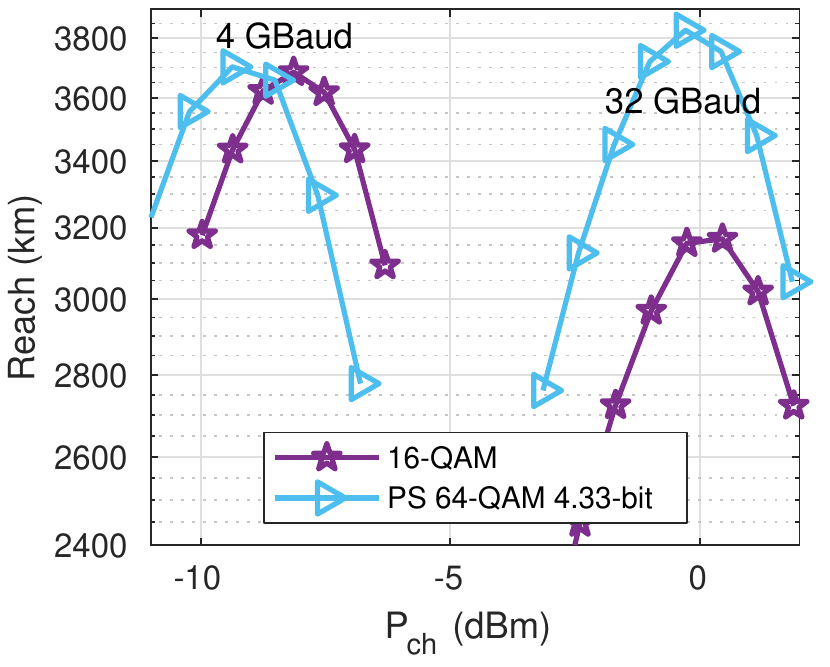}
		\caption{}
	\end{subfigure}
	\begin{subfigure}[b]{0.48\textwidth}
		\centering \includegraphics[width=\textwidth]{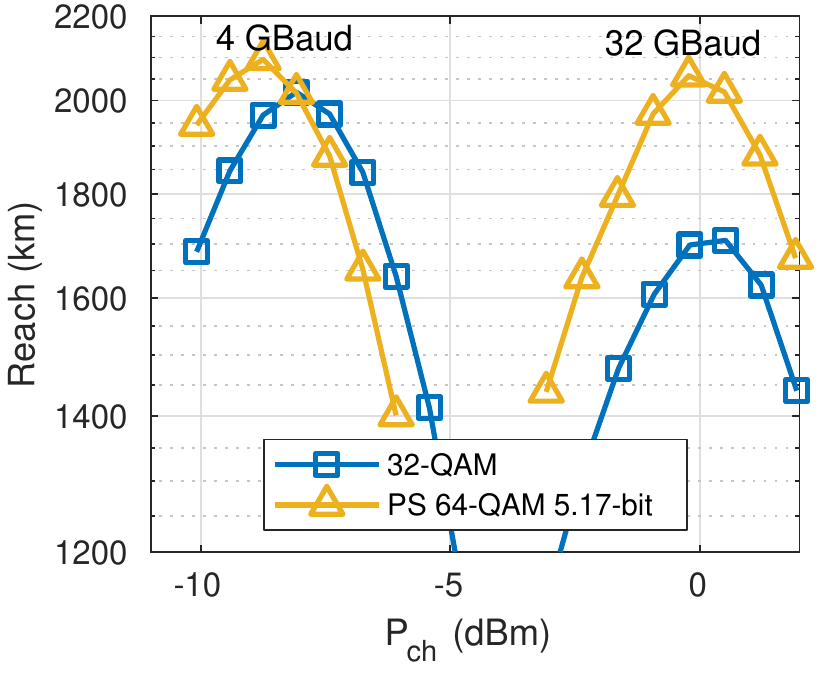}
		\caption{}
	\end{subfigure}
	\caption{Maximum reach of 16-QAM, 32-QAM and PS 64-QAM over $100$-km spans of SMF at different launch powers at 4 and 32 GBaud.}\label{fig:sropchsmf}
\end{figure}
More details on this performance loss can be seen in Fig. \ref{fig:sropchsmf}, where the maximum reach of systems at 4 and 32 GBaud are directly compared as a function
of the transmit power. At 32 GBaud, reach difference between PS 64-QAM and uniform constellations is constant for different powers. Indeed, the optimal
power is the same, which means that the effect of NLI (i.e. the $\eta$ coefficient) 
is the same for all constellations. However, at 4 GBaud, there is a significant reach loss when increasing the power, and
the optimal launch power is lower.

In conclusion, while a low symbol rate reduces the overall NLI, constellation-dependent NLPN is also stronger. While this effect is weak with standard 16-QAM and 32-QAM, with PS 64-QAM
this effect is particularly strong. This is due to the higher fourth-order moment, which generates more NLPN, and the CPE is not able to fully compensate for it.
Moreover, there may be an additional penalty, due to a suboptimal performance of the BPS phase recovery algorithm,
which has been shown to have a penalty when applied to PS constellations \cite{Mello:2018}.

\subsection{Low-dispersion fibers}\label{sec:7:lowdisp}
\begin{table}
	\centering
	\begin{tabular}{c c c c}
		\toprule
		Fiber & $\alpha$ &  $\beta_2$ & $\gamma$ \\
		& (dB/km) &  (ps$^2$/km) & (1/(W km)) \\
		\midrule
		PSCF (G.654) & 0.17 & -26.75 & 0.75 \\
		SMF (G.652) & 0.20 & -21.27 & 1.3 \\
		NZDSF (G.655) & 0.23 & 3.38 & 2 \\
		\bottomrule
	\end{tabular}
	\caption{Fiber parameters.}\label{tab:fibparams}
\end{table}
Modern deployed optical fiber is usually G.652.D (SMF) \cite{std:smf} for terrestrial links and G.654.D (PSCF) \cite{std:pscf} for submarine links \cite{Fischer:2018}. 
However, many networks have legacy fibers installed, 
such as G.652.A fiber \cite{std:smf,Fischer:2018} or G.655 (NZDSF) fiber \cite{std:nzdsf,Filer:2018}. While G.652.A is not particularly deleterious for coherent communications,
Non-Zero Dispersion-Shifted Fiber (NZDSF) generates a significantly higher amount of NLI, 
due to its low dispersion and small effective area. While many networks have few NZDSF, it is still widely
deployed in several countries such as Italy\footnote{E.g. Telecom Italia network scheme in \url{https://cordis.europa.eu/project/rcn/105820_en.html}, D1.1 document,
page 37.}, Japan, Brazil and many others.

To measure the impact of low-dispersion fibers on PS constellations, we used the same setup of Fig. \ref{fig:expsetuppspscf} (in Chapter \ref{ch:shaping}) and replaced the
fiber with $80$-km spools of NZDSF, whose parameters are summarized in Table \ref{tab:fibparams} \cite{Pilori:JLT2018}. The receiver used the same
BPS-ML phase recovery algorithm with $2\%$ pilot overhead.

\begin{table}
	\centering
	\begin{tabular}{c c c}
		\toprule
		Constellation & Entropy &  GMI threshold \\
		 & (bit/symbol) &  (bit/symbol) \\
		\midrule
16-QAM & $4$ & $3.6$ \\
32-QAM & $5$ & $4.5$ \\
PS 64-QAM & $13/3\approx4.33$ & $56/15\approx3.73$ \\
PS 64-QAM & $31.6\approx5.17$ & $137/30\approx 4.57$ \\		\bottomrule
	\end{tabular}
	\caption{GMI thresholds for an NGMI threshold of $0.9$.}\label{tab:ngmithresholds}
\end{table}
As an AIR (Sec. \ref{sec:6:pasdatarate}), instead of the MI, we used the normalized GMI, with a threshold of $0.9$ equal to all four constellations. This corresponds to the (un-normalized) GMI thresholds of Table \ref{tab:ngmithresholds}. 16-QAM and 32-QAM used a Gray mapping, while 32-QAM used the bit mapping defined in Appendix \ref{app:qamconst}.
Note that the GMI threshold of PS constellations is slightly higher than (uniform) QAM constellations at the same net data rate.
Nevertheless, the SNR gain in back-to-back is $0.65$ dB and $1.02$ dB with respect to, respectively, 16-QAM and 32-QAM.

\begin{figure}
	\begin{subfigure}[b]{0.48\textwidth}
		\centering 	\includegraphics[width=\textwidth]{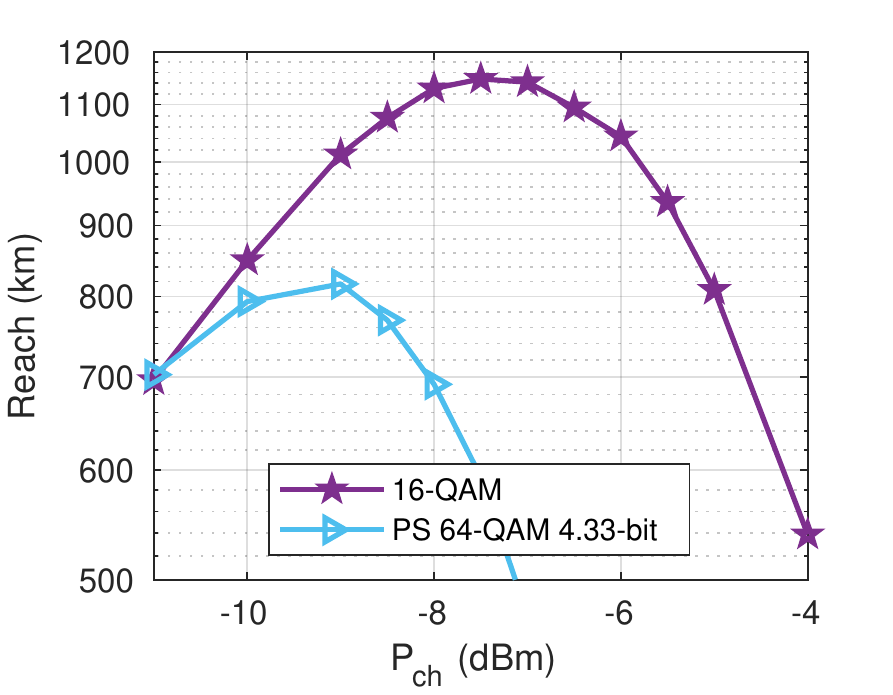}
		\caption{}
	\end{subfigure}
	\begin{subfigure}[b]{0.48\textwidth}
		\centering \includegraphics[width=\textwidth]{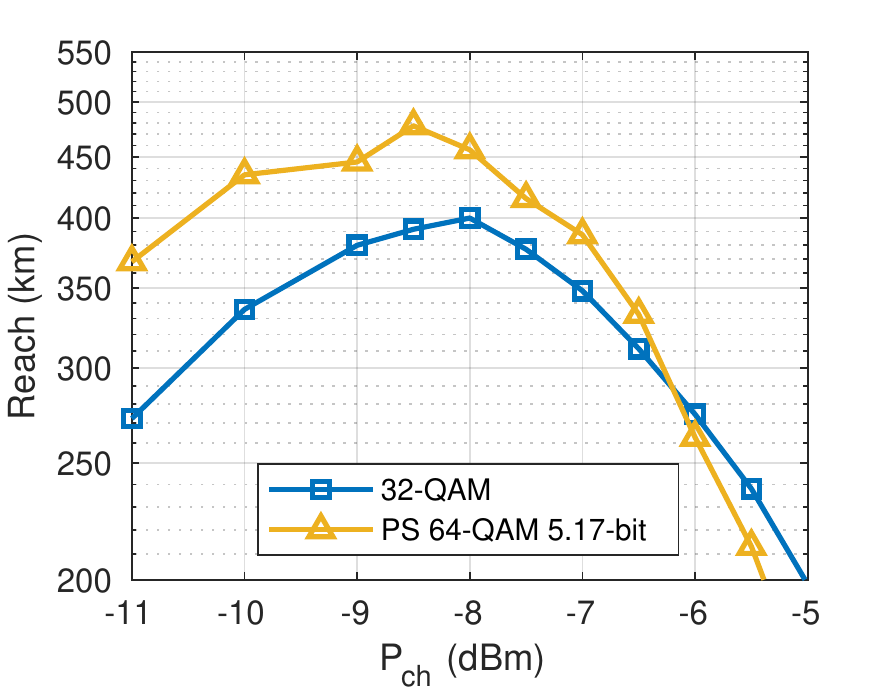}
		\caption{}
	\end{subfigure}
	\caption{Maximum reach of 16-QAM, 32-QAM and PS 64-QAM over $80$-km spans of NZDSF. Each figure compares constellation at the same net data rate
	with an NGMI threshold of 0.9.}\label{fig:nzdsfps}
\end{figure}
Results, expressed as maximum-reach as a function of the per-channel transmit power, are shown in Fig. \ref{fig:nzdsfps}. It is interesting to compare these results with the same
experiment on PSCF (Fig. \ref{fig:psqampscfpch}), even if the performance metric and thresholds were different. Results are especially different in Fig. \ref{fig:nzdsfps}a,
 where 16-QAM \emph{outperforms} PS 64-QAM, even if it requires an OSNR $0.65$ dB higher in back-to-back. 
 Optimal launch power, which is linked to the NLI strength coefficent $\eta$, is also reduced.
This effect is also present in Fig. \ref{fig:nzdsfps}b, but it is less pronounced. 

\begin{figure}
	\centering \includegraphics[width=0.6\textwidth]{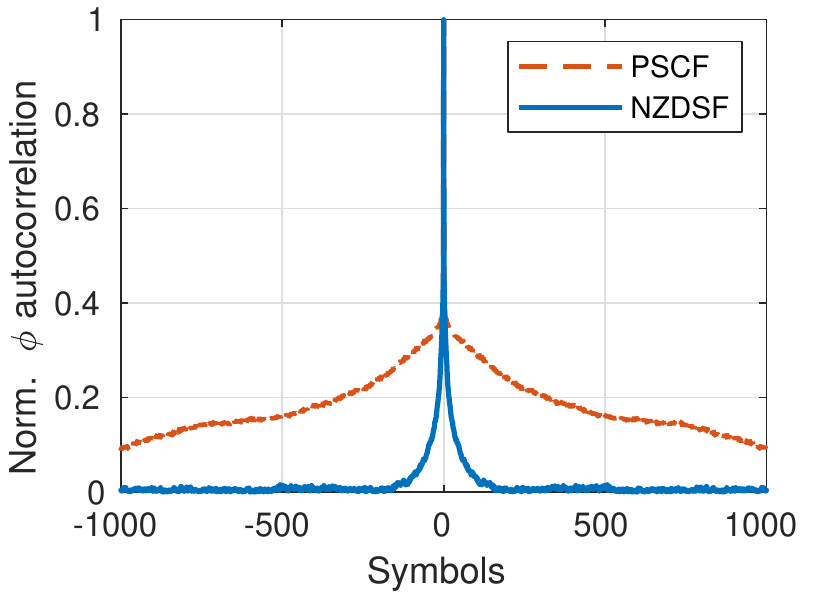}
	\caption{Autocorrelation of the NLPN in the PSCF and NZDSF experiments. Data obtained from noise-less SSFM simulations.}\label{fig:nzdsfmemory}
\end{figure}
This significant difference with respect to back-to-back (and PSCF propagation) is again explained by two-pulse collisions.
NLPN is both strong and fast, and BPS cannot fully compensate for it. 
To illustrate NLPN memory, Fig. \ref{fig:nzdsfmemory} shows its autocorrelation, 
extracted (using the method illustrated in \cite{Poggiolini:2017}) from noise-less SSFM simulations over PSCF and NZDSF. Simulation parameters were carefully chosen to match
the experiments. As shown in the Figure, memory of NLPN is significantly shorter in NZDSF. This means that the CPE is not able to fully estimate and
compensate for it, which explains the performance loss.

\section{Compensation techniques}
As shown in the previous section, there are system scenarios where NLPN is too strong and fast and cannot be compensated for by standard DSP techniques.
Therefore, specialized DSP techniques specifically targeted for NLPN may give a benefit in these scenarios.
 
Several works focused on finding optimal shaping techniques, different from the Maxwell-Boltzmann distribution, that have a smaller kurtosis, which generates less NLPN 
\cite{Dar:ISIT:2014,Geller:2016,Cho:2016,Sillekens:2018,Sillekens:ECOC:2018}.
However, the real-world gain of these techniques was found to be limited. Part of the reason lies in the fact that, if a distribution is different than Maxwell-Boltzmann, then it is not
optimal in an AWGN channel. If the distribution successfully reduces (or cancels) NLPN, the channel becomes AWGN, and the distribution is not optimal. Moreover, minimization of the kurtosis
does not allow distinguishing between short-memory and long-memory NLPN.

Other works, instead of preventing NLI generation, focused on compensation techniques at the receiver 
\cite{Yankov:2015,Eriksson:2016,Yankov:2017,Yankov:PJ:2017,Golani:2018}. If the compensation 
is successful, then the channel becomes approximately AWGN, allowing the use of AWGN-optimized constellations and techniques. 
In \cite{Yankov:2015}, the authors designed a phase-tracking  algorithm  based  on  the  assumption  that  phase  noise is  a  first-order  Wiener  process. A similar approach was taken in \cite{Golani:2018}, where the authors used a Kalman adaptive filter to track and compensate the time-varying ISI  effect of NLI. Both of these methods were found to be effective,  
and they do not need to be tailored for a specific links. However, they require a change in the standard DSP of a coherent receiver.
Moreover, the hardware implementation  complexity may be too high to be practically realized. 

\subsection{Modified decoding metric}
A soft decoder, as introduced in Sec. \ref{sec:5:bicm}, assumes some channel statistics $q(y|x)$ 
to perform decoding. While --- in the previous chapters --- the AWGN channel law was always assumed, a 
different law can be adopted. In Sec. \ref{sec:7:examples} some examples were shown where NLPN is significant; in those cases, the presence of NLPN makes channel statistics quite
different than AWGN. Therefore, a channel law $q(y|x)$ that takes into account phase noise is able to give better performances.

In general, NLPN is a random process with memory. Exploiting such memory at the receiver is, in principle, possible but very
difficult to implement \cite{Agrell:2018}.
However, we have seen that the phase recovery at the receiver can remove a large portion of NLPN. In particular, it removes the ``slow'' portion of NLPN \cite{Poggiolini:2017}.
This means that the remaining phase noise after CPE, called \emph{residual} NLPN, has a very short memory. 
We can furthermore assume that this NLPN is \emph{memoryless} (e.g. this can be achieved in practice with a symbol interleaver). With this key approximation,
a simple expression of $q(y|x)$ can be obtained. There are different methods to obtain such expression. 
A ``safe'' method would involve sending pilot symbols and estimating $p(y|x)$ using a histogram-based approach. However, this method is not very flexible, since the result
would be link-dependent, and it would require a large lookup-table at the receiver to store the histogram. An analytical expression would be therefore preferable.

\begin{figure}
\centering 	\includegraphics[width=0.6\textwidth]{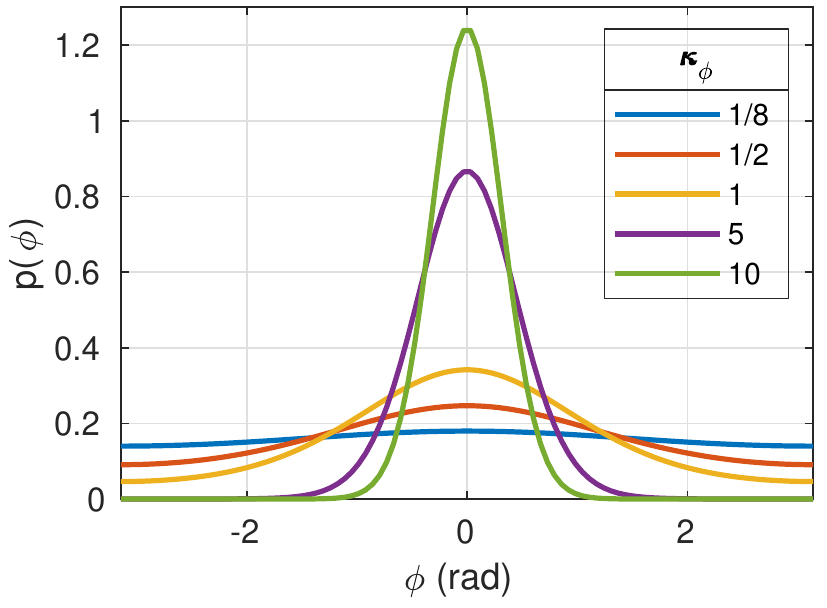}
\caption{Examples of Tikhonov distribution with different values of concentration.}\label{fig:tikhonovexample}
\end{figure}
In \cite{Kayhan:2014}, the authors developed an analytical expression for such case. The one-sample-per-symbol channel (Fig. \ref{fig:infothchannel}) is modeled as
\begin{equation}
y = xe^{j\phi} + n
\end{equation}
, where $n$ is an AWGN process with variance $\sigma^2_n$ and $\phi$ is a white (i.e. memoryless) random process that represents residual phase noise. 
Since $\phi$ is a phase, it is wrapped between $-\pi$ and $\pi$. Probability distribution functions with infinite support (like the Gaussian distribution) are therefore
 not suitable. A common distribution for phase noise is the Tikhonov (or Von Mises) distribution \cite{Foschini:1973}:
\begin{equation}
p(\phi) = \frac{e^{\kappa_\phi \cos\phi}}{2\pi \cdot I_0(\kappa_\phi)} \qquad \phi\in(-\pi,\pi]
\end{equation}
In this expression, $I_0(.)$ is the modified Bessel function of the first kind, and $\kappa_\phi$ is a parameter called \emph{concentration}. This distribution is an approximation
of the wrapped (between $-\pi$ and $\pi$) Gaussian distribution. For large values of $\kappa_\phi$ (i.e. small phase noise), this distribution resembles a Gaussian distribution
with zero mean and variance $1/\kappa_\phi$. Some examples of Tikhonov distributions for different values of concentration are shown in Fig. \ref{fig:tikhonovexample}.
With this distribution, the channel conditional probability can be expressed as \cite{Kayhan:2014}
\begin{equation}
p(y|x) \approx \sqrt{\frac{\kappa_\phi}{8\pi^3}}\frac{e^{-\kappa_\phi}}{\sigma^2_n}
\exp{\left(-\frac{|y|^2+|x|^2}{2\sigma^2_n}+\left\lvert \frac{y x^*}{\sigma^2_n}+\kappa_\phi \right\rvert\right)}
\label{eq:pnpdf}
\end{equation}
This result is only approximate, since the Bessel function has been approximated as $I_0(x)\approx e^x/\sqrt{2\pi x}$ (valid for large values of $x$).

This expression, provided that a correct estimation of $\kappa_\phi$ and $\sigma^2_n$ is performed, can be then used by the receiver to perform decoding. For instance, 
a BICM receiver can use it in the LLRs expression \eqref{eq:llrdef}.

\subsubsection{Example: hard decision metric}
\begin{figure}
	\centering
	\begin{subfigure}[b]{0.31\textwidth}
		\centering 	\includegraphics[width=\textwidth]{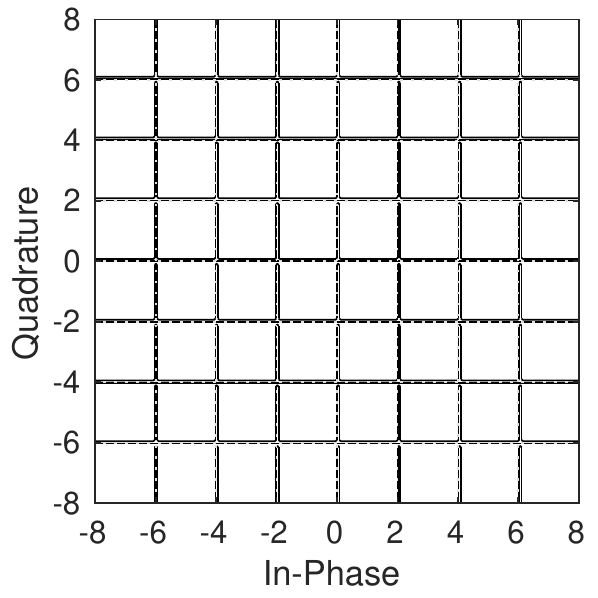}
		\caption{}
	\end{subfigure}
	\begin{subfigure}[b]{0.31\textwidth}
	\centering 	\includegraphics[width=\textwidth]{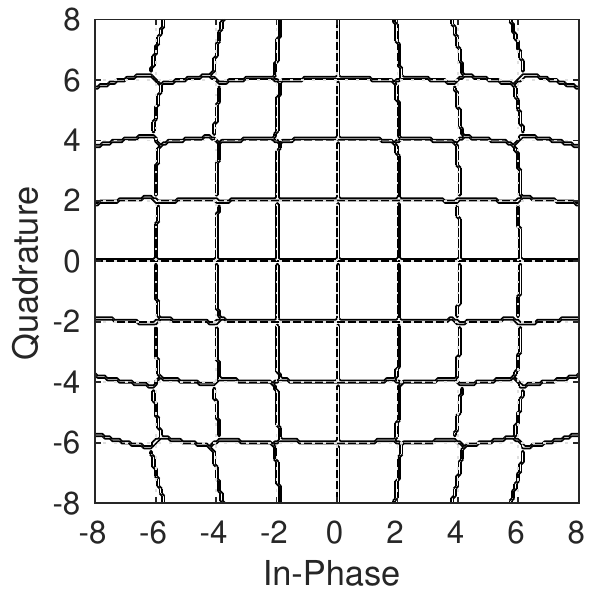}
	\caption{}
	\end{subfigure}
\begin{subfigure}[b]{0.31\textwidth}
	\centering 	\includegraphics[width=\textwidth]{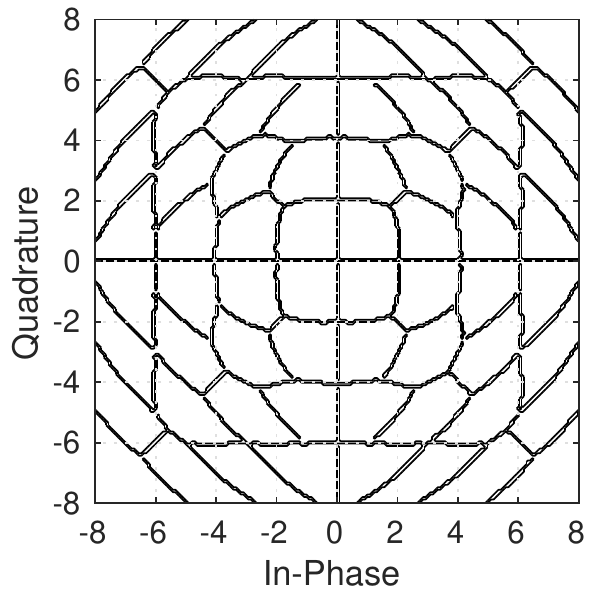}
	\caption{}
\end{subfigure}
	\caption{Hard decision regions for 64-QAM without phase noise (a), $1/\sqrt{\kappa_\phi} = 2^\circ$ (b) and $1/\sqrt{\kappa_\phi}=8^\circ$ (c). The SNR of
additive noise is $16$ dB.}\label{fig:pnregions}
\end{figure}
In this Part of the thesis, a soft-decision receiver was always assumed. However, the decision strategy for a soft 
decoder\footnote{See \cite{KlausOestreich:2018} for a nice demo of soft decoding.} is difficult to be ``graphically visualized'' . 
Hard decision metrics, instead, are simpler to visualize using the so-called decision (or Voronoi) regions \cite{proakis2007digital}.
Fig. \ref{fig:pnregions} shows the hard decision regions for a 64-QAM constellation with different concentrations of a Tikhonov-distributed NLPN. The AWGN variance was fixed to give
an SNR (neglecting NLPN) of $16$ dB. The regions were calculated with the maximum-likelihood criterion using \eqref{eq:pnpdf}. While the decision regions over a pure-AWGN
channel \ref{fig:pnregions}a are square, with phase-noise (b and c) the decision regions change shape to take into account its presence. The stronger the phase noise
(compared to AWGN), the more ``distorted'' are the decision regions. 

\subsubsection{Application to NZDSF results}
\begin{figure}
	\begin{subfigure}[b]{0.48\textwidth}
		\centering 	\includegraphics[width=\textwidth]{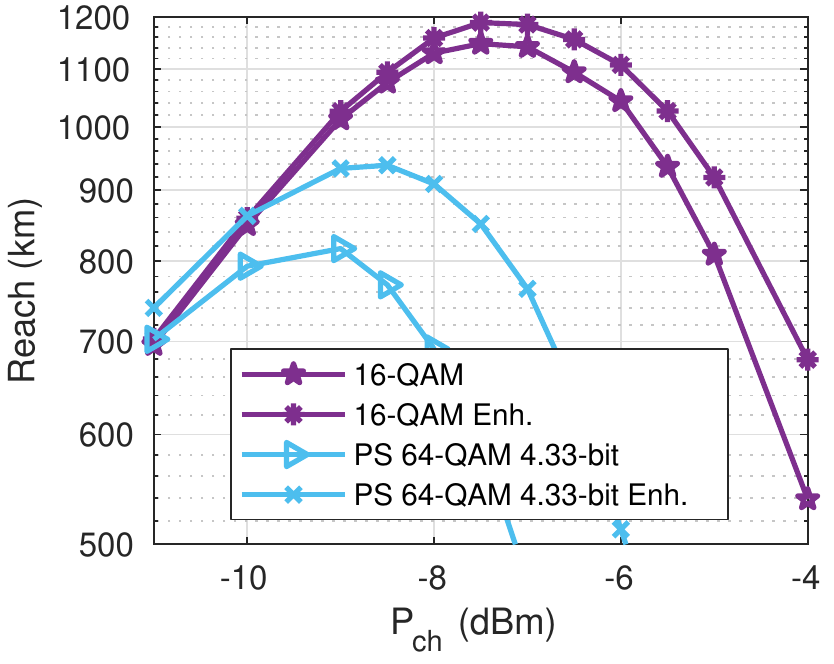}
		\caption{}
	\end{subfigure}
	\begin{subfigure}[b]{0.48\textwidth}
		\centering \includegraphics[width=\textwidth]{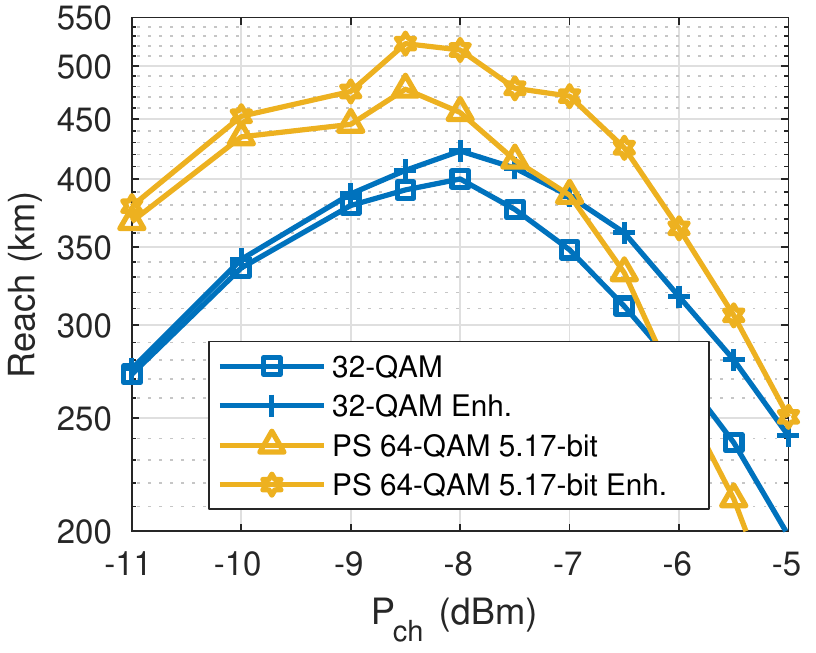}
		\caption{}
	\end{subfigure}
	\caption{Maximum reach of 16-QAM, 32-QAM and PS 64-QAM over $80$-km spans of NZDSF. Results were obtained both with a standard AWGN metric and the NLPN-aware (enhanced)
	metric.}\label{fig:nzdsfpspnaware}
\end{figure}
To test the effectiveness of the modified soft decoding metric \eqref{eq:pnpdf}, we re-processed the data we captured over low-dispersion fiber in
the experiment described in Sec. \ref{sec:7:lowdisp}. Results
are shown in Fig. \ref{fig:nzdsfpspnaware}, where they are compared to the results already presented of Fig. \ref{fig:nzdsfps}, where the receiver assumed (as for all
results presented so far) an AWGN channel. All constellations show a gain that increases for the increase of power, which means that the algorithm is effectively
compensating for NLPN. Gain is stronger for constellations with stronger NLPN, i.e. PS 64-QAM. In details, the maximum-reach gain was $+14.1\%$ and $+9.9\%$ for PS 64-QAM
with entropies $\sim4.33$ and $\sim5.17$ bit/symb, respectively. Uniform constellations exhibited more limited gains, $+4\%$ and $+5.6\%$ for 16-QAM and 32-QAM respectively.

In conclusion, the modified metric \eqref{eq:pnpdf} is able to partially compensate for NLPN, especially where NLPN is strong (e.g. PS 64-QAM over NZDSF).
Nevertheless, compared to other NLPN-compensation algorithms, this technique only slightly modifies the decoding metric already employed in soft decoders. Therefore,
receiver complexity is not expected to significantly increase.

\subsection{Geometric shaping}
Another possibility to compensate for NLPN is to adopt constellation shaping techniques optimized to a channel that is more realistic than the AWGN channel assumed
in Chapter \ref{ch:shaping}. This technique has been succesfully applied in the past \cite{Dar:ISIT:2014,Geller:2016,Cho:2016,Sillekens:2018,Sillekens:ECOC:2018}. However, as
discussed before, these methods mostly focused on generating less NLI. 
For this work, following \cite{Kayhan:2014}, we optimized the constellation using the NLPN-aware
metric of \eqref{eq:pnpdf}. In this case, assuming that the variance of NLPN does not significantly change by ``moving'' constellation points, 
this operation improves the performance of the NLPN-aware decoding previously presented.

Therefore, we used the simulated annealing algorithm \cite{Kirkpatrick:1983} to optimize constellation points, given
a specific SNR and NLPN concentration.

\subsubsection{Optimizing 32-QAM}
\begin{figure}
	\centering
	\begin{subfigure}[b]{0.45\textwidth}
		\centering 	\includegraphics[width=0.8\textwidth]{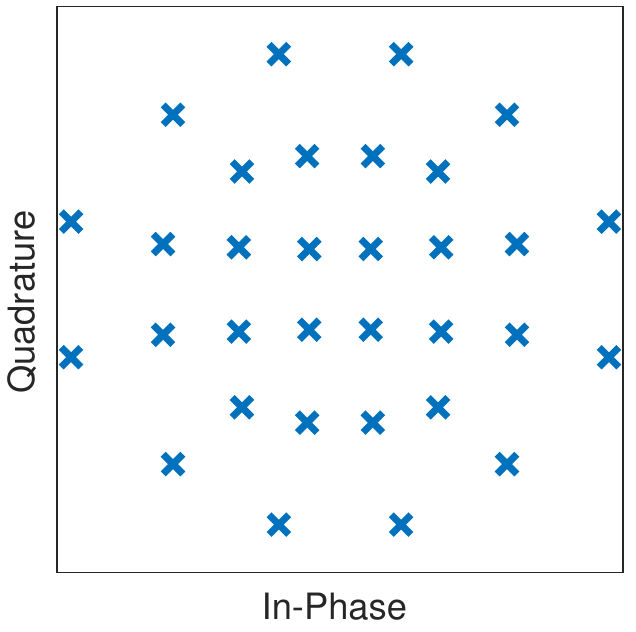}
		\caption{}
	\end{subfigure}
	\begin{subfigure}[b]{0.45\textwidth}
		\centering \includegraphics[width=\textwidth]{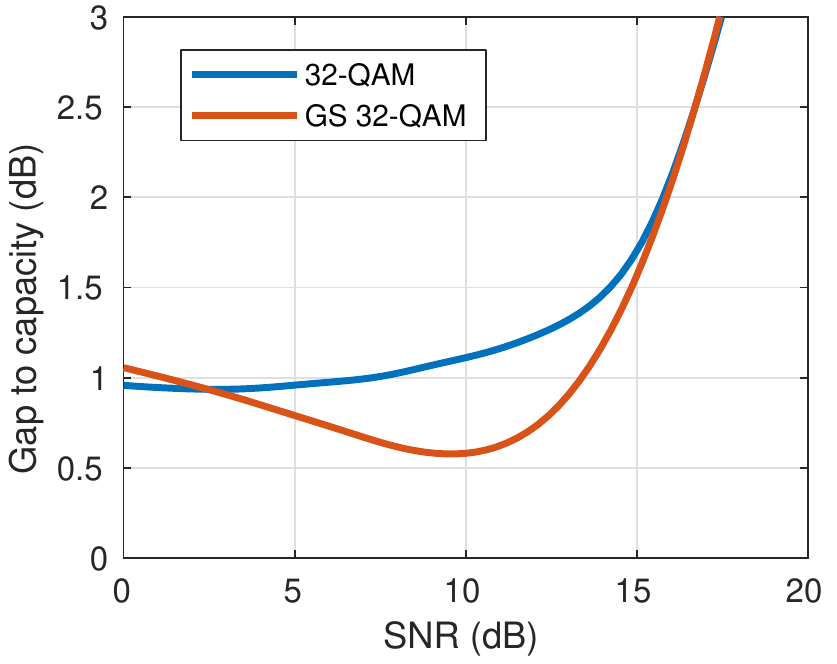}
		\caption{}
	\end{subfigure}
	\caption{Optimization of 32-QAM: (a) optimized constellation diagram, (b) performance over a BICM-AWGN channel.}\label{fig:gs32qam}
\end{figure}
As an example, we run the optimization algorithm over a 32-QAM constellation, obtaining the results shown in Fig. \ref{fig:gs32qam}. 
The constellation has been optimized with the simulated annealing algorithm over a channel with an SNR of $13$ dB (considering only additive noise) and a phase noise with concentration
$1/\sqrt{\kappa_\phi}=2.27^\circ$. A BICM receiver is assumed, i.e. the adopted performance metric was the GMI, and the LLRs were calculated using \eqref{eq:pnpdf}.
The resulting constellation is shown in Fig. \ref{fig:gs32qam}a, and its performance over an AWGN-BICM channel is shown in Fig. \ref{fig:gs32qam}b. 
This result is similar to APSK, shown in Fig. \ref{fig:64apskgmi}. The largest SNR gain is $\sim0.5$, which is obtained
for an SNR of approximately $10$ dB. Unfortunately, this SNR is too low for the considered NGMI thresholds. Nevertheless, as discussed in Chapter \ref{ch:shaping}, even if the gain is small, GS does not require any special scheme to combine FEC and shaping (like PAS), which reduces its implementation complexity.

A Table with the optimized constellation, along with its bit mapping, is shown in Appendix \ref{app:qamconst}.

\subsubsection{Experimental demonstration}
\begin{figure}
		\centering \includegraphics[width=0.7\textwidth]{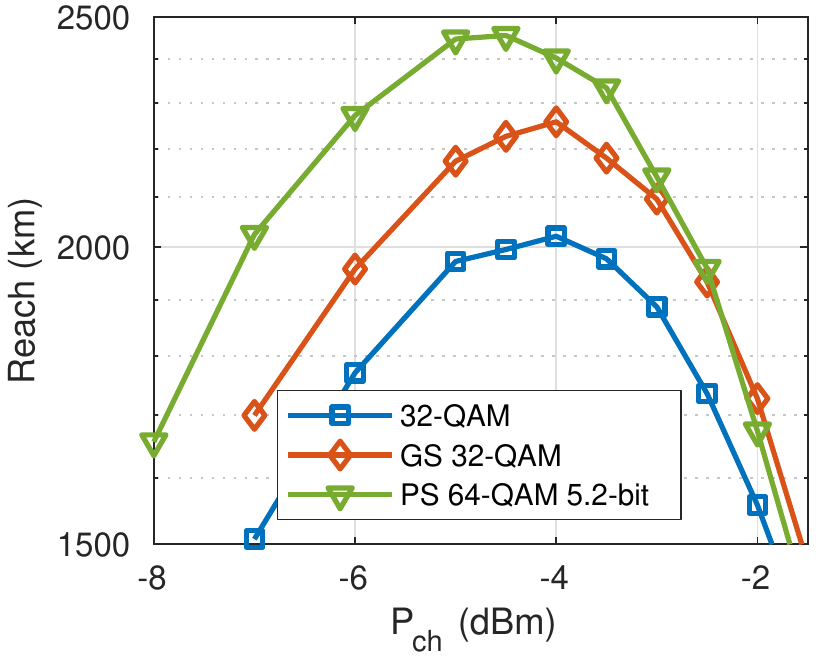}
	\caption{Experimental test of 32-QAM, GS 32-QAM and PS 64-QAM with entropy 5.2 bit/symb. Maximum reach at 16 GBaud over $80$-km spans of SMF
	with NGMI threshold of $0.86$.}\label{fig:gs32qamexperiment}
\end{figure}
The effectiveness of the NLPN-optimized GS 32-QAM was experimentally tested. The experimental setup is identical to Fig. \ref{fig:expsetuppspscf} in Chapter \ref{ch:shaping}, where
fiber was replaced with $80$-km spans of SMF. The constellations under test were three: 32-QAM (unshaped), GS 32-QAM and PS 64-QAM with entropy $5.2$ bit/symb. This value
was chosen to obtain the same net data rate of 32-QAM (both shaped and unshaped) with a $4/5$-rate FEC, corresponding to a $25\%$ overhead. For this experiment, similarly
to Sec. \ref{sec:7:lowdisp}, we used the NGMI performance metric, with an NGMI threshold of $0.86$. This corresponds to a GMI threshold of $4.3$ bit/symb for 32-QAM (shaped
and unshaped) and $4.36$ for PS 64-QAM 5.2-bit. The theoretical SNR gains at the NGMI threshold with respect to 32-QAM are $0.37$ dB and $1.00$ dB for GS 32-QAM and PS 64-QAM, respectively.

Results are shown in Fig. \ref{fig:gs32qamexperiment}. As expected, since signals are propagating over SMF, strong NLPN is not expected. Indeed, PS 64-QAM has the highest
reach, followed by GS 32-QAM. However, looking in detail at the Figure, there is a slight decrease of optimal power with PS 64-QAM, which is not experienced by GS 32-QAM.
This slightly reduces the reach of PS 64-QAM.

\subsection{Joint subcarrier phase recovery}
\begin{figure}
\centering \includegraphics[width=0.6\textwidth]{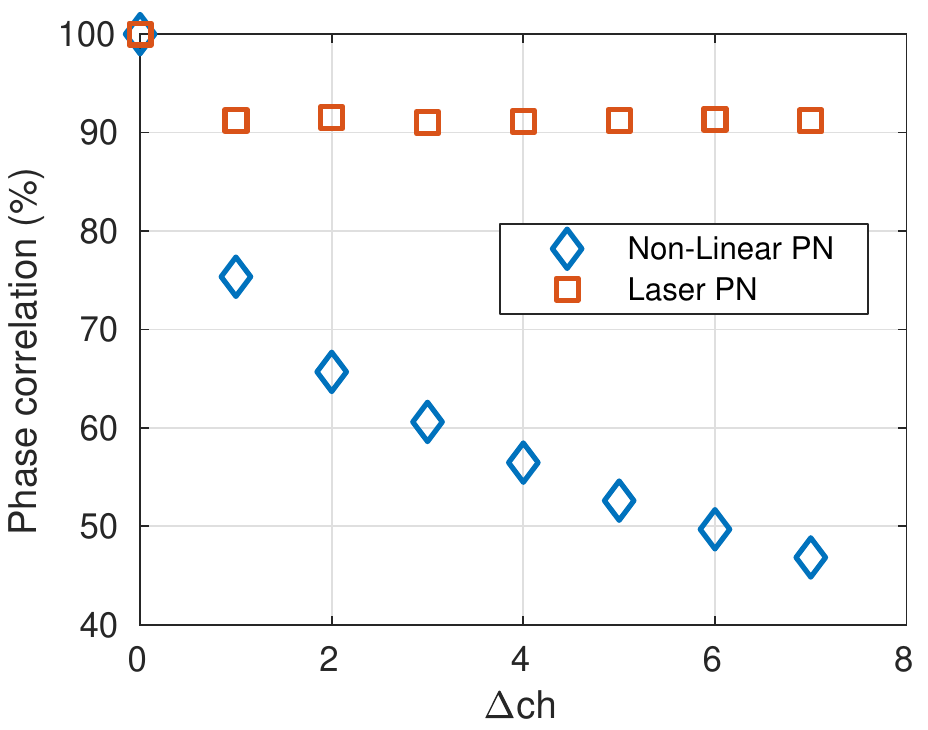}
\caption{Correlation of NLPN among subcarriers at $4$ GBaud with PS 64-QAM 4.33-bit, as a function of the relative frequency separation between subcarriers
	 $\Delta\mathrm{ch}$.}\label{fig:nlpncorrelation}
\end{figure}
When using SCM, as shown in Sec. \ref{sec:7:lowrs}, NLPN was found to be both stronger and faster, reducing the maximum-reach with PS constellations. However, in SCM systems,
several subcarriers are processed by the same DSP. This allows using algorithms that reduce NLI by jointly processing several subcarriers. 
However, to jointly process several subcarriers, the NLPN process that is affecting them must be approximately the same. Otherwise, the effectiveness would be limited. 

Therefore, we extracted the NLPN process from $8$ subcarriers at the center of the WDM comb from the 4-GBaud simulation (without ASE and laser phase noise) of Sec. \ref{sec:7:lowrs}. The constellation under test was PS 64-QAM 4.33-bit, and the distance-launch power was set to its maximum reach. Then, we calculated the correlation (using Pearson's correlation coefficent) between those processes. To check the effectiveness of the method, we executed the same process over a linear fiber with laser phase noise.
Results of these two tests are shown in Fig. \ref{fig:nlpncorrelation}. Correlation is measured as a function of the relative frequency spacing between subcarriers.
While with laser phase noise correlation is always very high, with NLPN correlation decreases with
the frequency distance between subcarriers. Therefore, a joint algorithm may be designed, but it will not be trivial.

In conclusion, future work should design specialized algorithms that are able to exploit such correlation to increase the effectiveness of SCM systems. In that case, the gain of symbol-rate
optimization can be combined with the gain of PS, allowing significant reach enhancements \cite{Guiomar:2018}.

\section{Conclusion}
This Chapter described the phase-noise process that is generated by fiber Kerr non-linearities, with a specific focus on systems using constellation shaping. 
The most important parameter of NLPN is its memory, i.e. how fast it changes over time. Large-memory NLPN is already compensated by receiver phase recovery algorithms, which is always
present in digital receivers to compensate for laser phase noise. On the other end, short-memory NLPN directly affects receiver performance, reducing the reach if its variance
is too high.

The variance of NLPN depend on the fourth-order moment of the constellation, which is stronger if constellation shaping is applied. While 
in ``standard'' conditions (Chapter \ref{ch:shaping}) this is not an issue, over low-dispersion fibers and at low symbol rates NLPN gives a significant maximum-reach reduction to 
systems using constellation shaping. This effect is theoretically explained by the time-domain NLI theory (pulse-collision theory).

However, there are low-complexity techniques that are able to partially alleviate this issue. In detail, we investigated a modified soft-decoding metric, NLPN-aware geometric shaping
and joint subcarrier processing for SCM. All of these techniques were found to give some improvements. However, more sophisticated techniques may give even higher
improvements, albeit with a higher complexity.

Future work should focus on the improvement of these algorithms. Most importantly, a joint phase recovery for SCM systems will allow combining symbol-rate
optimization (SRO) gains with the advantages of constellation shaping.

\chapter{Conclusion}

    \graphicspath{{Chapter8/}}

In this thesis, we applied DSP techniques to different optical communications applications.

In Part \ref{part:dd}, we focused on direct-detection systems, which are adopted for short-reach links. In Chapter \ref{ch:bidir} we focused on $<2$ km interconnects
between servers inside the same data-center. For this application, we proposed a novel bi-directional architecture. Thanks to receiver adaptive equalization
and a slight frequency difference between transceivers, we were able to significantly reduce penalties caused by back-reflections. This remarkable
result has been explained using first a theoretical model, simulations and finally an experimental setup. 
In Chapter \ref{ch:ssb} we instead focused on $\sim80$ km links, used to connect different data-centers in a region. For this application, we presented SSB modulation
as a hybrid solution between traditional IM/DD systems and coherent. Using a theoretical model and numerical simulations, we showed that SSB is more resistant to chromatic
dispersion than IM/DD systems, allowing the removal of optical dispersion compensation over these links. This will ease the future transition to fully coherent systems.

Part \ref{part:coh} focused instead on coherent long-distance links. For this part, we mainly focused on constellation shaping techniques, which have been recently
applied in optical communications. However, the coherent long-haul channel is non-linear,
which can potentially reduce the advantages of constellation shaping. In Chapter \ref{ch:shaping}, we assumed that fiber non-linear effects act only as an AWGN source. In that
case, we found that constellation shaping can give full data-rate flexibility with a maximum reach gain of $\sim25\%$. However, in Chapter \ref{ch:phnoise} we showed
that in some situations non-linear effects generate a strong phase noise component. This phase noise is particularly harmful to constellation shaping; in fact, we showed that in some
specific scenarios -- small symbol rates and low dispersion fibers --  all the advantage of shaping is erased. Consequently, we proposed and presented 
some compensation techniques which can potentially mitigate this effect.

\appendix
\chapter{The Fast Fiber Simulator Software}\label{ch:ffss}
    \graphicspath{{Appendix0/}}
\section{Introduction}
As shown in Chapter \ref{ch:coherent}, accurate simulation of a fiber-optic communication system requires
numerical integration of the NLSE \cite{Agrawal:nonlinear}. However, full integration of the NLSE
is quite complicated, since it needs to take into account polarization-related effects such as
birefringence and PMD. Thus, most of fiber propagation analyses rely on the so-called Manakov equation \cite{Marcuse:1997}, which averages
out polarization-related effects, obtaining two scalar equations. While the Makakov equation is formally correct only for small optical
bandwidths, it has been shown with experiments \cite{Pastorelli:2012,Saavedra:2017} and simulations \cite{Cantono:OFC2018} that it can be used
for very wide bandwidth coherent optical links. Therefore, for our SSFM implementation, called the Fast Fiber Simulator Software (FFSS)  \cite{Pilori:FFSS2017}, we implemented the Manakov
equation.

\begin{figure}
	\centering \includegraphics[width=0.8\textwidth]{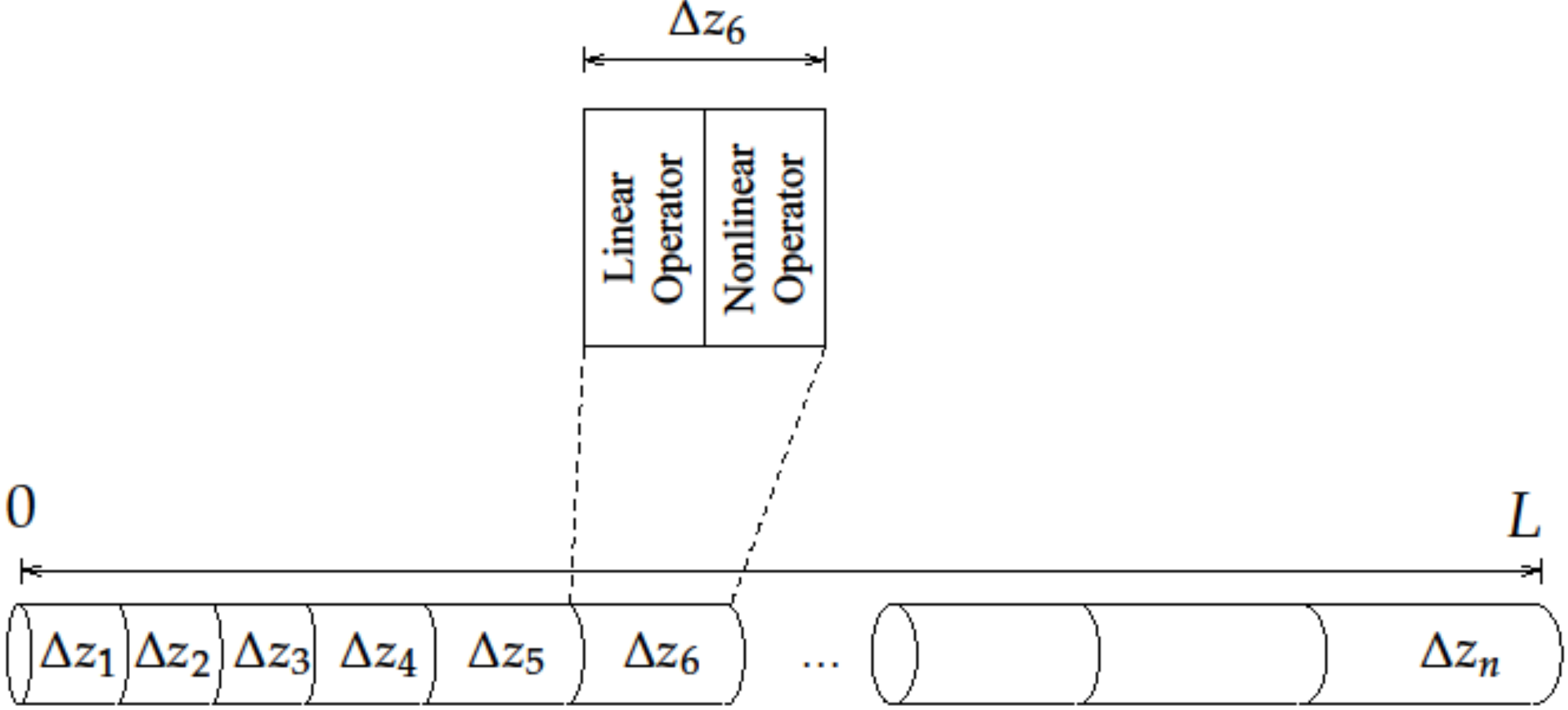}
	\caption{Basic block diagram of the SSFM scheme.}\label{fig:ssfmscheme}
\end{figure}

Assuming propagation of an electric field $\mathbf{E}(z,t)$ with Fourier transform $\tilde{\mathbf{E}}(z,\Omega)$, where $\Omega = \omega-\omega_0$ and
$\omega_0=2\pi f_0$ is the central pulsation of the signal, the Manakov equation can be expressed as
\begin{equation}
\frac{\partial \tilde{\mathbf{E}}(z,\Omega)}{\partial z} =
- \left[ \alpha(z,\Omega)+j\beta_2(\Omega)  \right] \tilde{\mathbf{E}}(z,\Omega)
-j \frac{8}{9}\gamma \mathcal{F}\left\{ \mathbf{E}(z,t) |\mathbf{E}(z,t)|^2 \right\}
\end{equation}
In this equation, $\beta_2(\Omega)$ represents (frequency-dependent) dispersion, $\alpha(z,\Omega)$ frequency- and space-dependent attenuation. Looking
in details at the equation, on its right-hand-side there are two main terms: a linear term (attenuation and dispersion), applied in frequency domain, and a non-linear
term (Kerr effect), applied in time domain. The SSFM method exploits this fact by dividing the fiber inside \emph{small} pieces of length $\Delta z$. For each piece, these two operators
(linear and non-linear) are subsequently applied. A pictorial representation of this procedure is shown in Fig. \ref{fig:ssfmscheme}.

The choice of the length of each step is crucial for simulation accuracy. The most widely adopted method is based on the ``non-linear phase shift'' \cite{Carena:1997}, over which step
length is adaptive depending on the total optical power. Therefore, step length increases during propagation due to attenuation. However, recent works \cite{Musetti:2018} found that
there are more accurate methods to calculate step length for a given simulation accuracy. Nevertheless, for the time being, the FFSS simulator adopted
the non-linear phase shift criterion for step length calculation.

\section{Benchmarks and results}
SSFM simulations are, in general, time-consuming. This is due to the large optical bandwidth that needs to be simulated, which requires many samples per symbol and
small steps length. For this reason, General-Purpose GPU (GPGPU) computations can be highly beneficial. Therefore, we used the MATLAB\textregistered Parallel Computing
Toolbox GPU capabilites to improve the speed of FFSS. Afterwards, we ran some benchmarks by simulating propagation of a random signal over a $100$-km span of SMF.

The benchmarks were tested over two different computers. The first computer is a stardard desktop computer, with a quad-core Intel\textregistered Core i7-6700 processor
and a consumer-grade NVIDIA\textregistered GeForce GTX 1070 graphics card. The second computer is a sever, with two 14-core Intel\textregistered Xeon E5-2690 v4 processors
and a sever-grade NVIDIA\textregistered Tesla P100 GPU. This GPU costs approximately 10 times with respect to the desktop-class GPU.
Simulations were run both with single-precision (i.e. 32-bit) and double-precision (i.e. 64-bit) floating point numbers.
Since consumer-grade GPUs are not designed for double-precision computing, we expect a larger penalty compared to the Tesla GPU.

\subsection{Simulation speed} 
\begin{figure}
	\begin{subfigure}[b]{0.48\textwidth}
		\centering 	\includegraphics[width=\textwidth]{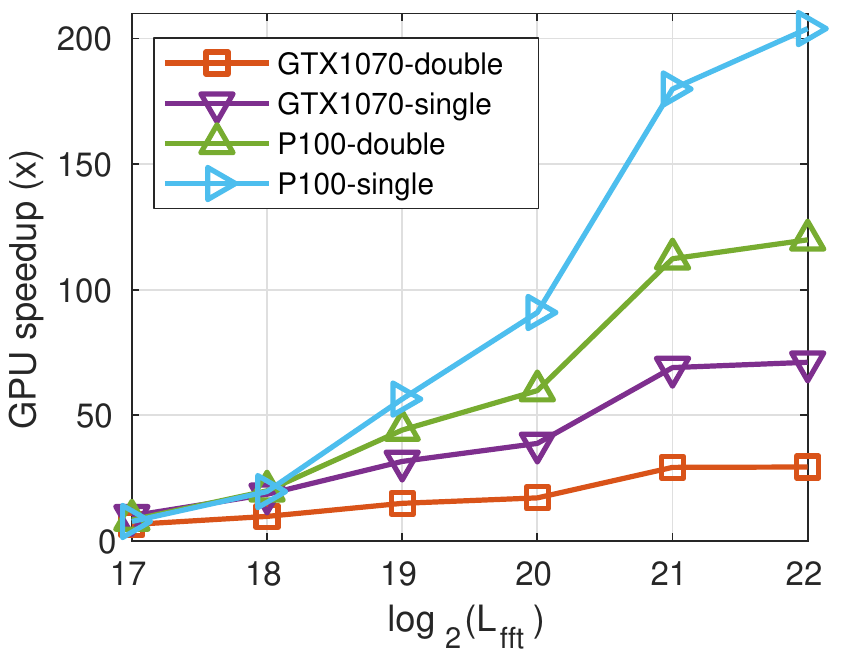}
		\caption{}
	\end{subfigure}
	\begin{subfigure}[b]{0.48\textwidth}
		\centering \includegraphics[width=\textwidth]{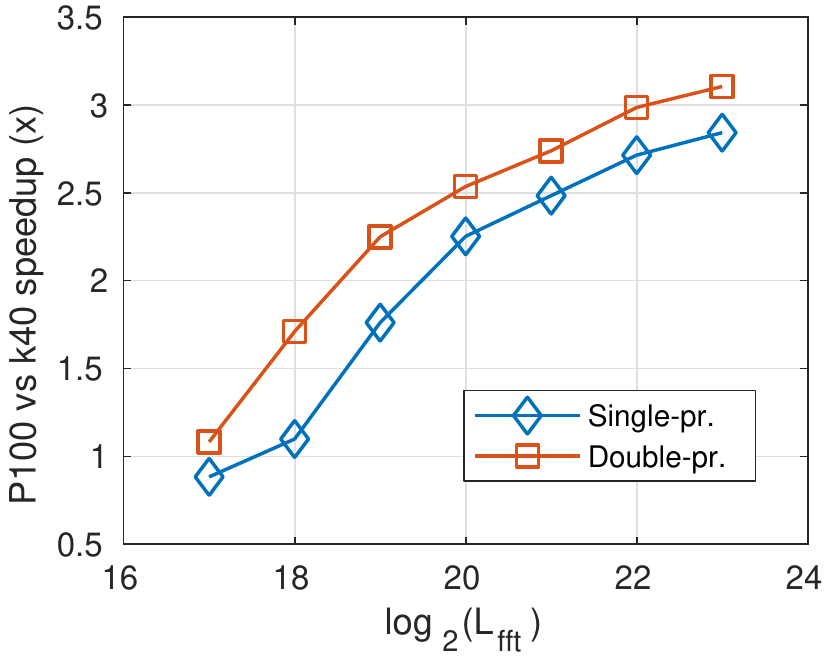}
		\caption{}
	\end{subfigure}
	\caption{GPU speedup compared to server CPU (a), and relative speedup between desktop-class and server-class GPU in single and double precision (b).}\label{fig:ffssbench}
\end{figure}
Fig. \ref{fig:ffssbench} shows some benchmarks, in terms of relative speed increase of the GPU compared to CPU-only simulation. In all the cases, the speedup is significant, and it increases
with the increase of the block length. At the largest block size, the Tesla P100 is $\sim200$ times faster (in single precision) than CPU-only simulation. This enormous speed increase
allows simulation of very large optical bandwidths that are not possible to simulate with CPUs. Comparing the desktop-class GPU and the server-class GPU ($\sim10\times$ price difference),
at low block sizes performance is similar. However, at large block sizes the difference becomes larger (approximately $3\times$). Given the price difference, desktop-class GPUs
are interesting for small-bandwidth simulations.

\subsection{Examples} 
\begin{figure}
\centering 	\includegraphics[width=0.7\textwidth]{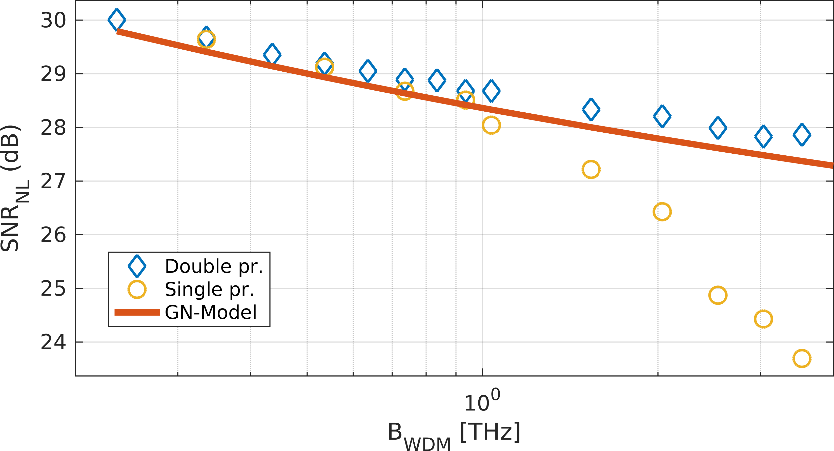}
	\caption{Evaluation of the non-linear SNR in split-step simulations with the increase of modulated WDM bandwidth.}\label{fig:ffssresults}
\end{figure}
We evaluated the accuracy of FFSS in single precision. We simulated a standard WDM system (Fig. \ref{fig:longhaulmodel}), where the channel under test
was PM-QPSK and the interfering WDM channels were Gaussian. Symbol rate was 32 GBaud, and channel spacing 50 GHz.
They were propagated over 30 100-km spans of SMF, with a transmit power per channel of -6 dBm. We then increased the number of transmitted channels, up to
a total modulated optical bandwidth of 4 THz, which correspond to an almost-full C-band.

Results in single and double precision are shown in Fig. \ref{fig:ffssresults}, where they are compared with the GN-model. Since, except for the central channel, all channels
were Gaussian, we expect an excellent agreement with the GN-model, which assumes that all channels are Gaussian. In fact, results in double precision closely follow the GN-model
for all optical bandwidths. Results in single precision, instead, are accurate up to $\sim1$ THz. Afterwards, NLI is overestimated. Therefore, we can conclude
that FFSS in double-precision is very accurate for large bandwidths, while single-precision is accurate only for small optical bandwidths.

We used FFSS also for C+L systems \cite{Cantono:JOCN:2019,Ferrari:OFC2019}, finding a good agreement with the NLI models, such as the GGN model \cite{Cantono:2018}.

\chapter{Derivation of theoretical formulas}
\section{PAM-$M$ detection in the presence of coherent interference}\label{app:paminterf}
Let us consider a transmitted PAM-$M$ signal. This signal has the following transmit electric field:
\begin{equation}
E_\textup{tx}(t) = \sqrt{\sum_{i=-\infty}^{+\infty} a_i \rect\left( \frac{t-iT}{T} \right)} = \sqrt{x_\textup{sig}(t)}
\end{equation}
$a_k$ are PAM-$M$ symbols, $T$ is the symbol duration and $\rect(.)$ is the rectangular pulse-shaping function.
To this signal, another interfering PAM-$M$ signal is added $E_\textup{int}(t)=\sqrt{x_\textup{int}(t)}$. This signal is
scaled by the Signal to Interference power Ratio (SIR), and multiplied by a complex exponential that represents the difference
(in phase and frequency) between the two lasers. Therefore, the electric field immediately before the photodiode is:
\begin{equation}
E_\textup{rx}(t) = \sqrt{x_\textup{sig}(t)} + \sqrt{\frac{x_\textup{int}(t)}{\mathrm{SIR}}}e^{j2\pi\Delta f t}e^{j\phi(t)}
\end{equation}
$\Delta f$ is the frequency separation between the lasers, and $\phi(t)$ represents phase noise.
Without loss of generality, photodiode responsivity is set to $1$. Therefore, the photocurrent is:
\begin{equation}
i(t) = x_\textup{sig}(t) + \frac{x_\textup{int}(t)}{\mathrm{SIR}} +
2\sqrt{\frac{x_\textup{int}(t) x_\textup{sig}(t)}{\mathrm{SIR}}}\cos[2\pi\Delta f t + \phi(t)]
\end{equation}
Then, \eqref{eq:paminterf} is obtained by filtering this signal with the matched filter and sampling every $T$ seconds.
The matched filter for a rectangular pulse is simply an integral over $T$:
\begin{equation}
y(kT) = \frac{1}{T}\int_{ (k-1)T }^{kT} i(t)\diff t =
 a_k + \frac{b_k}{\mathrm{SIR}} + 2\sqrt{\frac{a_k b_k}{\mathrm{SIR}}}\cos(\phi)\frac{\sin(\pi\Delta f T)}{\pi\Delta f T}
\end{equation}

\chapter{Non-square QAM constellations}\label{app:qamconst}
\begin{table}[h]
	\centering
	\begin{tabular}{c c c}
		\toprule
Bit mapping (decimal) & Real part & Imaginary Part \\
		\midrule
0 & -3 & -5 \\ 
1 & -1 & -5 \\ 
2 & -3 & 5 \\ 
3 & -1 & 5 \\ 
4 & -5 & -3 \\ 
5 & -5 & -1 \\ 
6 & -5 & 3 \\ 
7 & -5 & 1 \\ 
8 & -1 & -3 \\ 
9 & -1 & -1 \\ 
10 & -1 & 3 \\ 
11 & -1 & 1 \\ 
12 & -3 & -3 \\ 
13 & -3 & -1 \\ 
14 & -3 & 3 \\ 
15 & -3 & 1 \\ 
16 & 3 & -5 \\ 
17 & 1 & -5 \\ 
18 & 3 & 5 \\ 
19 & 1 & 5 \\ 
20 & 5 & -3 \\ 
21 & 5 & -1 \\ 
22 & 5 & 3 \\ 
23 & 5 & 1 \\ 
24 & 1 & -3 \\ 
25 & 1 & -1 \\ 
26 & 1 & 3 \\ 
27 & 1 & 1 \\ 
28 & 3 & -3 \\ 
29 & 3 & -1 \\ 
30 & 3 & 3 \\ 
31 & 3 & 1 \\ 
		\bottomrule
	\end{tabular}
	\caption{Constellation and bit mapping of 32-QAM.}\label{tab:32qammapping}
\end{table}

\begin{table}[h]
	\centering
	\begin{tabular}{c c c}
		\toprule
		Bit mapping (decimal) & Real part & Imaginary Part \\
		\midrule
0 & 1.426 & 0.359 \\ 
1 & -1.426 & 0.359 \\ 
2 & 1.426 & -0.359 \\ 
3 & -1.426 & -0.359 \\ 
4 & 0.163 & 0.214 \\ 
5 & -0.163 & 0.214 \\ 
6 & 0.163 & -0.214 \\ 
7 & -0.163 & -0.214 \\ 
8 & 0.938 & 0.240 \\ 
9 & -0.938 & 0.240 \\ 
10 & 0.938 & -0.240 \\ 
11 & -0.938 & -0.240 \\ 
12 & 0.536 & 0.223 \\ 
13 & -0.536 & 0.223 \\ 
14 & 0.536 & -0.223 \\ 
15 & -0.536 & -0.223 \\ 
16 & 0.324 & 1.247 \\ 
17 & -0.324 & 1.247 \\ 
18 & 0.324 & -1.247 \\ 
19 & -0.324 & -1.247 \\ 
20 & 0.174 & 0.706 \\ 
21 & -0.174 & 0.706 \\ 
22 & 0.174 & -0.706 \\ 
23 & -0.174 & -0.706 \\ 
24 & 0.885 & 0.926 \\ 
25 & -0.885 & 0.926 \\ 
26 & 0.885 & -0.926 \\ 
27 & -0.885 & -0.926 \\ 
28 & 0.520 & 0.625 \\ 
29 & -0.520 & 0.625 \\ 
30 & 0.520 & -0.625 \\ 
31 & -0.520 & -0.625 \\ 
		\bottomrule
	\end{tabular}
	\caption{Constellation and bit mapping of GS-32-QAM.}\label{tab:gs32qammapping}
\end{table}

\backmatter%

\printnomencl

\printbibliography[heading=bibintoc]

\printindex

\end{document}